\documentclass{aa}  
\usepackage{graphicx}
\usepackage{comment}
\usepackage{subcaption}
\usepackage{xcolor}
\usepackage{dblfloatfix}
\usepackage{txfonts}
\newcommand{\DD}    {\mathcal{D}}
\newcommand{\dd}    {{\text d}}
\newcommand{\LL}    {\mathcal{L}} 
\newcommand{\btheta}{{\boldsymbol \theta}}
\newcommand{\xk}    {x_k}
\newcommand{\dxk}   {\delta x_k}
\newcommand{\yk}    {y_k}
\newcommand{\dyk}   {\delta y_k}
\newcommand{\sigmak}{\sigma_k}
\newcommand{\ohhdp}{$o$-H$_2$D$^+$\:}
\newcommand{\X}{$X$($o$-H$_2$D$^+$)\:}

\newcommand{\logten}{\log_{10}}
\newcommand{\dGC}   {D_{\rm GC}}
\newcommand{\Reff}  {R_{\rm eff}}
\newcommand{\Mclump}{M_{\rm clump}}
\newcommand{\Lbol}  {L_{\rm bol}}
\newcommand{\Tdust} {T_{\rm dust}}

\begin{document} 

   \title{Survey of ortho-H$_2$D$^+$ in high-mass star-forming regions}
   %OR(why not?
   %\title{\ohhdp and cosmic-ray ionization rate in a sample of high-mass clumps}
   %\title{Estimates of the cosmic-ray ionization rate from \ohhdp observations}
   % \subtitle{I. H$_2$D$^+$ as new confirmed chemical-clock}
   \author{G.~Sabatini
          \inst{1,2,3}\fnmsep\thanks{Marco Polo fellow -- University of Bologna.}
          \and
          S.~Bovino\inst{2}
          \and
          A.~Giannetti\inst{3} 
          \and
          F.~Wyrowski\inst{4}
          \and
          M.~A.~\'Ordenes\inst{2} 
          \and
          R.~Pascale\inst{5}
          \and
          T.~Pillai\inst{6}
          \and 
          M.~Wienen\inst{4}
          \and 
          T.~Csengeri\inst{7}
          \and
          K.~M.~Menten\inst{4}
          }

      \institute{Dipartimento di Fisica e Astronomia, Universit\'a degli Studi di Bologna, Via Gobetti 93/2, I-40129 Bologna, Italy\\
          \email{giovannisabatini4@unibo.it}
         \and
         Departamento de Astronom\'ia, Facultad Ciencias F\'isicas y Matem\'aticas, Universidad de Concepci\'on, Av. Esteban Iturra s/n Barrio Universitario, Casilla 160, Concepci\'on, Chile
        %$\email{}
         \and
         INAF - Istituto di Radioastronomia - Italian node of the ALMA Regional Centre (It-ARC), Via Gobetti 101, I-40129 Bologna, Italy
         \and 
         Max-Planck-Institut f\"ur Radioastronomie, Auf dem H\"ugel, 69, 53121, Bonn, Germany         
         \and 
         INAF - Osservatorio di Astrofisica e Scienza dello Spazio di Bologna, Via Gobetti 93/3, I-40129 Bologna, Italy
     % \email{wuchterl@amok.ast.univie.ac.at}
         \and
         Institute for Astrophysical Research, 725 Commonwealth Ave, Boston University Boston, MA 02215, USA
         \and 
         Laboratoire d'Astrophysique de Bordeaux, Univ. Bordeaux, CNRS, B18N, all\'ee Geoffroy Saint-Hilaire, 33615 Pessac, France
         \\
         %    %%\email{c.ptolemy@hipparch.uheaven.%space}
        %   %  \thanks{The university of %heaven temporarily does not
        %             accept e-mails}
             }

   \date{Received July 23, 2020; accepted September 22, 2020}

% \abstract{}{}{}{}{} 
% 5 {} token are mandatory
 
  \abstract
   {Deuteration has been suggested to be a reliable chemical clock of star-forming regions due to its strong dependence on density and temperature changes during cloud contraction. In particular, the H$_3^+$ isotopologues (e.g. ortho-H$_2$D$^+$) seem to act as good proxies of the evolutionary stages of the star formation process. While this has been widely explored in low-mass star-forming regions, in the high-mass counterparts only a few studies have been pursued, and  the reliability of deuteration as a chemical clock remains inconclusive.}
   {We present a large sample of \ohhdp observations in high-mass star-forming regions and discuss possible empirical correlations with relevant physical quantities to assess its role as a chronometer of star-forming regions through different evolutionary stages.} 
   {APEX observations of the ground-state transition of \ohhdp were analysed in a large sample of high-mass clumps selected from the ATLASGAL survey at different evolutionary stages. Column densities and beam-averaged abundances of \ohhdp with respect to H$_2$, \X, were obtained by modelling the spectra under the assumption of local thermodynamic equilibrium.}
   {We detect 16 sources in \ohhdp and find clear correlations between \X and the clump bolometric luminosity and the dust temperature, while only a mild correlation is found with the CO-depletion factor. In addition, we see a clear correlation with the luminosity-to-mass ratio, which is known to trace the evolution of the star formation process. This would indicate that the deuterated forms of H$_3^+$ are more abundant in the very early stages of the star formation process and that deuteration is influenced by the time evolution of the clumps. In this respect, our findings would suggest that the \X abundance is mainly affected by the thermal changes rather than density changes in the gas. We have employed these findings together with observations of H$^{13}$CO$^+$, DCO$^+$, and C$^{17}$O to provide an estimate of the cosmic-ray ionisation rate in a sub-sample of eight clumps based on recent analytical work.}
   {Our study presents the largest sample of \ohhdp in star-forming regions to date. The results confirm that the deuteration process is strongly affected by temperature and suggests that \ohhdp can be considered  a reliable chemical clock during the star formation processes, as proved by its strong temporal dependence.}

   \keywords{Astrochemistry --
            Star: formation --
            ISM: molecules --
            Molecular processes
               }
   \maketitle
%and
%-------------------------------------------------------------------

\section{Introduction}

    Crucial to the study of star formation is the measurement of timescales, which may be mass- and density-dependent (e.g. \citealt{Urquhart18}; see also \citealt{Motte18} for a review). The duration of the star formation process is crucial to distinguishing between competing theories, which predict a slow or a fast evolution towards the formation of stars (e.g. \citealt{Mouschovias06} and \citealt{Hartmann12}), estimating its fundamental energy budget, and  understanding, and thus  predicting, its impact on the chemical enrichment of the interstellar medium (ISM).
    
    Chemistry is a powerful tool for infering how long the star-forming gas remains dense and cold, and thus is capable of actually forming stars. In addition, chemical signatures can help to characterise evolved sources, due to the change in temperature and density, and their effects on the chemistry.

    In the last few decades, among the chemical processes that occur under the typical conditions prevalent in Dark Clouds, the study of freeze-out (e.g. \citealt{Caselli98}, \citealt{Chen10}, \citealt{Hernandez11}, \citealt{Giannetti14}, \citealt{Feng19} and \citealt{Bovino19}) and deuterium fractionation processes (e.g. \citealt{Millar89}, \citealt{Pagani92_model,Pagani92_obs}, \citealt{Ceccarelli07}, \citealt{Chen10}, \citealt{Sipila10}, \citealt{Fontani11}, \citealt{Kong15} and \citealt{Koertgen18}) have certainly played a central role in astrochemical research.
    Especially during the first phases of star formation, the volume density of molecular hydrogen, $n$(H$_2$), and the temperature of the clouds, $T$, reach the ideal conditions (i.e. $n$(H$_2$) $>$ few $\times 10^4$ cm$^{-3}$ and  $T < 20$ K) to favour freeze-out (or depletion), by which several C-, N-, and O-bearing species,  in particular carbon monoxide (CO), are efficiently removed from the gas phase and trapped on the surface of the dust grains (e.g. \citealt{Kramer99}, \citealt{Bergin02}, \citealt{Caselli08}, \citealt{Fontani12}, \citealt{Wiles16}, and \citealt{Sabatini19}; see also \citealt{BerginTafalla07} for a review). 
    The CO {adsorption} onto the grain surfaces also favours the progressive enrichment of deuterium atoms compared to hydrogen in the molecules, through the   deuterium fractionation process (e.g. \citealt{Ceccarelli14}). In this context, the main driver of deuteration,  the trihydrogen cation H$_3^+$,  leads to the enrichment of deuterated molecules like H$_2$D$^+$ and N$_2$D$^+$ in the gas phase (e.g. \citealt{Pagani09}, \citealt{Vastel12}, and \citealt{Caselli19}).
    
    Using cold-gas chemistry to measure the age of low-mass cores (i.e. M $<8$ M$_\odot$) has been exploited via different tracers (e.g. via H$_2$D$^+$ by \citealt{Brunken14}), while in the high-mass regime age estimates are much more complex, in particular due to the uncertainties in the dynamical and chemical initial conditions.
   
    A first attempt to adopt this kind of analysis to the high-mass regime was carried out by \cite{Kong16} for cores of $\sim 15-60$ M$_{\odot}$ by measuring the deuterium fraction via N$_2$H$^+$. However, it was shown by \citet{Pillai12} that this species may not be ideal for tracing the first stages of the high-mass star formation process. They obtained maps for part of the DR21 complex of the $ortho-$H$_2$D$^+$ $J_{\rm {K_a, K_c}} = 1_{10} - 1_{11}$ (hereafter \mbox{\ohhdp})  and N$_2$D$^+$ $J = 3-2$ transitions, observed with the {\it James Clerk Maxwell Telescope} (JCMT; \citealt{Holland99}) and the {\it Submillimiter Array} (SMA; \citealt{Ho04}), respectively. The data reveal very extended \ohhdp emission, in agreement with the results of \cite{Vastel06} for the low-mass source L1544, and that this species mainly traces gas that is not seen in dust continuum emission or in the interferometric N$_2$D$^+$ data. H$_2$D$^+$ may thus be sensitive to gas that eludes detection in the most commonly used tracers, and can represent an even earlier stage in the process of star formation.

    In a pilot study to this work, with the {\it Atacama Pathfinder EXperiment 12-meter submillimeter telescope} (APEX; \citealt{Gusten06}),  \cite{Giannetti19}  detected \ohhdp and N$_2$D$^+$ in three high-mass star-forming clumps for the first time, opening the possibility of investigating their abundance variation in sources at different evolutionary stages (G351.77–0.51 complex; e.g. \citealt{Leurini19}). They considered sources ranging from a clump that is still quiescent at 70 $\mu$m to one that hosts luminous young stellar objects (YSOs), adopting the classification from \cite{Giannetti14}, \cite{Konig17}, and \cite{Urquhart18}.
    \cite{Giannetti19} observed that the abundance of N$_2$D$^+$ progressively increases with evolution, showing a difference of a factor of $\sim 2$ between the least and the most evolved clump. Particularly relevant is also that the \ohhdp abundance decreases in the two most evolved clumps (by a factor of $\sim 10$), likely due to the chemical conversion of H$_2$D$^+$ into D$_2$H$^+$ and D$^+_3$ (which would also boost the production of N$_2$D$^+$) or to the destruction by desorbed CO in the presence of a protostellar object.
    Similar results were also reported by \cite{Kong16}, who obtained upper limits for abundances for \ohhdp of $\sim 10^{-11}$ in two high-mass star-forming regions associated with outflow activity (see \citealt{Tan16}), and high abundances of N$_2$D$^+$, confirming the fact that N$_2$D$^+$ is forming at later times compared to \ohhdp and under different physical conditions.
    
    Attempts to demonstrate this hypothesis have been recently reported by \cite{Miettinen20}, using APEX observations in three prestellar and three protostellar cores in Orion-B9. Although observing a downward trend in \ohhdp  abundance of about a factor of 4 with evolution, the anti-correlation with N$_2$D$^+$ was not confirmed. However, as pointed out by the authors, their spatial offset between the N$_2$D$^+$ and the \ohhdp observations ($\sim 10''$ on average) might be the cause of the observed behaviour (i.e. the two molecules sample different regions).
    
    With the present work we  extend the number of high-mass star-forming regions with \ohhdp detections, assembling a sample of 106 massive clumps selected from several spectral line surveys carried out on clumps as part of the ATLASGAL survey (see Sect.~\ref{sec2:sample} for  details). Our goal is to test the trend reported by \cite{Giannetti19} for \textit{o}-H$_2$D$^+$, on a sample that could be statistically representative of regions with ongoing massive star formation activity in order to exclude possible bias induced by particular initial chemical conditions that may have affected previous (much smaller) samples.
    
    This paper has the following structure{: w}e describe the sample and its selection in Sect.~\ref{sec2:sample}. In Sect.~\ref{sec3:data} we discuss how the spectra have been obtained and reduced, and we summarise the physical properties of the sources in our sample. The information on how the column densities and the dynamical parameters of the sources have been calculated is presented in Sect.~\ref{sec4:analisys}. In Sect.~\ref{sec5:results} we present the comprehensive set of correlations we found between the \ohhdp abundance and the other physical quantities, and we present the  new estimates for the cosmic-ray ionisation rate (CRIR) for hydrogen molecules, based on the same data. Following the evolutionary stages of massive clumps, in Sect.~\ref{sec6:dicussion} we discuss how \ohhdp can be used as a powerful chemical clock to follow the evolution of high-mass star-forming regions. Finally, in Sect.~\ref{sec7:conclusion} we summarise our conclusions.

\section{Sample}\label{sec2:sample}
            \begin{table*}
        \caption{\label{tab:obsprop}Summary of the physical
properties of the ATLASGAL sources  in our sample.}
        \setlength{\tabcolsep}{4.5pt}
    \renewcommand{\arraystretch}{1.1}
    \centering
        \begin{tabular}{l|cccccccccccc}
        \hline\hline
    ATLASGAL-ID&$d_{\odot}$&D$_{GC}$&R$_{eff}$\tablefootmark{b}&$T_{dust}$&$\Delta T_{dust}$\tablefootmark{c}&v$_{lsr}$&$M_{clump}$&$L_{bol}$&N(H$_2$)&n(H$_2$)&Class\tablefootmark{b,c}\\
          &(kpc)&(kpc)&(pc)&(K)&(K)&(km s$^{-1}$)&($10^2 M_{\odot}$)&($10^2 L_{\odot}$)&log$_{10}$(cm$^{-2}$)&log$_{10}$(cm$^{-3}$)& \\
    \hline                                                                          

    G08.71--0.41\tablefootmark{a}  &4.8& 4.0 & 1.0 &11.8&0.3& 39.4 &16.6&  5.0&22.8& 4.0 & IRw\\
    G13.18+0.06\tablefootmark{a}   &2.4& 5.9 & 0.5 &24.2&0.8& 49.9 & 3.7& 83.2&22.9& 4.3 & 70w\\
        G14.11--0.57\tablefootmark{a}  &2.6& 6.9 & 0.5 &22.4&0.8& 20.8 & 3.5& 31.8&22.9& 4.4 & IRw\\
        G14.49--0.14\tablefootmark{a}  &3.9& 5.4 & 0.8 &12.4&0.4& 39.5 &19.2&  7.5&23.1& 4.4 & 70w\\
        G14.63--0.58\tablefootmark{a}  &1.8& 6.9 & 0.4 &22.5&0.4& 18.5 & 2.5& 27.8&23.0& 4.6 & IRw\\
        G18.61--0.07\tablefootmark{a}  &4.3& 5.3 & 0.7 &13.8&0.3& 46.6 & 8.8&  5.9&22.8& 4.2 & IRw\\
        G19.88--0.54\tablefootmark{a}  &3.7& 5.3 & 0.6 &24.2&1.4& 44.9 & 8.0&124.0&23.1& 4.5 & IRb\\
        G28.56--0.24\tablefootmark{a}  &5.5& 4.8 & 1.3 &11.7&0.1& 87.3 &54.1& 17.7&23.1& 4.2 & IRw\\
        G333.66+0.06\tablefootmark{a}  &5.3& 4.5 & 1.1 &17.8&0.3& -85.1&14.2& 42.7&22.7& 3.9 & 70w\\
        G351.57+0.76\tablefootmark{a}  &1.3& 7.0 & 0.3 &17.0&0.1& -3.2 & 1.6&  4.3&22.7& 4.4 & 70w\\
        G354.95--0.54\tablefootmark{a} &1.9& 7.4 & 0.4 &19.1&1.3& -5.4 & 1.5&  4.8&22.6& 4.2 & 70w\\
    \hline                                
        G12.50-0.22\tablefootmark{b}   &2.6& 5.9 &0.4&13.0&0.2& 35.5  & 1.2&  1.6&22.8& 4.4 & 70w\\  
        G14.23--0.51\tablefootmark{b}  &1.5& 6.9 &0.5&17.2&2.7& 19.5  & 7.2& 14.8&23.3& 4.8 & HII \\   
        G15.72--0.59\tablefootmark{b}  &1.8& 6.6 &0.2&12.1&0.5& 17.8  & 1.7&  0.3&22.8& 4.7 & IRw\\   
        G316.76-0.01\tablefootmark{b}  &2.5& 6.7 &0.2&18.9&0.4& -39.9 & 4.7& 24.2&23.1& 5.0 & IRb \\
        G351.77-CL7 &1.0& 7.4 & 0.1\tablefootmark{d}&13.0&0.3\tablefootmark{e}&-3.2\tablefootmark{d}& 1.2&  0.2&22.9& 5.2 & 70w \\

    \hline
        \end{tabular}
        \tablefoot{Physical properties (observed and derived) of our sample. {\it Top row}: TOP100 sources; {\it Bottom row}: ATLASGAL sources not in the TOP100; {T$_{dust}$ and $L_{bol}$ are obtained from a SED fit described in Appendix D in \citealt{Urquhart18}, while their uncertainties are derived as discussed in Sect.~\ref{sec2:sample}; R$_{eff}$ values are evaluated from the dust continuum at 870 $\mu$m, while N(H$_2$) and $M_{clump}$ come from the 870 $\mu$m flux peaks and the integrated flux density over R$_{eff}$, respectively; n(H$_2$) values are lower limits derived here as n(H$_2$) = N(H$_2$)/2R$_{eff}$;}\\
        \tablefoottext{a}{$d_\odot$, M$_{clump}$, $L_{bol}$, and N(H$_2$) from \citet{Giannetti17_june}; R$_{eff}$ and $T_{dust}$ from \cite{Konig17}; D$_{GC}$ from \citet{Urquhart18}; v$_{lsr}$ from \cite{Giannetti14} and derived from the C$^{17}$O $J = 3-2$;}
        \tablefoottext{b}{distances, R$_{eff}$, v$_{lsr}$, $M_{clump}$, $L_{bol}$, N(H$_2$), and classification from \citet{Urquhart18}: the error associated with N(H$_2$) is 20\% for each source; v$_{lsr}$ values are derived from the NH$_3$ ($1,1$) data published in \cite{Wienen12};}
        \tablefoottext{c}{Classification scheme from \citet{Konig17};}
        \tablefoottext{d}{R$_{eff}$ and v$_{lsr}$ are  from \citet{Leurini11}; v$_{lsr}$ is derived from the C$^{18}$O $J = 2-1$;}
        \tablefoottext{e}{T$_{dust}$ standard deviation computed on the clump region defined in \citet{Leurini19}.}}
        \end{table*}
        
   \begin{figure*}
   \centering
   \includegraphics[width=0.435\hsize]{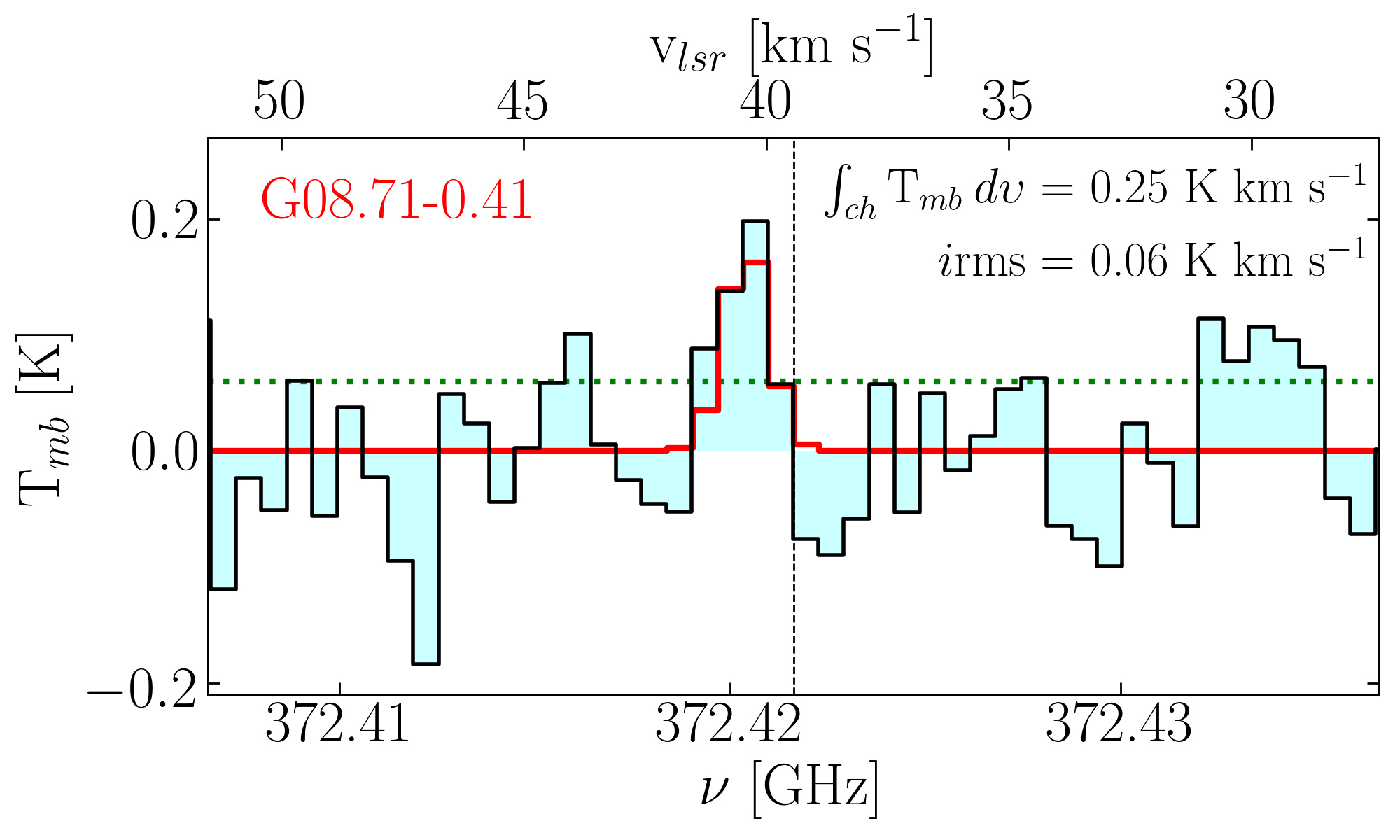}\quad\quad
   \includegraphics[width=0.435\hsize]{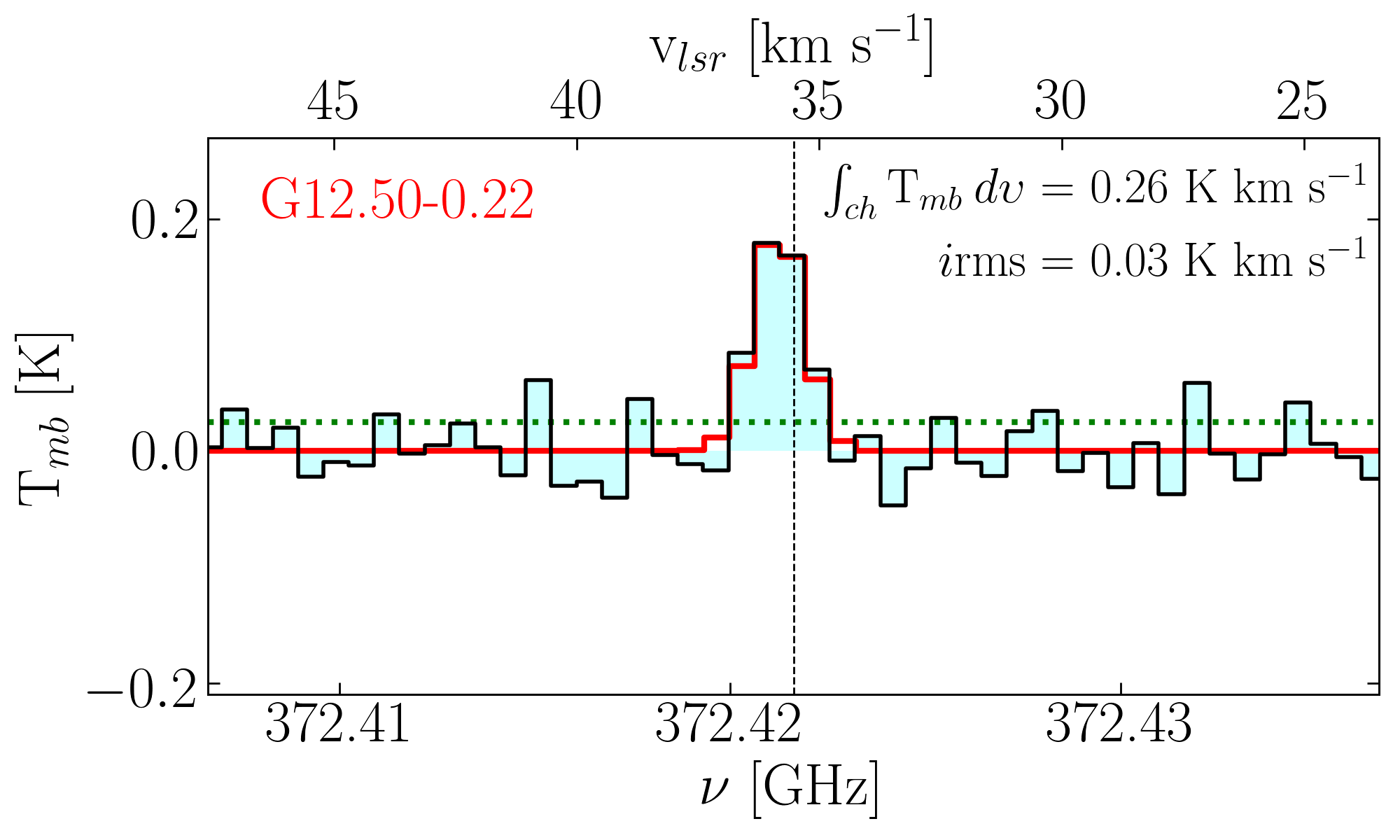}
   \includegraphics[width=0.435\hsize]{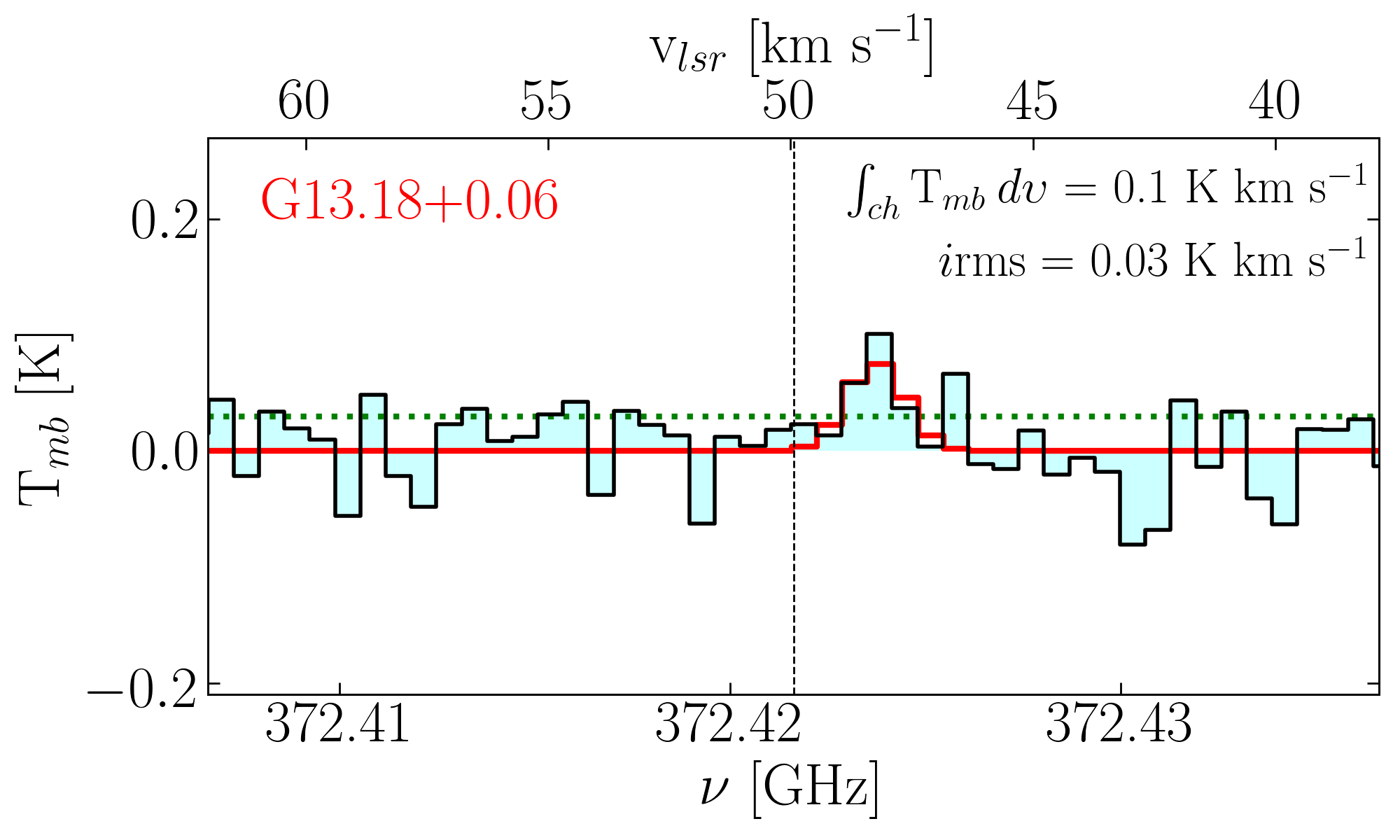}\quad\quad
   \includegraphics[width=0.435\hsize]{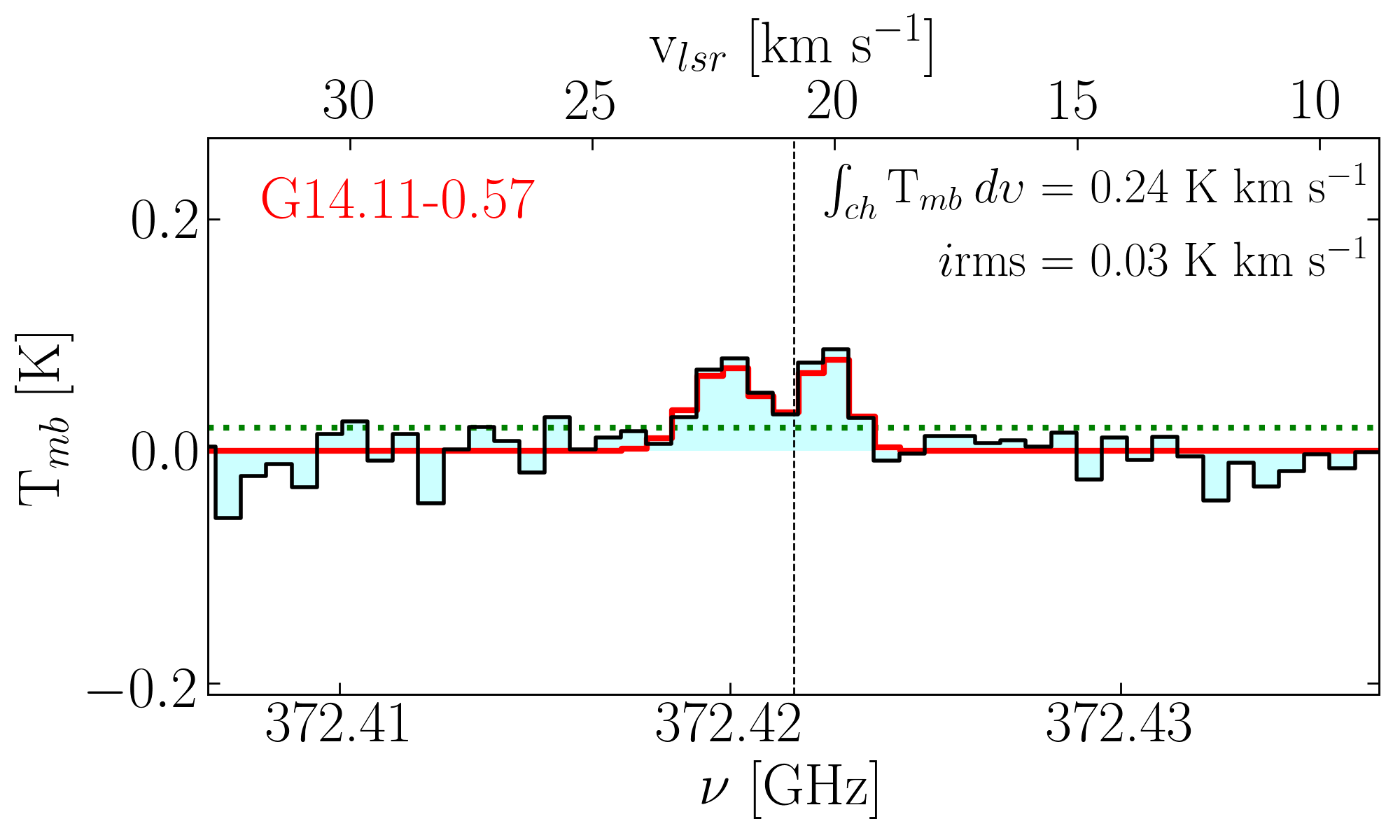}
   \includegraphics[width=0.435\hsize]{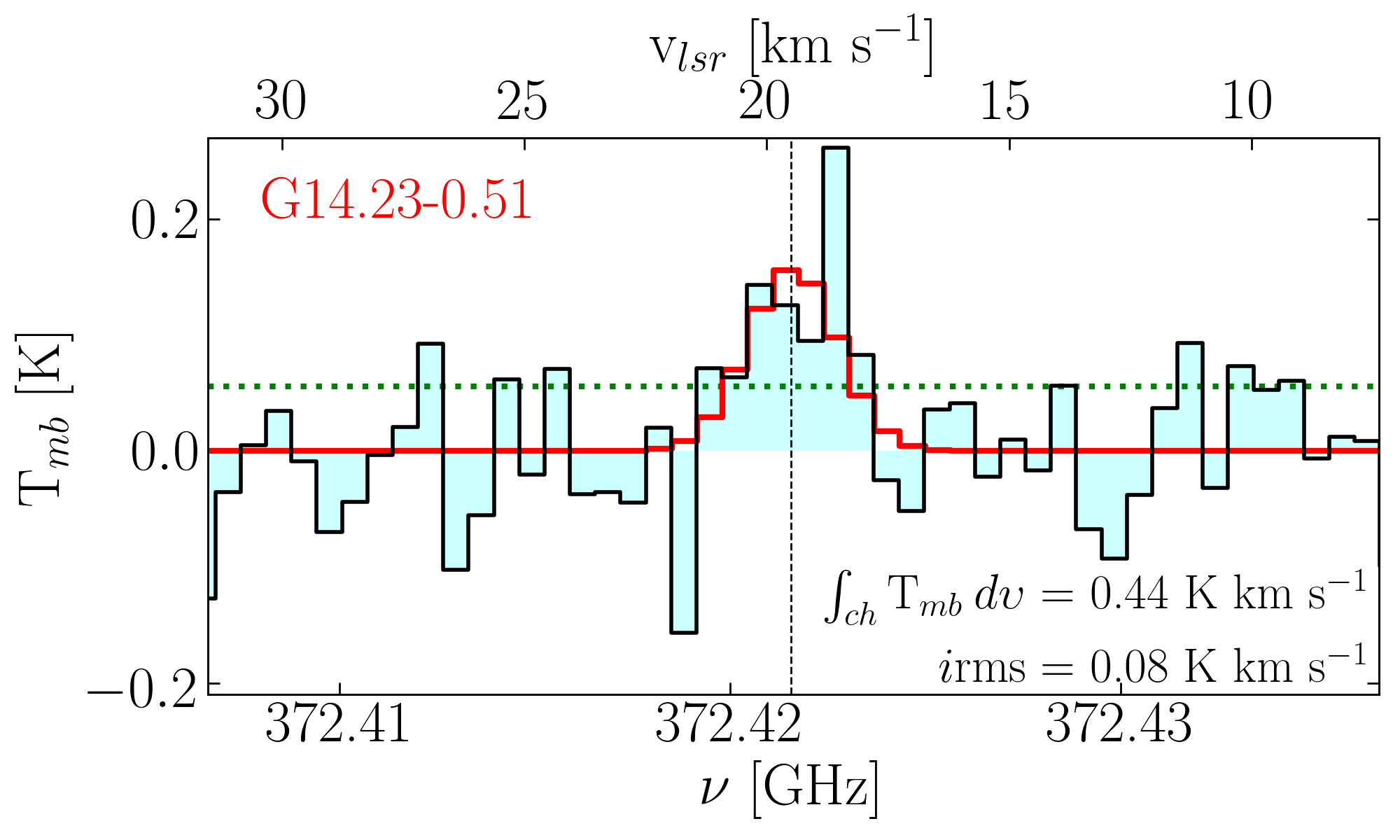}\quad\quad
   \includegraphics[width=0.435\hsize]{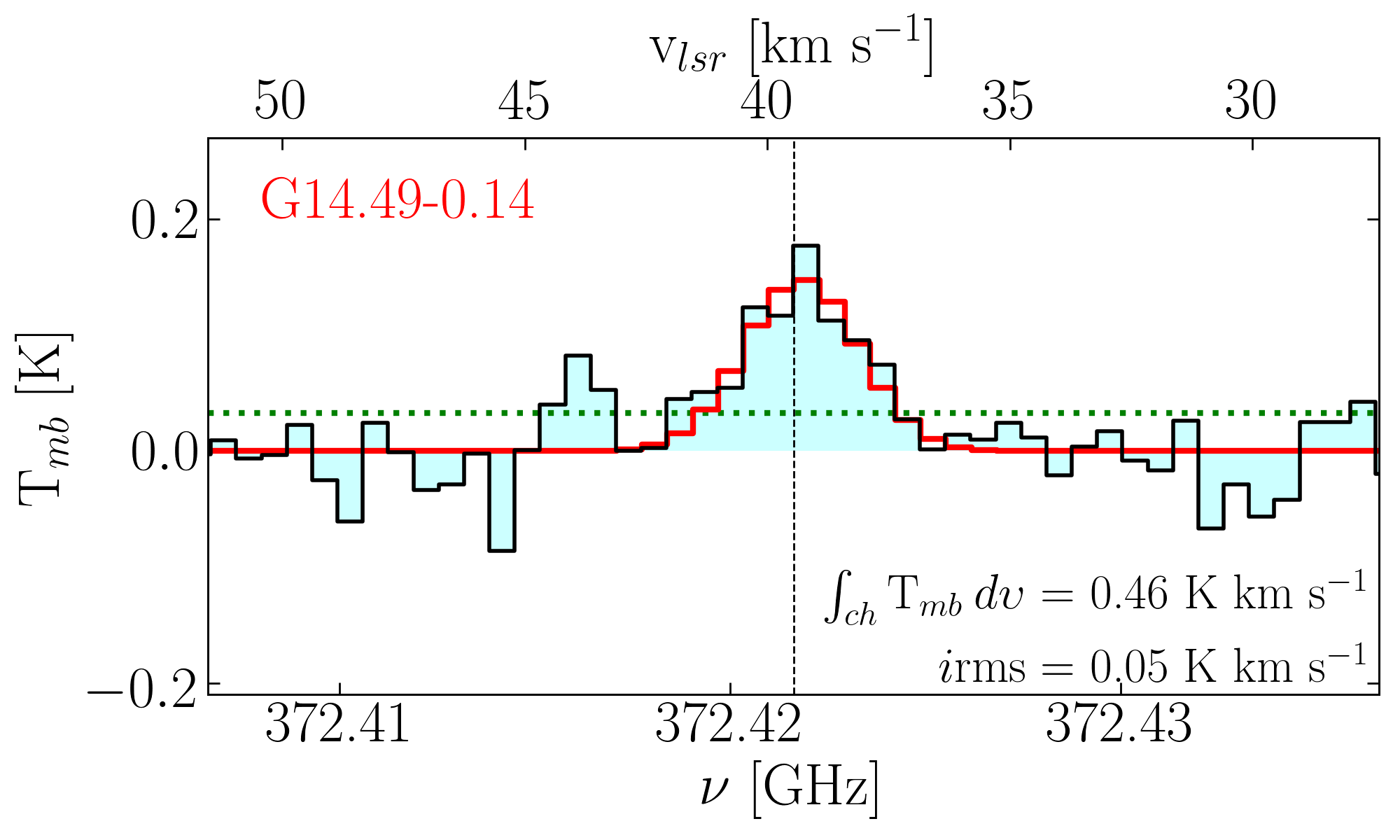}
   \includegraphics[width=0.435\hsize]{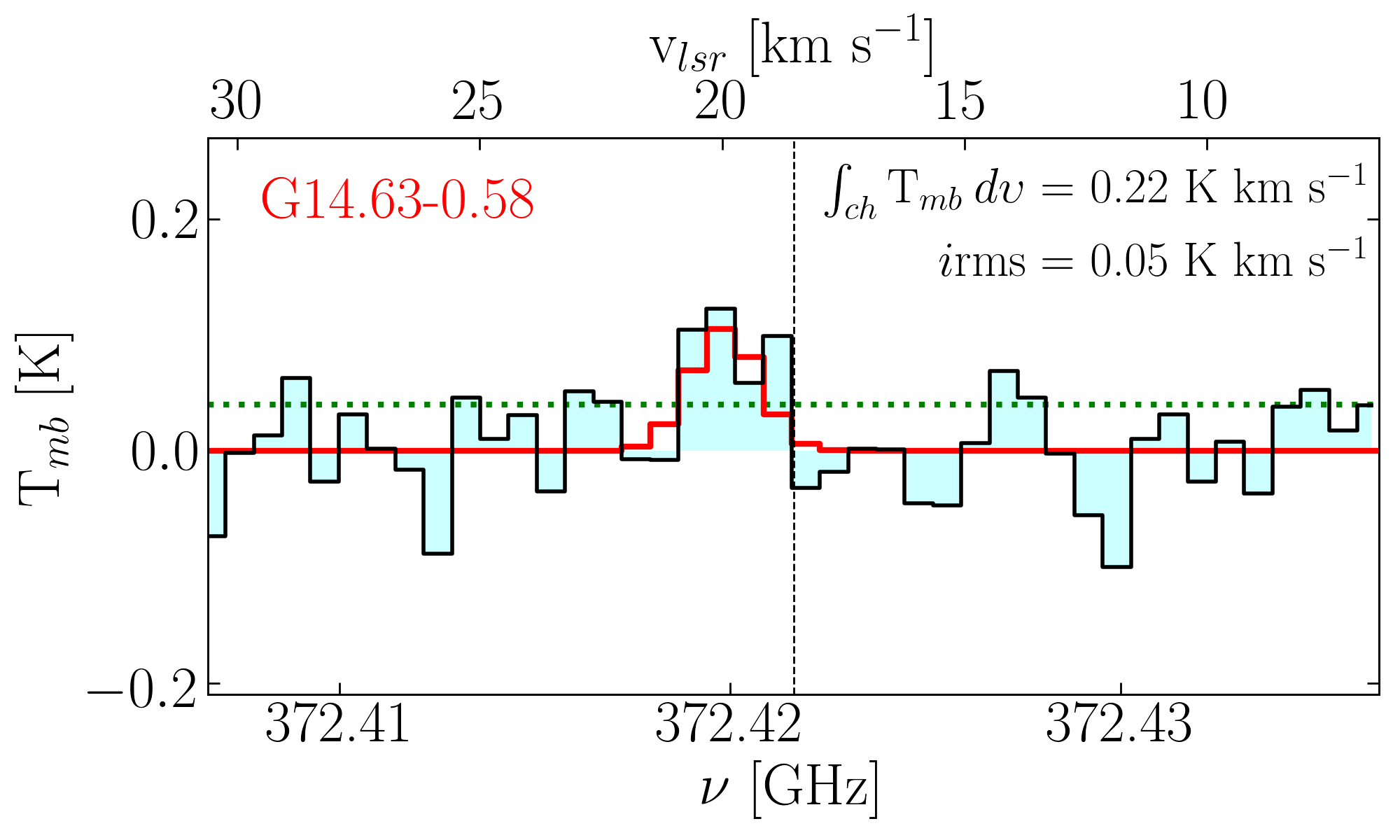}\quad\quad
   \includegraphics[width=0.435\hsize]{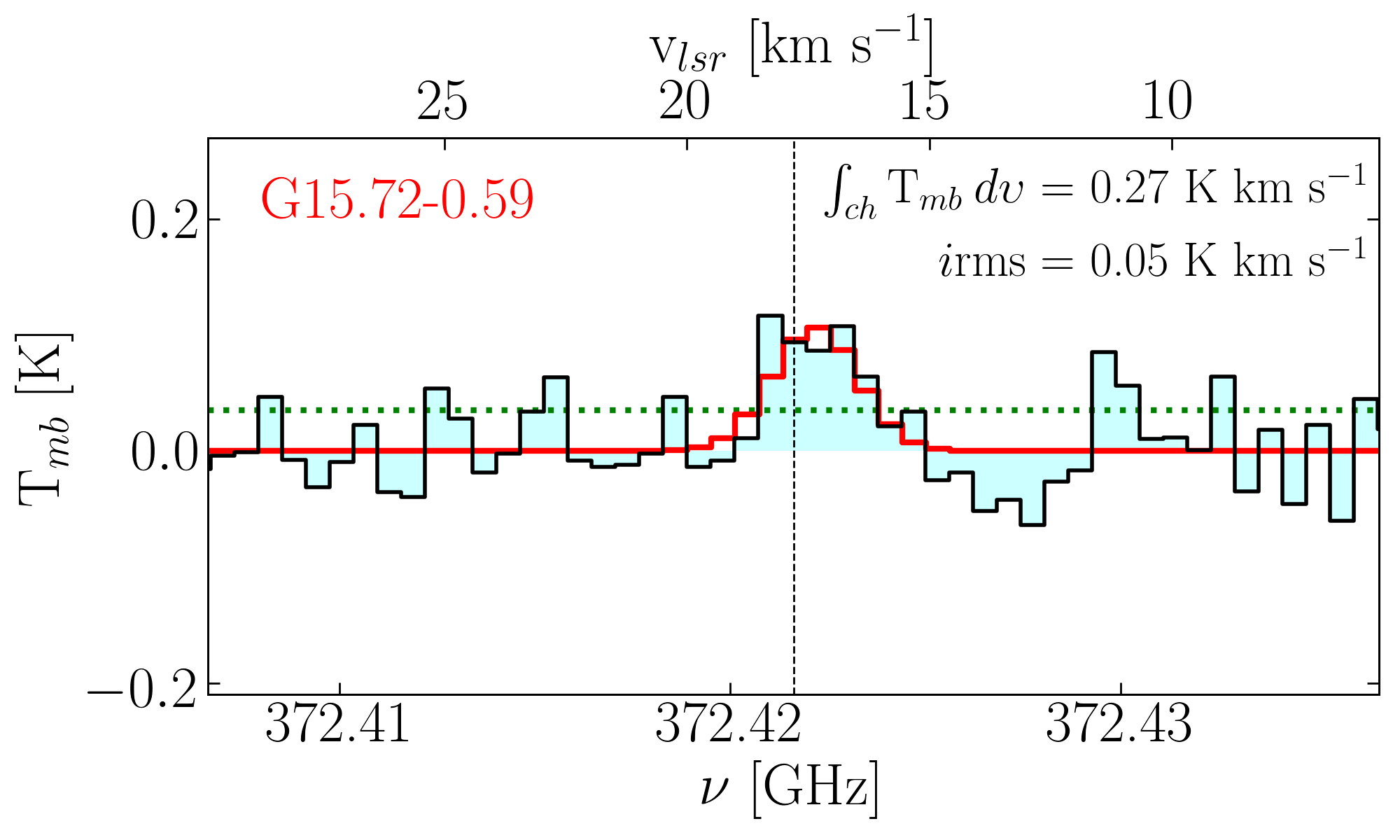}

   \caption{\ohhdp spectra for half the sources of our sample (cyan). In each panel the source name is shown on the left in red, while the integrated main-beam temperature, $\int_{ch} T_{mb}\:d\upsilon$, and rms ($i$rms) are  on the right in black. The green dotted line represents the 1$\sigma$ noise levels in T$_{mb}$. The frequency and velocity axes are reported and cut around the line, while the intensity axis is fixed to the highest T$_{mb}$ detected (observed in G14.23-0.51). The black dashed lines indicate the v$_{lsr}$ of each source derived from the C$^{17}$O $J = 3-2$ and the NH$_3$ ($1,1$) lines (see Table~\ref{tab:obsprop}). The model obtained with MCWeed is superimposed in red.\label{fig:H2DpspectraA}}%
    \end{figure*}
    \begin{figure*}
   \centering
   \includegraphics[width=0.435\hsize]{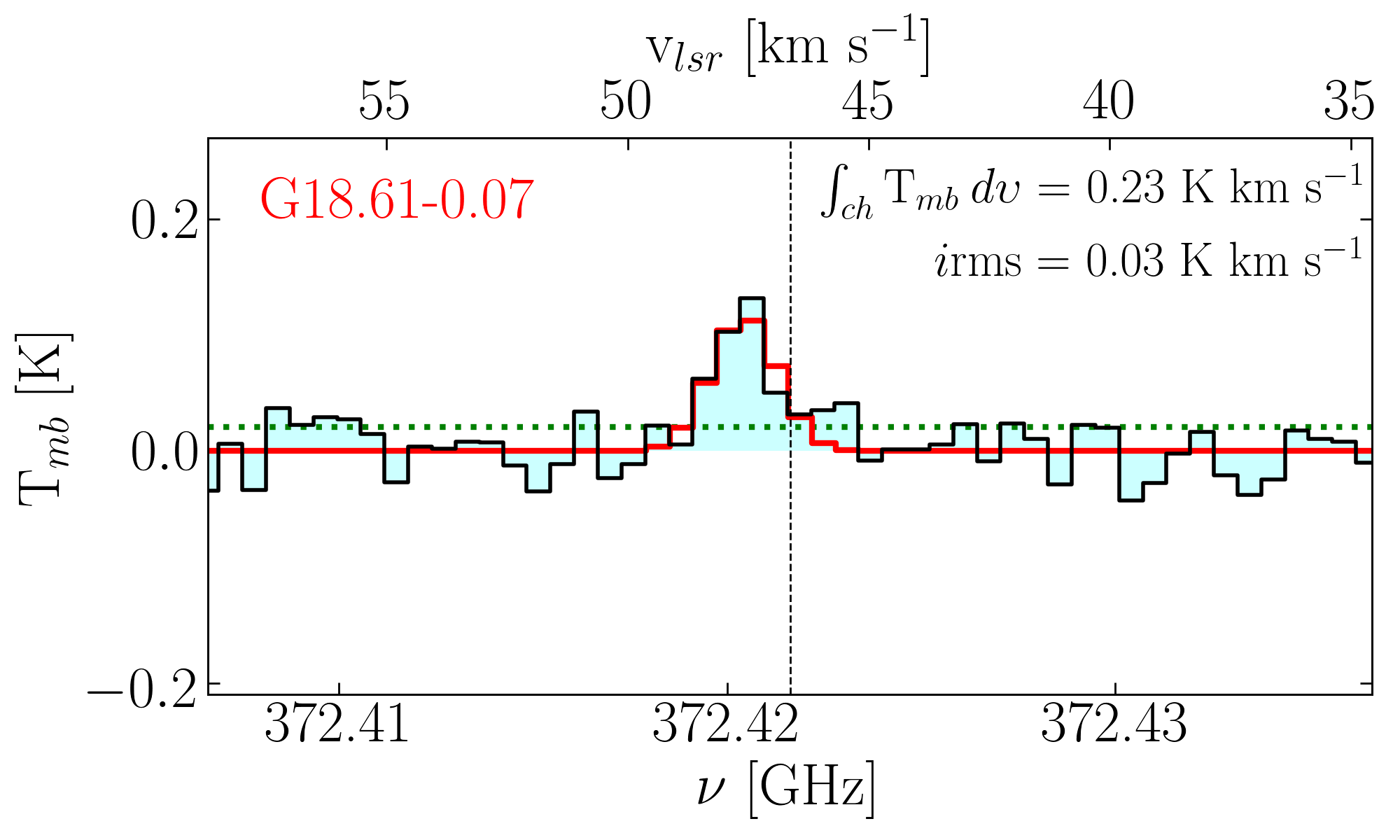}\quad\quad
   \includegraphics[width=0.435\hsize]{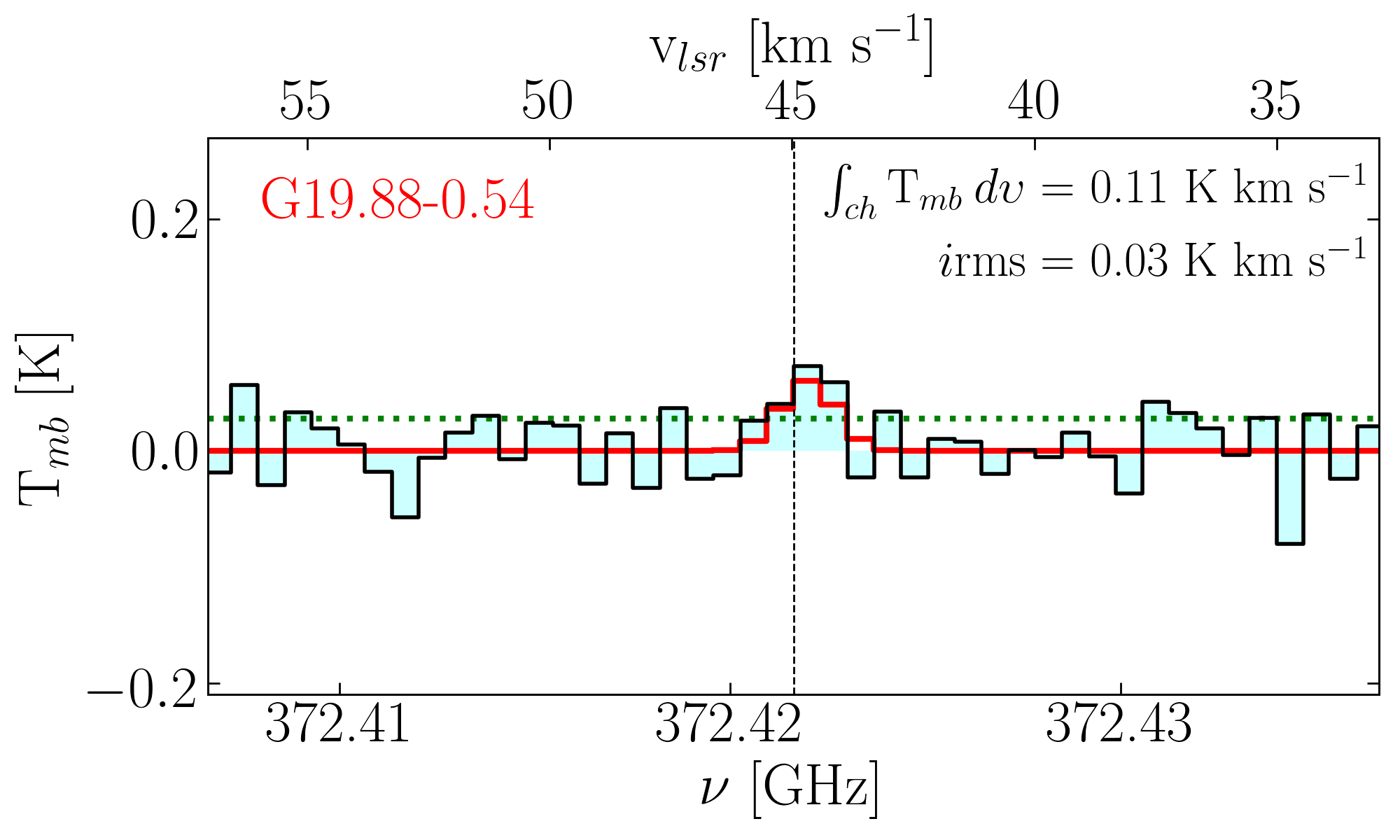}
   \includegraphics[width=0.435\hsize]{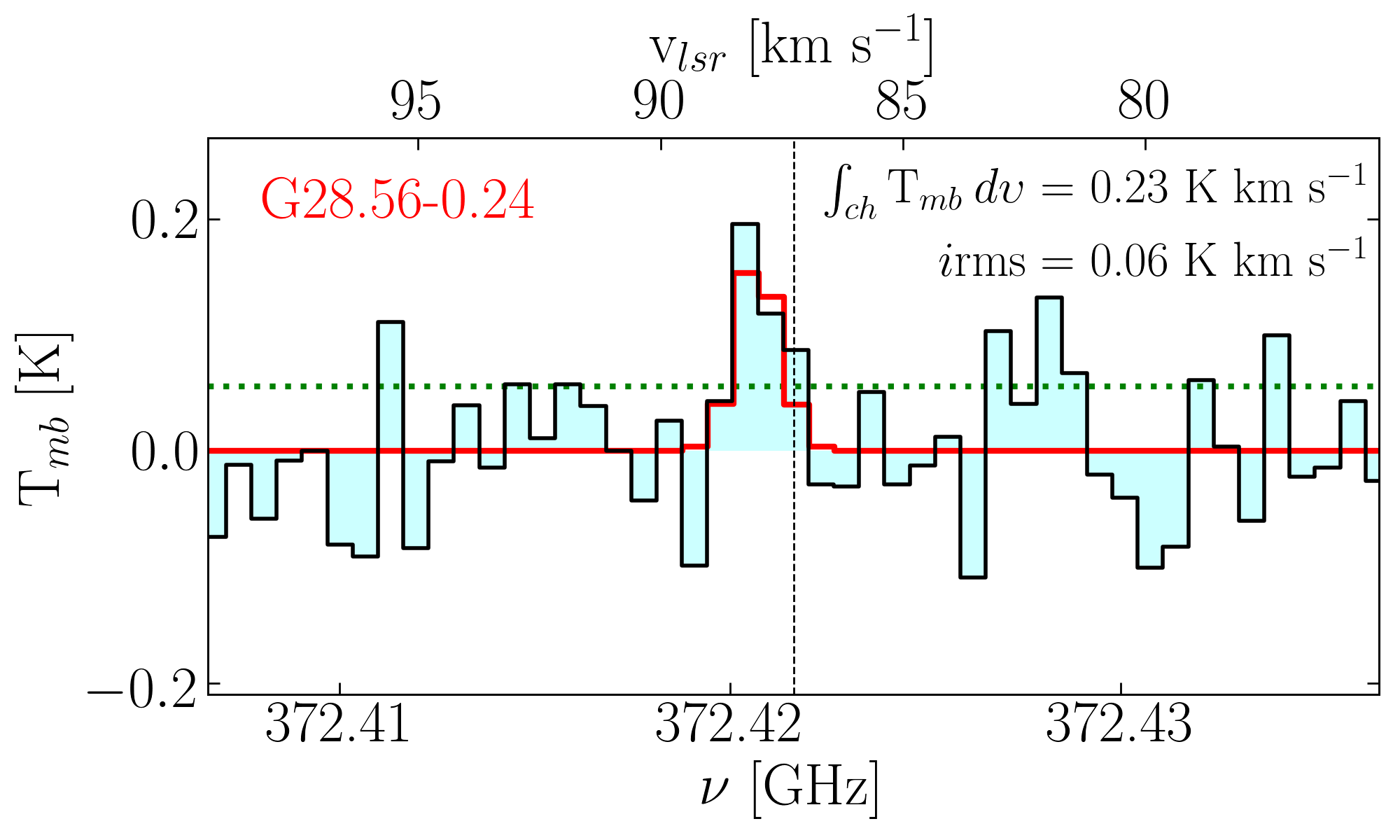}\quad\quad
   \includegraphics[width=0.435\hsize]{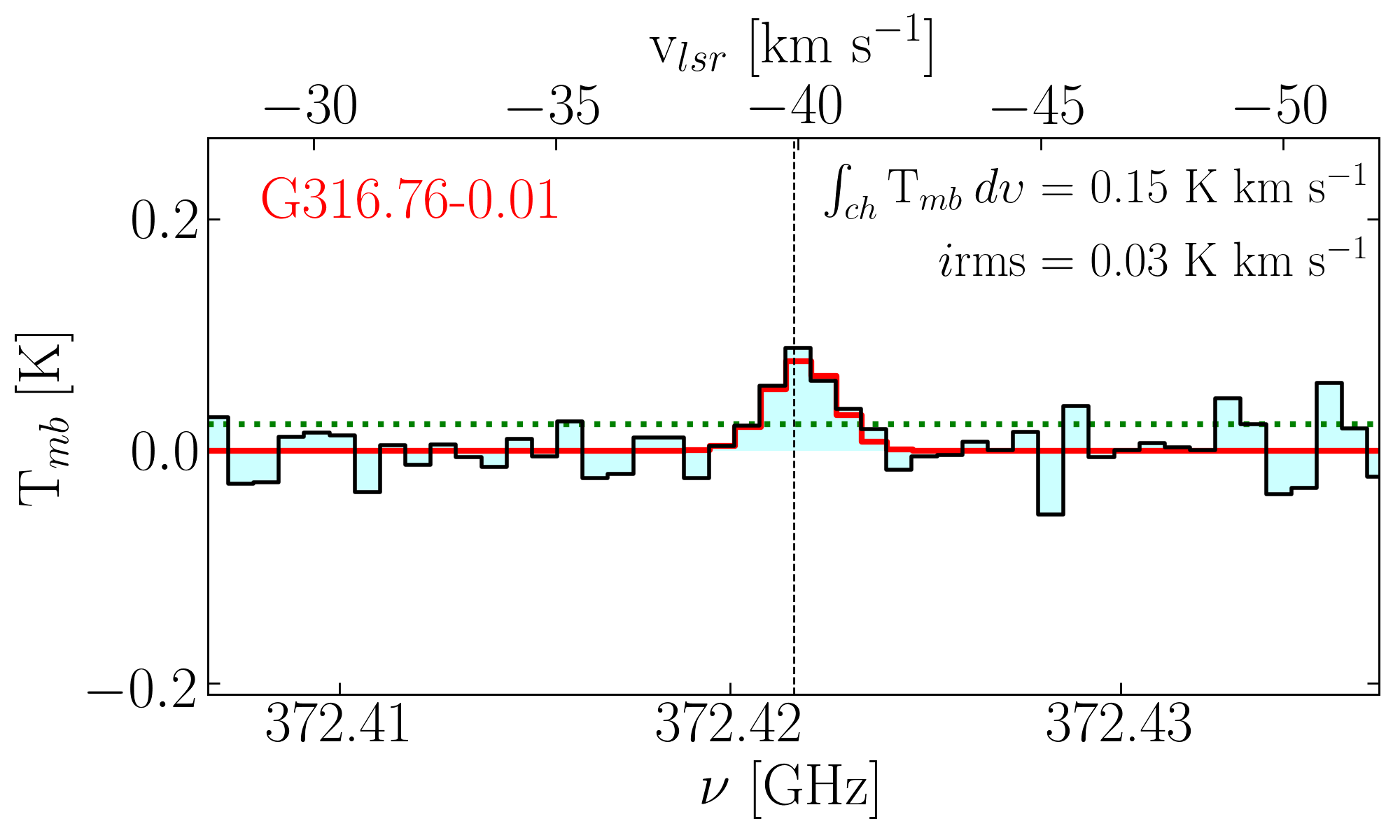}
   \includegraphics[width=0.435\hsize]{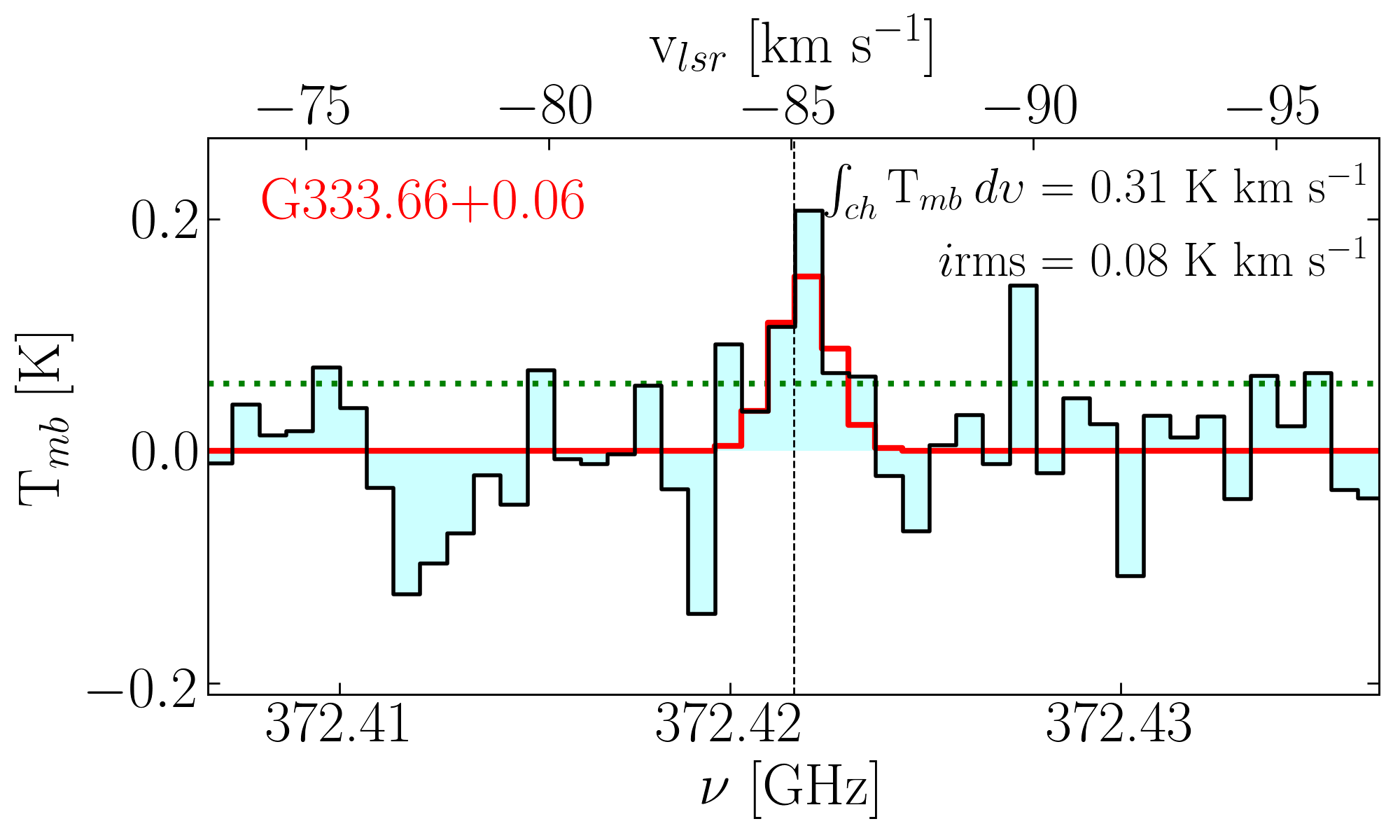}\quad\quad
   \includegraphics[width=0.435\hsize]{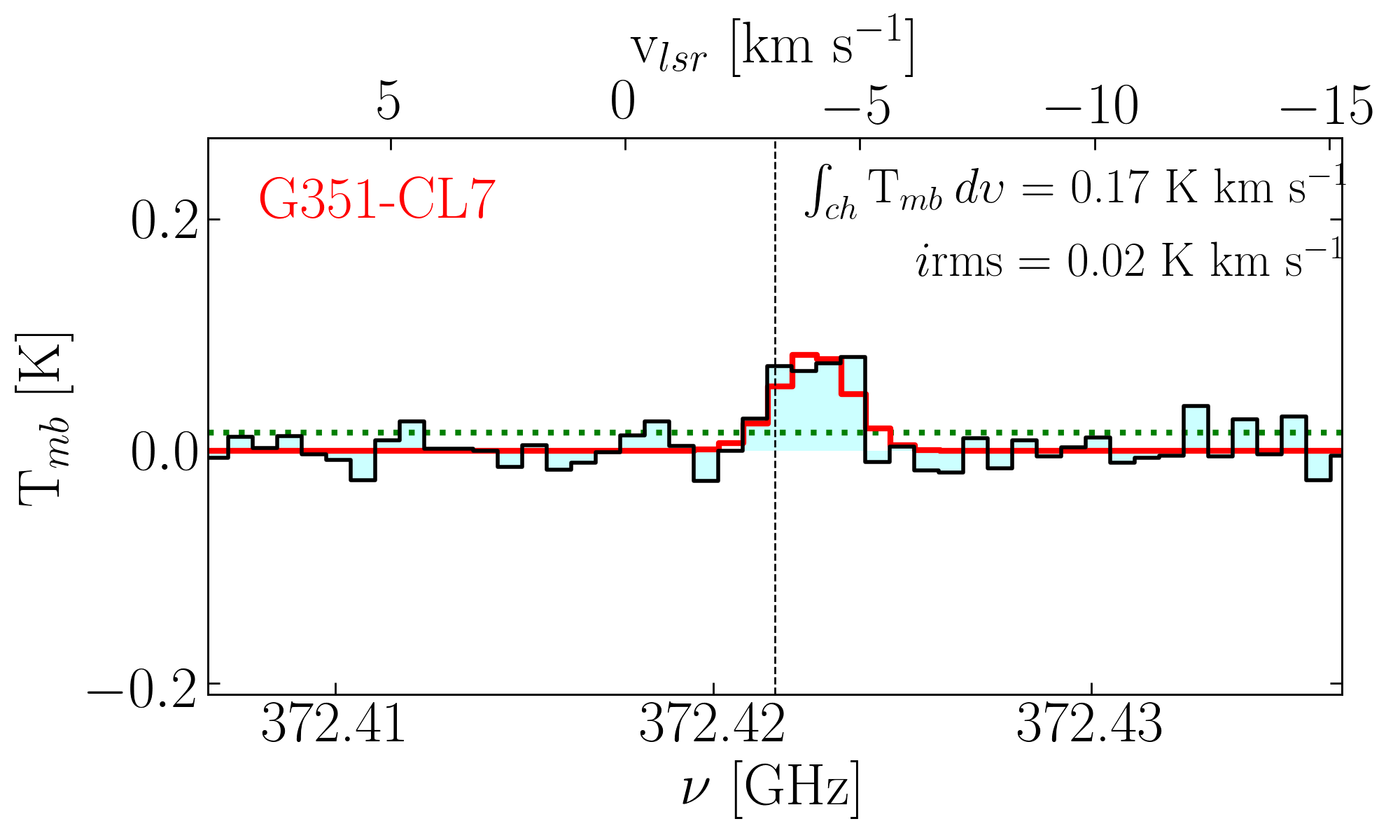}
   \includegraphics[width=0.435\hsize]{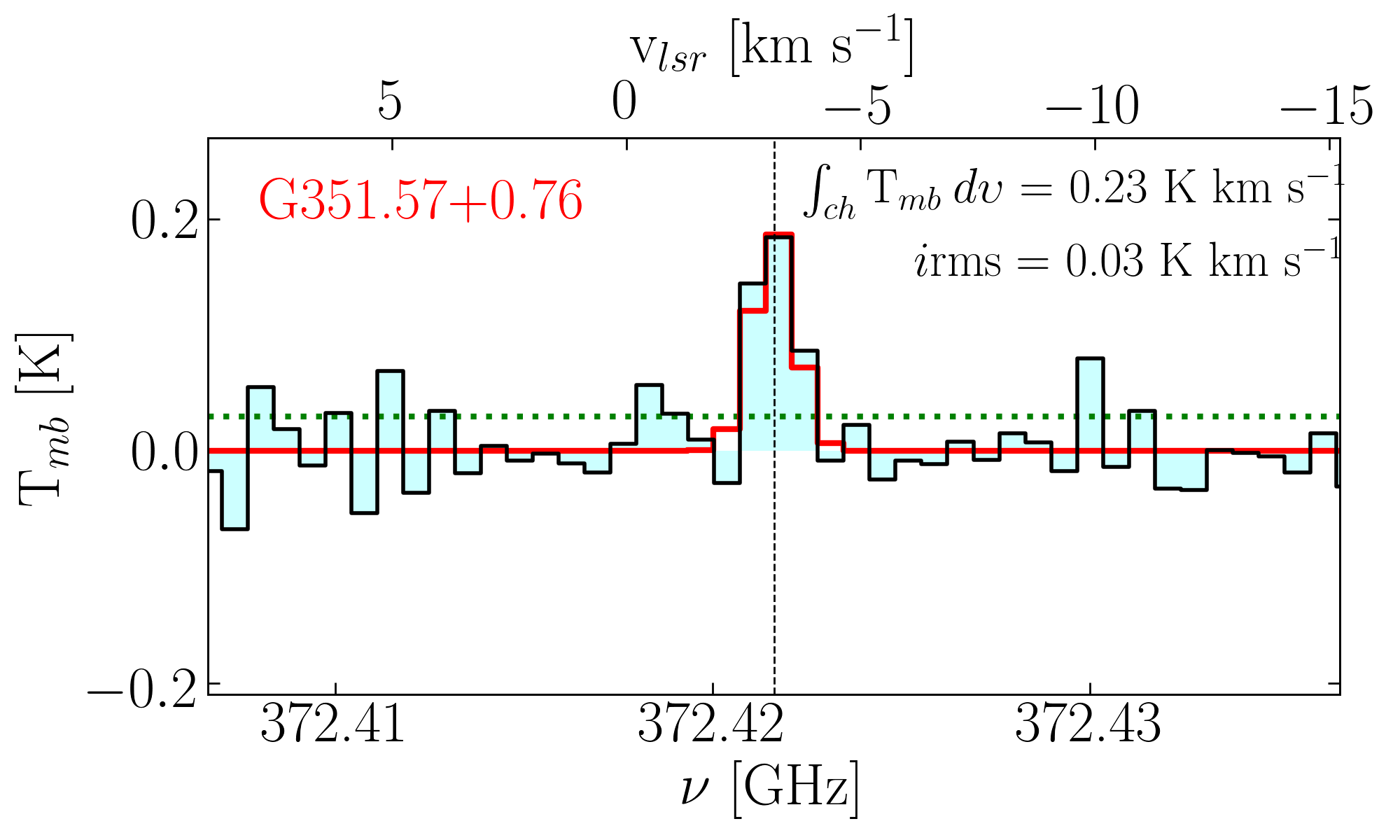}\quad\quad
   \includegraphics[width=0.435\hsize]{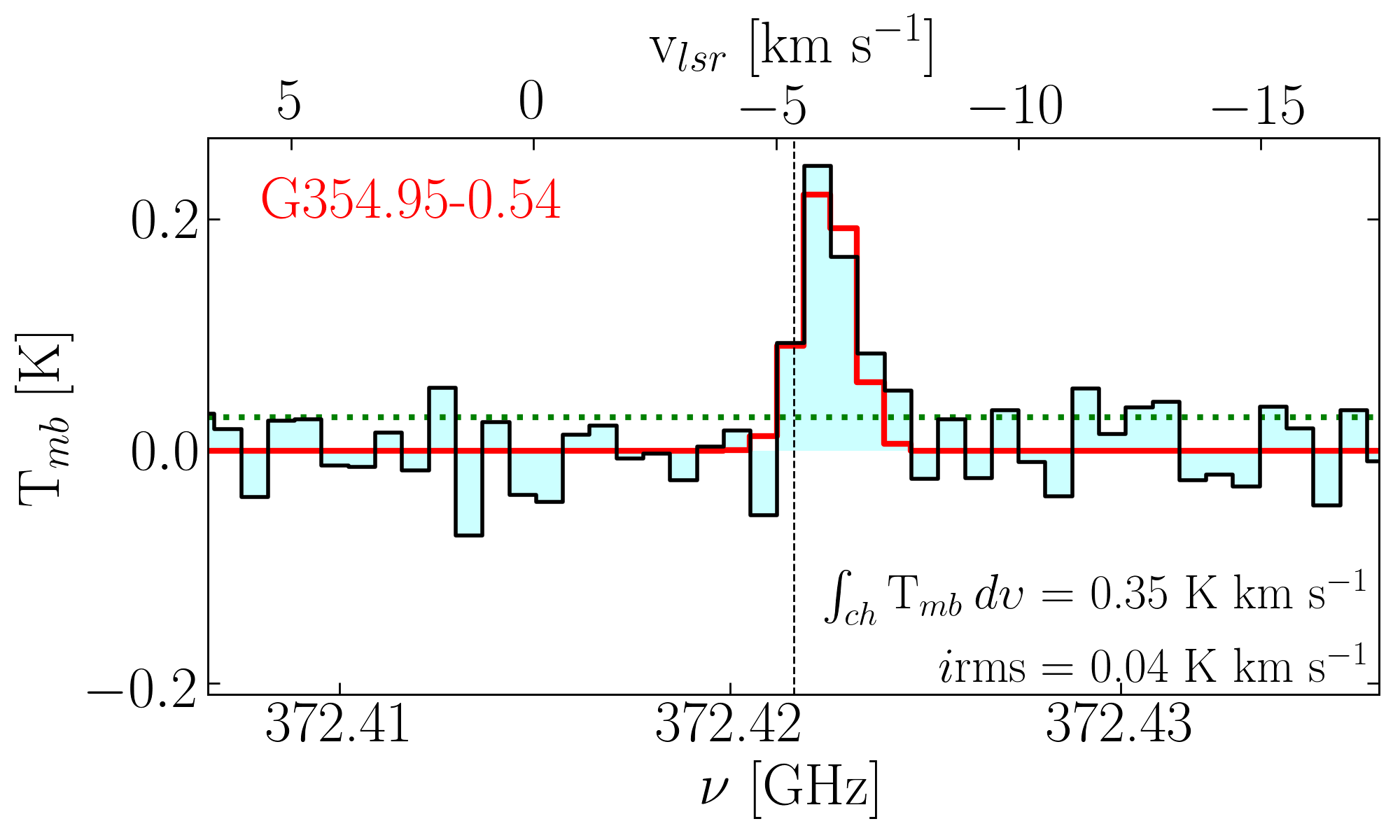}
   \caption{Same as Fig.~\ref{fig:H2DpspectraA}, but for the second half of the sample.}
              \label{fig:H2DpspectraB}%
    \end{figure*}
    
        The APEX {\it Telescope Large Area Survey of the Galaxy} (ATLASGAL; \citealt{Schuller09}, \citealt{Csengeri14} and \citealt{Li16}) provides an ideal basis for detailed studies of large numbers of massive clumps in different stages of the evolutionary sequence of high-mass star-forming regions  (\citealt{Molinari08}). The ATLASGAL {\it Compact Source Catalog} has delivered $\sim 10^4$ clumps (e.g. \citealt{Contreras13}, \citealt{Urquhart14a}, and \citealt{Csengeri14}), with reliable estimates of kinematic distances, masses, luminosities, and dust temperatures and distribution (\citealt{Urquhart14c}, \citealt{Wienen15} and \citealt{Urquhart18}). The ATLASGAL-TOP100 sample (hereafter TOP100; see \citealt{Giannetti14}) includes 111 clumps;  it was selected from ATLASGAL as a flux-limited sample, using additional infrared (IR) criteria in order to include sources in different evolutionary stages (see \citealt{Konig17}). Molecular line surveys have been carried out on the TOP100 with APEX-12m, Mopra-22m, and the IRAM-30m single-dish telescopes, between 80 and 345 GHz, covering more than 120 GHz of bandwidth in three spectral windows\footnote{Not yet completely published.} that contain a multitude of emission lines of both simple and complex molecules. A detailed analysis of a small sample of chemical species allowed us to estimate their excitation parameters, and to derive accurate column densities (e.g. \citealt{Giannetti14}, \citealt{Csengeri16}, \citealt{Giannetti17_june}, and \citealt{Tang18}). 
        From the analysis of this  data set \cite{Giannetti17_june}  demonstrate that temperature, column densities of several tracers, and H$_2$ volume density increase with time, as a function of evolutionary class and luminosity-to-mass ratio ($L/M$). These trends were interpreted as the signature of the initial compression phase of the clump material and the progressive accretion onto the forming YSOs. Therefore, the evolutionary sequence defined for the TOP100 is statistically valid, and the TOP100 can be considered representative of the Galactic protocluster population through all the evolutionary stages.
        
        \subsection{Observed \ohhdp sub-sample}
        In this work, we present new observations of the \ohhdp ground-state transition $J_{\rm {K_a, K_c}} = 1_{10} - 1_{11}$, based on two spectral line surveys of the ATLASGAL sources. The first set of observations\footnote{Sixteen objects in total, not fully contained in the TOP100.} is composed of a sample of young massive clumps, selected from previous observations that showed a high degree of deuterated ammonia (\citealt{WienenSUB}).
        The second set of observations were part of a survey of [CI] $^3P_1$ -- $^3P_0$ fine structure line at 492 GHz (\citealt{LeePREP}) carried out on the TOP100 sample. These were observed simultaneously with the dual-band FLASH345/460 GHz receiver on APEX, tuned to 372 GHz to observe the N$_2$H$^+$ $J = 4-3$ molecular line, which is very close to the \ohhdp transition. Since the main goal of that survey was the observation of the [CI] line, the noise levels were often not adequate for an \ohhdp detection. For this reason many of the sources observed in this project were used here to calculate \ohhdp detection limits (we postpone further discussions to Sect.~\ref{sec4:analisys}).
        
        In Table~\ref{tab:obsprop}, we summarise the physical parameters, for each clump with detection, from the dust continuum taken from \cite{Giannetti17_june}, \cite{Konig17}, and \cite{Urquhart18}. The heliocentric distances, $d_\odot$, of our sample are taken from \cite{Konig17} and \cite{Urquhart18}, and range between 1 and 5.5 kpc; the galactocentric radii, D$_{GC}$, are between 4.5 and $\sim 7.5$ kpc with a mean associated error of 0.3 kpc, and are taken from \cite{Urquhart18} assuming a distance to the Galactic Center of 8.35 kpc (\citealt{Reid14}); the effective radii, R$_{eff}$, range from 0.1 to 1.3 pc; the radial velocities, v$_{lsr}$, are taken from \cite{Giannetti14} and \cite{Wienen12}, based on the C$^{17}$O $J = 3-2$ and the NH$_3$ ($1,1$) lines, respectively\footnote{ See also the ATLASGAL Database Server at \url{https://atlasgal.mpifr-bonn.mpg.de/cgi-bin/ATLASGAL_DATABASE.cgi}}. We note that a few TOP100 sources show an offset of $\sim 1$ km s$^{-1}$ between the v$_{lsr}$ derived from C$^{17}$O and the central velocity of the \ohhdp lines. However, the  v$_{lsr}$ derived from \ohhdp are consistent with the values from N$_2$H$^+$ $J = 4-3$ at $\sim372.672$ GHz. Such offsets are potential indications of strong internal motions driven by contraction, consistent with our finding that these clumps {are on the verge of collapse} (see Sect.~\ref{sect:dinamical_q}).
        
        The dust temperatures, $T_{dust}$, cover the typical range of values between $\sim 10$ and $\sim 25$ K, while $\Delta T_{dust}$ is the error derived from the covariance matrix of the Spectral Energy Distribution (SED) fit performed using a Levenberg-Marquardt least-squares minimisation (see Appendix D in \citealt{Urquhart18}). We note that a wider exploration of the parameter space used to describe the dust properties, would lead to an error of $\sim 10\% $ in $T_{dust}$ \citep[e.g.][]{Schisano20}. \\
        The SED fit assumes a dust absorption coefficient at 870 $\mu$m, $\kappa_{870} = 1.85$ \mbox{cm$^2$ g$^{-1}$}, and a dust emissivity index $\beta=1.75$. The clump masses, $M_{clump}$, and bolometric luminosities, $L_{bol}$, are within $\sim 100-5400$ $M_\odot$ and $\sim 10^{-1}-10^{4}$ $L_\odot$, respectively. The molecular hydrogen column density, N(H$_2$), spans less than an order of magnitude among the different sources, with values of log$_{10}$(N(H$_2$) [cm$^{-2}$])   between 22.7 and 23.3. The sample includes clumps associated with all evolutionary classes. We refer to \cite{Giannetti17_june}, \cite{Konig17}, and \cite{Urquhart18} for the details on how each parameter and the relative error were derived.\\
        \indent Since more than 2/3 of the sample in Table~\ref{tab:obsprop} is part of the TOP100, we employ the evolutionary classes defined in \cite{Giannetti14}, \cite{Csengeri16}, and \cite{Konig17} assigned in the TOP100, where four stages were defined: ($1$) 70w, i.e. 70 $\mu$m weak, which represents the earliest stage of massive-star formation and includes starless or prestellar cores; ($2$) IRw, i.e. 24 $\mu$m weak, but bright at 70 $\mu$m, associated with sources in an early star formation stage, with clumps likely dominated by cold gas; ($3$) IRb, bright both at 70 and 24 $\mu$m, but not in radio continuum at 5-9 GHz and associated with the high-mass protostellar stage; ($4$) HII, as the compact HII region phase, bright also in radio continuum. We extended the same classification to the sources not included in the TOP100 by following the classes defined in \cite{Urquhart18} and double-checking the 70 $\mu$m and 24 $\mu$m maps\footnote{\url{https://atlasgal.mpifr-bonn.mpg.de/cgi-bin/ATLASGAL_DATABASE.cgi}}$^,$\footnote{\url{http://www.alienearths.org/glimpse/glimpse.php}} according to the TOP100 classification: ($i$) Quiescent $\rightarrow$ 70w; ($ii$) Protostellar $\rightarrow$ IRw; ($iii$) YSOs $\rightarrow$ IRb; ($iv$) MSF $\rightarrow$ HII.
        
\section{Observations and data reduction}\label{sec3:data}

\subsection{{\rm \ohhdp} detections}\label{sec3.1:h2dp}
    The \ohhdp $J_{\rm {K_a, K_c}} = 1_{10} - 1_{11}$ spectra (rest frequency 372.4214 GHz; \citealt{Amano05})  were observed using the on-off (ONOFF) observing mode with the FLASH+ dual-frequency MPIfR principal investigator (PI) receiver (\citealt{Klein14}), mounted at the APEX telescope (\citealt{Gusten06}).
    
\begin{table*}
\caption{Summary of the observations.}\label{tab:observations}
\setlength{\tabcolsep}{5pt}
\renewcommand{\arraystretch}{1.2}
\centering
\begin{tabular}{l|ccccccc}
\hline\hline
Molecules       & Quantum  & Frequencies & Telescopes & Beam FWHM & spectral resolution & rms noise\tablefootmark{a}& References\tablefootmark{b}\\
                & numbers  &    (GHz)    &            & ($arcsec$)&   (km s$^{-1}$)     &   (K)     &             \\
\hline                                                                                             
\ohhdp             & 1$_{10}$ - 1$_{11}$ & 372.4 &   APEX   & 17 & 0.5-0.6 & 0.02-0.05 & {\it this work} \\
H$^{13}$CO$^+$     & 1-0                 & 86.8  & IRAM-30m & 24 &   0.75  &   0.02    & {\it this work} \\
DCO$^+$            & 1-0                 & 72.0  & IRAM-30m & 28 &   0.75  &   0.02    & {\it this work} \\ 
C$^{17}$O (TOP100) & 1-0                 & 112.4 & IRAM-30m & 21 &   0.5   &   0.05    & [1] \\   
C$^{17}$O          & 1-0                 & 112.4 & IRAM-30m & 21 &   0.6   &   0.06    & [2] \\   

\hline
\end{tabular}
\tablefoot{\tablefoottext{a}{The temperatures are reported on the main-beam temperature scale.\tablefoottext{b}{[1]~\cite{Giannetti14}; [2]~\cite{Csengeri16};}}
}
\end{table*}  

    The complete set of observations consists of 106 sources belonging to the ATLASGAL survey. Among them, 99 targets are also contained in the TOP100 sample, observed in three projects (IDs: 0101.F-9517, M-097.F-0039-2016, and M-098.F-0013-2016; PI: F. Wyrowski), from July 2017 to December 2018, with a total of 11568 spectra (with about 20-30 s   of integration time each). The receiver bandpass is covered with overlapping FFTS backends of 2.5 GHz widths (\citealt{Klein12}), one of them covering from 372.250 to 374.750 GHz with a native spectral resolution of 0.038 km s$^{-1}$ and with final mean system temperatures, $T_{sys}$, ranging between $\sim 700$ and $1300$ K, depending on the weather conditions. At $\sim$372 GHz, the APEX telescope has an effective  {full width at half maximum} (FWHM) beam size of 16.8'', with a corresponding main-beam efficiency of $\sim$0.73\footnote{\url{http://www.apex-telescope.org/telescope/efficiency/}}. The observations were reduced using a GILDAS class\footnote{\url{https://www.iram.fr/IRAMFR/GILDAS/}} Python interface pipeline.
    
    To discard the low-quality spectra throughout the survey, the first step of the {procedure} was to evaluate the $T_{sys}$ in each spectrum. 
    We generated $T_{sys}$ distributions source-by-source to avoid possible anomalies between different APEX projects. Only a few spectra (138 in total) are associated with clear $T_{sys}$ outliers and were therefore discarded.
    
    As a second step the main-beam efficiency was applied to the spectra after setting the reference frequency to 372.421 GHz. From each spectrum we subtracted a first-order polynomial baseline around the line, properly masked. In some rare cases, where a first order was not enough, a second-order (or third-order)  baseline was used. 
    In the last step, all spectra related to each source were averaged (stitched) with a spectral resolution between 0.5 and 0.6 km s$^{-1}$ to reduce the noise, generating one final spectrum per source.

    Despite the relatively high Einstein-A coefficient (i.e. log$_{10}$($A_{ul}$ [$s^{-1}$])$=-3.96$; CDMS\footnote{{\it Cologne Database for Molecular Spectroscopy} (CDMS): \url{https://cdms.astro.uni-koeln.de/cdms/portal/}};  \citealt{Muller01}), the $J_{\rm {K_a, K_c}} = 1_{10} - 1_{11}$ transition of \ohhdp is usually faint due to the low relative abundance of this species with respect to H$_2$. Typical observed abundances range between $\sim 10^{-10}-10^{-12}$ in many low-mass star-forming regions (e.g. \citealt{Vastel06}, \citealt{Harju08}, \citealt{Caselli08}, \citealt{Friesen10, Friesen14} and \citealt{Miettinen20}) and few high-mass regime counterparts (e.g. \citealt{Harju06}, \citealt{Swift09}, \citealt{Pillai12} and \citealt{Giannetti19}). For this reason, in this work we evaluated the significance of each detection considering the {integrated signal-to-noise} ratio of the line ({iS/N}) and taking each line with {iS/N} > 3 as a detection. For reference, the integrated main-beam temperature, $\int_{ch} T_{mb}\:d\upsilon$ , and the integrated root mean square noise, $i$rms\footnote{$i$rms = $\sqrt{N_{ch}}~d\upsilon~\sigma$, where $N_{ch}$ is the number of channels included in the integration, $d\upsilon$ is the velocity resolution, and $\sigma$ is the rms per channel.}, of each line is reported in the spectra shown in Figures~\ref{fig:H2DpspectraA} and~\ref{fig:H2DpspectraB}.  
    
    Sixteen ATLASGAL sources with \ohhdp detection\footnote{One of them showing a double velocity component.} are reported, 11 of which belong to the TOP100 sample. All the evolutionary classes are contained in our sample allowing us to study (within statistical limits) how the \ohhdp emission changes through the massive star formation process, and how these changes correlate with the other observed or derived quantities.

\begin{table*}
\caption{\label{tab:dervprop}Summary of derived properties of the ATLASGAL-sources in our sample.}
\setlength{\tabcolsep}{5pt}
\renewcommand{\arraystretch}{1.15}

\centering
\begin{tabular}{l|ccccccccc}
\hline\hline
ATLASGAL-ID&$L/M$\tablefootmark{a}&[$f_D$]\tablefootmark{b}&N(\ohhdp)\tablefootmark{c}&\X&$\Delta \upsilon_{\rm obs}$\tablefootmark{c} & $\alpha$\tablefootmark{d} & $\mathcal{M}$\tablefootmark{e} \\
          &$L_{\odot} M_{\odot}^{-1}$&($\chi^{obs}_{\rm C^{17}O}/\chi^{exp}_{\rm C^{17}O}$)&log$_{10}$(cm$^{-2}$)&log$_{10}$(N[$o$-H$_2$D$^+$]/N[H$_2$])&(km s$^{-1}$)& & \\ 
          
\vspace{-8pt}\\
\hline 
\vspace{-8pt}\\
G08.71--0.41     &0.3 &8.2& 12.7$^{+0.2}_{-0.2}$& -$10.1\pm 0.2$ &1.0$^{+0.7}_{-0.7}$& 0.1 &1.9\\ 
G13.18+0.06      &22.5&6.6& 12.2$^{+0.3}_{-0.3}$& -$10.6\pm 0.3$ &1.4$^{+1.2}_{-1.0}$& 0.5 &1.9\\
G14.11--0.57 (C1)&9.1 &2.1& 12.1$^{+0.3}_{-0.3}$& -$10.8\pm 0.3$ &1.0$^{+0.7}_{-0.7}$& 0.3 &1.3\\
G14.11--0.57 (C2)&9.1 &2.1& 12.3$^{+0.3}_{-0.3}$& -$10.6\pm 0.3$ &1.6$^{+1.1}_{-0.9}$& 0.9 &2.3\\
G14.49--0.14     &0.4 &9.0& 13.0$^{+0.1}_{-0.1}$& -$10.1\pm 0.1$ &2.7$^{+0.9}_{-0.8}$& 0.6 &5.4\\
G14.63--0.58     &11.1&2.0& 12.4$^{+0.2}_{-0.2}$& -$10.6\pm 0.2$ &1.7$^{+1.0}_{-0.7}$& 1.0 &2.4\\
G18.61--0.07     &0.7 &4.2& 12.6$^{+0.1}_{-0.1}$& -$10.2\pm 0.1$ &1.6$^{+0.7}_{-0.6}$& 0.5 &3.0\\
G19.88--0.54     &15.5&2.2& 12.1$^{+0.3}_{-0.3}$& -$11.0\pm 0.3$ &1.3$^{+1.0}_{-0.9}$& 0.3 &1.7\\
G28.56--0.24     &0.3 &6.3& 12.7$^{+0.3}_{-0.4}$& -$10.4\pm 0.4$ &1.0$^{+0.6}_{-0.7}$& 0.1 &1.9\\
G333.66+0.06     &3.0 &4.0& 12.6$^{+0.3}_{-0.4}$& -$10.1\pm 0.4$ &1.3$^{+1.0}_{-1.0}$& 0.2 &2.1\\
G351.57+0.76     &2.7 &2.4& 12.6$^{+0.2}_{-0.2}$& -$10.1\pm 0.2$ &1.0$^{+0.5}_{-0.4}$& 0.4 &1.5\\
G354.95--0.54    &3.2 &3.8& 12.7$^{+0.1}_{-0.1}$& -$9.9\pm 0.1$ &1.2$^{+0.4}_{-0.4}$ & 0.8 &1.8\\
\hline                                                                           
\vspace{-8pt}\\                         
G12.50-0.22   &1.3 &            --           & 12.8$^{+0.1}_{-0.1}$& -$10.0\pm 0.1$ &1.2$^{+0.4}_{-0.3}$&1.0&2.2\\
G14.23--0.51  &2.1 &            --           & 12.8$^{+0.2}_{-0.2}$& -$10.5\pm 0.2$ &2.1$^{+0.8}_{-0.8}$&0.6&3.5\\
G15.72--0.59  &0.2 &            --           & 12.8$^{+0.2}_{-0.2}$& -$10.0\pm 0.2$ &1.9$^{+1.0}_{-1.0}$&1.0&3.8\\
G316.76-0.01  &5.2 &            --           & 12.3$^{+0.2}_{-0.2}$& -$10.8\pm 0.2$ &1.6$^{+0.6}_{-0.5}$&0.2&2.5\\
G351.77-CL7   &{0.7}\tablefootmark{f}&3.4\tablefootmark{g}&12.6$^{+0.1}_{-0.1}$&-$10.3\pm 0.1$&1.8$^{+0.6}_{-0.6}$& 0.5 &3.5\\
\hline 

\end{tabular}
\tablefoot{Derived properties of our sample. Top panel: TOP100 sources; Bottom panel: ATLASGAL sources not in the TOP100. The symbol ``--'' means or that this value is not available from the literature (for $f_D$) or that it could not be calculated (for G14.11-0.57).\\
\tablefoottext{a}{data from columns (5) and (6) in Table~\ref{tab:obsprop};}
\tablefoottext{b}{we associate with the CO depletion factor, $f_D$, a conservative error of 15\%, considering recent results discussed in \citet{Sabatini19}, where we presented a detailed CO-depletion study in a local filament of massive star-forming regions;}
\tablefoottext{c}{computed using MCweeds (\citealt{Giannetti17_june}) assuming $T_\mathrm{ex}$ = $T_\mathrm{dust}$;}
\tablefoottext{d}{virial parameters derived following \cite{MacLaren88}, see Sect.~\ref{sect:dinamical_q};}
\tablefoottext{e}{Mach numbers derived following eq.~\ref{eq:mach} in Sect.~\ref{sect:dinamical_q}, where $m(\mathrm{H_2D^+})=6,692 \times 10^{-24}$ $g$ is the mass of the H$_2$D$^+$ molecule;}
\tablefoottext{f}{this value was calculated from \citet{Giannetti19} as the integrated values on the APEX-beam at 230 GHz (i.e. 28 arcsec);}
\tablefoottext{g}{data from \citet{Sabatini19}; see their Sect. 4.1 for more details on how this value is computed.}
}
\end{table*}

    \subsection{Additional tracers: {\rm H$^{13}$CO$^+$, DCO$^+$ and C$^{17}$O}}\label{sec:additional_tracers}
   For one-half of the sources with \ohhdp detection, we  collected further observations of H$^{13}$CO$^+$, DCO$^+$, and C$^{17}$O that we employ to estimate the cosmic-ray ionisation rate (explained in Sect. \ref{sec:CRIR}).
   We simultaneously observed the rotational transition of both H$^{13}$CO$^+$ and DCO$^+$ in 30 ATLASGAL sources, using the {\it Eight MIxer Receiver} E90 (EMIR; \citealt{Carter12}) at the IRAM-30m single-dish telescope\footnote{IRAM is supported by INSU/CNRS (France), MPG (Germany), and IGN (Spain).} (IRAM project-id 107-15). From the study of \cite{WienenSUB}, the 30 brightest clumps in deuterated ammonia were selected. The receiver was tuned at 75 GHz, allowing the simultaneous detection of H$^{13}$CO$^+$ and DCO$^+$ at 87 and 72 GHz, respectively, in $2\times8$ GHz sidebands with a velocity resolution of 0.75 km s$^{-1}$. The sources were observed with an average integration time of $\sim 1$ h in February 2016. The average $T_{sys}$ are $\sim 130$ K for the H$^{13}$CO$^+$ lines and $210$ K for the DCO$^+$ lines. The data were originally calibrated to antenna temperature, $T_A^*$, and then converted to $T_{mb}$ using the tabulated values for the IRAM-30m forward efficiency, $\eta_{eff} = 0.95$, and  main-beam efficiency, $\eta_{mb} = 0.74$. The final averaged rms noise was 0.02 K for both lines.
    
    The C$^{17}$O $J = 1-0$ observations at 112 GHz, were taken from \cite{Giannetti14}  for the TOP100 sources  and from \cite{Csengeri16}  for the two sources not included in the TOP100,  to which we refer for a full description of the data sets.
    
    The full set of observations used in this work is summarised in Table~\ref{tab:observations}, where the details on the observed molecules (columns 1-3), the telescope setup used (columns 4-7) and the references from which the data were selected (column 7) are reported.
    
    \section{Analysis}\label{sec4:analisys}
    The spectra for all the sources with a reliable \ohhdp detection  are presented in Figures~\ref{fig:H2DpspectraA} and~\ref{fig:H2DpspectraB}, where the 1$\sigma$ noise level is indicated by the green dotted line, and the integrated main-beam temperature and rms of each line are shown for each spectrum. 
    
    The \ohhdp line emission is clearly visible with a single velocity component in all the sources, with the exception of G14.11-0.57 (see Fig.~\ref{fig:H2DpspectraA}), which shows two line components separated by $\sim 2.5$ km s$^{-1}$. Both the components have {iS/N} $\gg 3$, and for this reason in the following analysis they are treated separately (i.e. C1 and C2). The average noise of the spectra is almost constant at $\sim 0.02-0.03$ K (T$_{mb}$-scale) throughout the sample. In a few cases the noise is slightly higher (i.e. $\sim 0.05$ K). The FWHM line widths are between 1 and 3 km s$^{-1}$, in agreement with those reported by \cite{Giannetti14} and \cite{Wienen12} for the C$^{17}$O $J = 3-2$ and the NH$_3$ ($1,1$), respectively.%, invoking - in some cases - the presence of a non-thermal component to justify values larger than the expected one of 1-1.5 km s$^{-1}$.

    \subsection{Column densities}\label{sec4.1:column_densities}
    
    Column densities are obtained by fitting the observed spectra by employing MCWeeds (\citealt{Giannetti17_june}), an external interface between Weeds (\citealt{Maret11}), simple and fast in building synthetic spectra assuming LTE, and the Bayesian statistical models and adaptation algorithms of PyMC (\citealt{Patil10}). Model fits are shown in red in Figures~\ref{fig:H2DpspectraA} and \ref{fig:H2DpspectraB}. 
    For the 16 sources with an \ohhdp detection, the results of the fit are summarised in Table~\ref{tab:dervprop}, computed assuming that the excitation temperature of the transition $T_\mathrm{ex}$ = $T_\mathrm{dust}$ (e.g. \citealt{Giannetti19}). 
    This assumption implies that the dust and \ohhdp are well mixed, which might not be the case for every source as the average volume densities (see Table \ref{tab:dervprop}) are slightly lower than the \ohhdp critical density, n$_{cr}\sim10^5$ cm$^{-3}$ (e.g. \citealt{Caselli08} and \citealt{Vastel12}). By following \cite{Caselli08}, who reported excitation temperatures up to 40\% lower than  the dust values, we re-performed the column density calculations taking $T_\mathrm{ex}$ = $0.6\:T_\mathrm{dust}$. We find that in most of the cases, the \ohhdp column densities are well within the uncertainties reported in  Table~\ref{tab:dervprop}, so that our assumption of $T_\mathrm{ex}$ = $T_\mathrm{dust}$ does not affect the final results. In this range of temperatures the line optical depths, $\tau$, computed as in eq.~(5) of \cite{Caselli08}, are $\lesssim 0.1$, implying optically thin regimes.

    We also assumed extended emission relative to the FWHM APEX-beam to compute the column densities, supported by the results of \cite{Pillai12}, where the emission scale of the \ohhdp was observed to be  close to the typical clump size. We note that for all the clumps with a clear \ohhdp detection, the angular size corresponding to the effective radii (R$_{eff}$; \citealt{Konig17} and \citealt{Urquhart18}) is always larger than the APEX beam sizes, except for G316.76-0.01 (\citealt{Pillai12}). 
    
    Fits were performed assuming the \ohhdp molecular line parameters provided by the CDMS database: an Einstein-$A$ coefficient log$_{10}$($A_{ul}$ [s$^{-1}$]) $=-3.96$, a statistical weight of g$_{ul}= 9$, an energy gap between the two quantum levels E$_{ul} = 17.9$ K, and the partition function, $Q(T)$ for \ohhdp in the relevant temperature range [9.375 K: 10.3375, 18.750 K: 12.5068, 37.500 K: 15.5054] (see also \citealt{Giannetti19}).
    
    Throughout the sample the \ohhdp column density varies by less than an order of magnitude, from $1.3 \times 10^{12}$ to $10^{13}$~cm$^{-2}$, while its abundance \X = N[$o$-H$_2$D$^+$]/N[H$_2$] is in the range $\sim (1$--$12.6)\times 10^{-11}$. The abundances were derived with respect to the H$_2$ column densities reported by \cite{Giannetti17_june} and \cite{Urquhart18}. The final \X values are slightly lower than those found in many low-mass pre- and protostellar regions (e.g. \citealt{Vastel06}; \citealt{Harju08}; \citealt{Caselli08}; \citealt{Friesen10, Friesen14}; and \citealt{Miettinen20}), but in agreement with those derived for the few high-mass star-forming regions observed by \cite{Harju06}, \cite{Pillai12}, and \cite{Giannetti19};  the only exception is the significantly lower values found by \cite{Swift09}, who reports an abundance of $\sim (3-5) \times 10^{-13}$ in the infrared dark clouds (IRDCs) G030.88+00.13 and G028.53-00.25. 
    
    The \ohhdp detection limits were calculated for sources with no detection. For each evolutionary class, we  selected the observations with rms within a factor of 2 with respect to the average noise of the spectra with detection (see Appendix~\ref{sec:appB} and Table~\ref{tab:limits} for  details). This selection criteria leads to a loss of 44 sources. Each class has a similar number of objects (70w:8, IRw:16, IRb:9, and HII:13 sources), which makes detection limits comparable through the progressive evolution of the clumps, with a \ohhdp detection rate of 47\% in 70w, 27\% in IRw, 18\% in IRb, and 7\% in HII.
    The detection limits were calculated for each selected source in order to obtain the column density value that would correspond to a 3$\sigma$ detection. The assumptions made for the column density calculations are the same as for the sources with detection. We used the typical FWHM line width of 1.5 km s$^{-1}$, which corresponds to the average value of the detection sources. We note that this approach provides only qualitative information on the upper limits of column densities of \ohhdp throughout the sample as it is mainly influenced by the quality of observations rather than by the physics of the sources. Nevertheless, we limit this problem by selecting spectra with noise levels comparable to those of detected sources. The (upper limit) relative abundances retrieved in this way are in the range of the typically observed values. 

\subsection{Estimates of dynamical quantities}\label{sect:dinamical_q}
    In addition to the N($o$-H$_2$D$^+$) value, MCWeeds also provides an estimate of the FWHMs ($\Delta \upsilon_{obs}$ in Table~\ref{tab:dervprop}) of each line, allowing us to derive two important dynamical quantities: the virial parameter, $\alpha$, and the Mach number, $\mathcal{M}$. \
    The first  gives us an assessment of the dynamical state of each source, hence how far a source is from {virial equilibrium} (\citealt{Chandrasekhar53}), while the second  tells us about the turbulence of the gas,  hence whether the clump is supported mainly by dynamical or by thermal motions.
    
    The estimates of $\alpha$ were made following the classic definition of \cite{Bertoldi92}:  

    \begin{equation}\label{eq:alpha}
       \alpha = \frac{M_{vir}}{M_{clump}}\:\:\:{\rm with}\:\:\:M_{vir} = k_2\:\left(\frac{{\rm R}_{clump}}{{\rm pc}}\right)\:\left(\frac{\Delta \upsilon_{dyn}}{{\rm km\:s}^{-1}}\right)^2. 
    \end{equation}
    
    \noindent
    Here $M_{vir}$ and $M_{clump}$ are the source virial mass and total mass, respectively; R$_{clump}$ is the clump size in parsec, here assumed equal to the R$_{eff}$; and $k_2=210$ for the homogeneous core assumption we made (\citealt{MacLaren88}). The line width $\Delta \upsilon_{dyn}$ in eq.~\ref{eq:alpha} is the combination of the thermal motions of the particle of mean mass, $\Delta \upsilon_{th, \langle m \rangle}$; the thermal gas motions, $\Delta \upsilon_{th}$; and the observed FWHM. Using the values of $\Delta \upsilon_{obs}$ in Table~\ref{tab:dervprop}, we derived $\Delta \upsilon_{dyn}$ by following \cite{Kauffmann13} (see their Sect.4): 
    
    \begin{equation}
    \Delta \upsilon_{dyn} = \sqrt{\Delta \upsilon_{obs}^2 + \Delta \upsilon_{th,\langle m \rangle}^2 - \Delta \upsilon_{th}^2}.
    \end{equation}

    The values of $\alpha$ reported in Table~\ref{tab:dervprop} suggest that the sample in this study is heterogeneous, but mainly composed of marginal to highly unstable sources ($\alpha < 1$). We also found agreement with the results of \cite{Kauffmann13}, who discussed the implications of the low virial parameters estimated in a large sample of high-mass star-forming regions, finding that for $M_{clump} \gtrsim 10\:M_\odot$, $\alpha\ll2$ (see their Fig. 1). Such low $\alpha$ values might be the symptomatic consequence of a rapid collapse.

    The Mach number has been computed from the \ohhdp line width provided by MCWeeds as 
    \begin{equation}\label{eq:mach}
        \mathcal{M} = \frac{\sigma_{turb}}{\sigma_{\rm H_2}},
    \end{equation}
    with% $\sigma_{turb} = (\sigma_{obs}^2-\sigma_{th}^2)^{1/2}$ and $\sigma_{\rm H_2} = \Delta \upsilon_{th,\langle m \rangle}/(2\sqrt{2ln2}$)}.
    \begin{equation}\label{eq:sigma_turb}
         \sigma_{turb} = \sqrt{\sigma_{obs}^2-\sigma_{th}^2} \quad
        {\rm and} \quad \sigma_{\rm H_2} = \frac{\Delta \upsilon_{th,\langle m \rangle}}{2\sqrt{2ln2}}. 
    \end{equation}
    In eq.~\ref{eq:sigma_turb}, $\sigma_{obs} = \Delta \upsilon_{obs}/(2\sqrt{2ln2}$) is the observed velocity dispersion of \ohhdp lines, while $\sigma_{th}=[k_B\:T_{dust}/m(\mathrm{H_2D^+})]^{1/2}$ is the gas thermal component at $T_{dust}$, where $k_B$ is the {\it Boltzmann constant} and $m(\mathrm{H_2D^+})$ is the mass of the H$_2$D$^+$ molecule. The Mach number range is $1.3-5.4$, implying supersonic turbulent motions for the gas involved in all the sources.
    
   \begin{figure*}
   \centering
   \includegraphics[width=0.48\hsize]{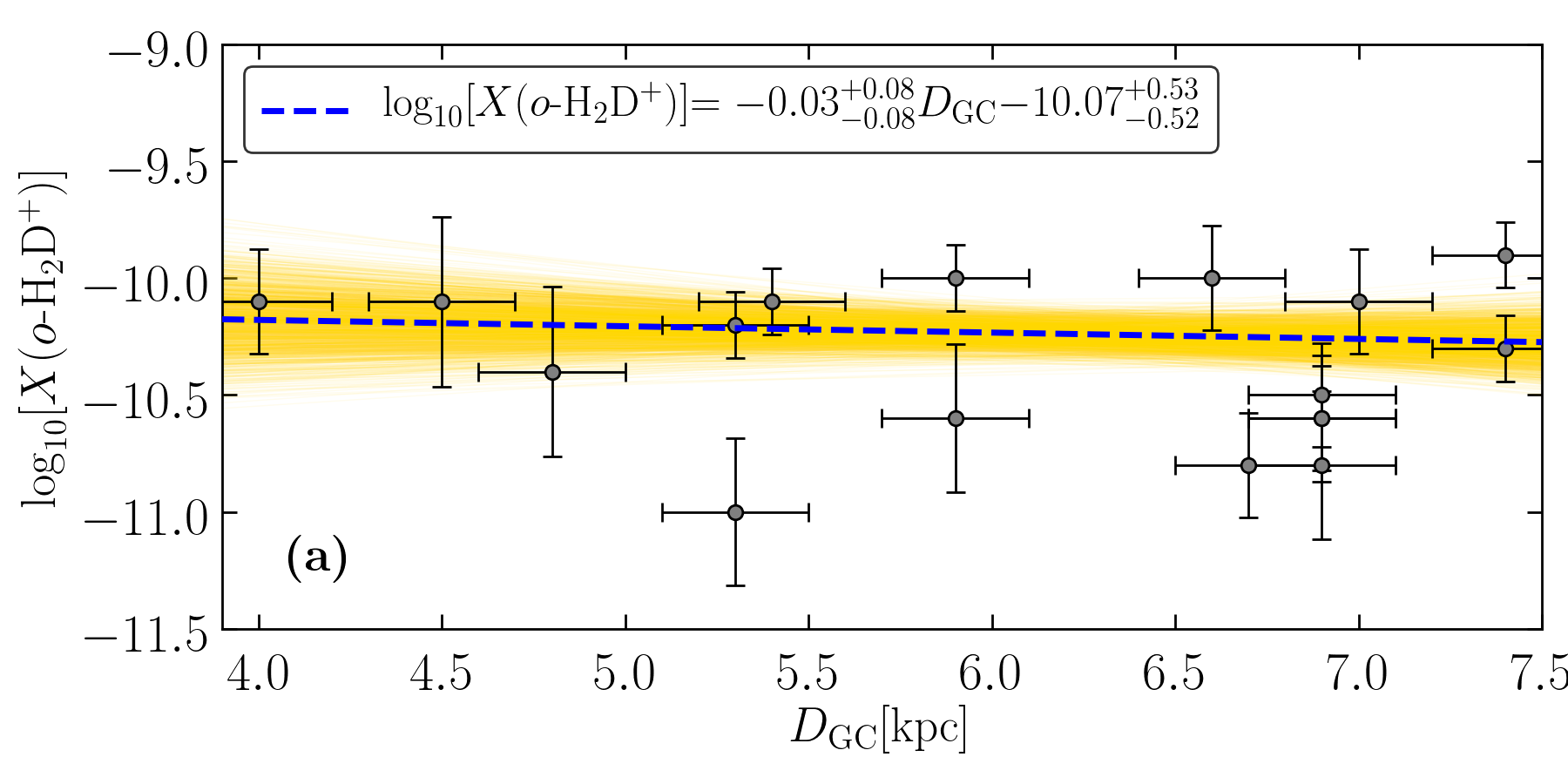}
   \includegraphics[width=0.48\hsize]{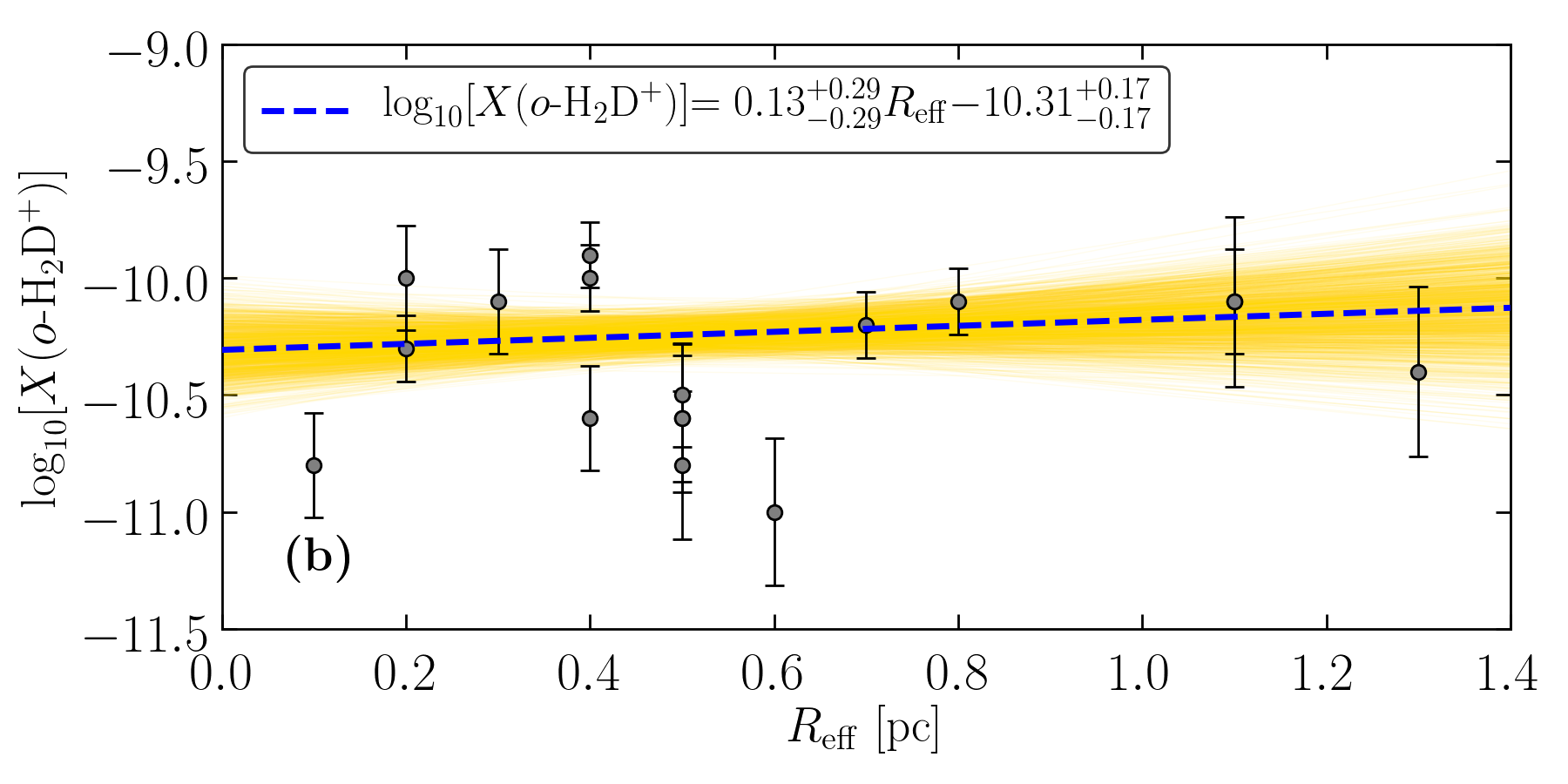}
   \includegraphics[width=0.48\hsize]{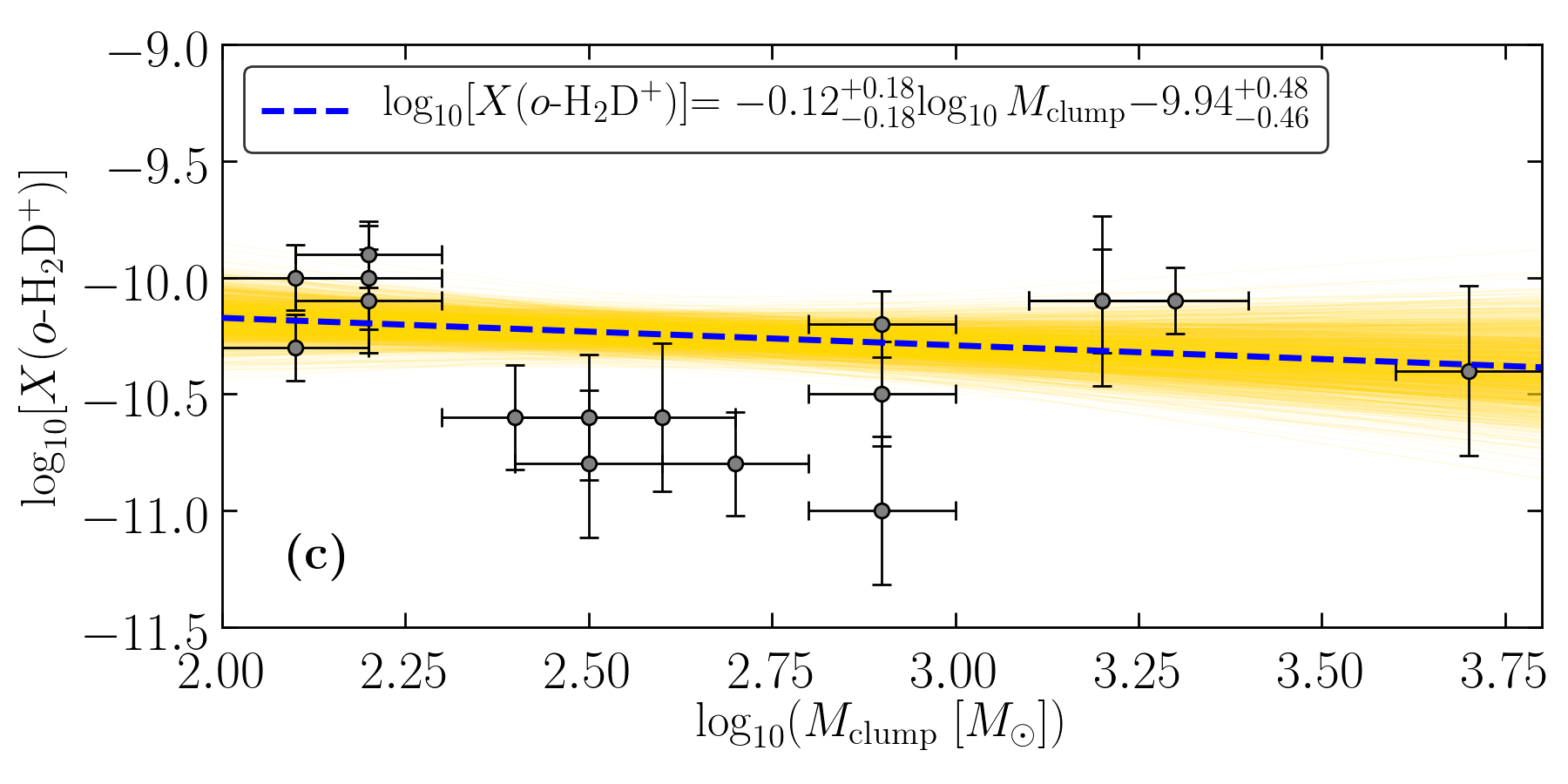}
   \includegraphics[width=0.48\hsize]{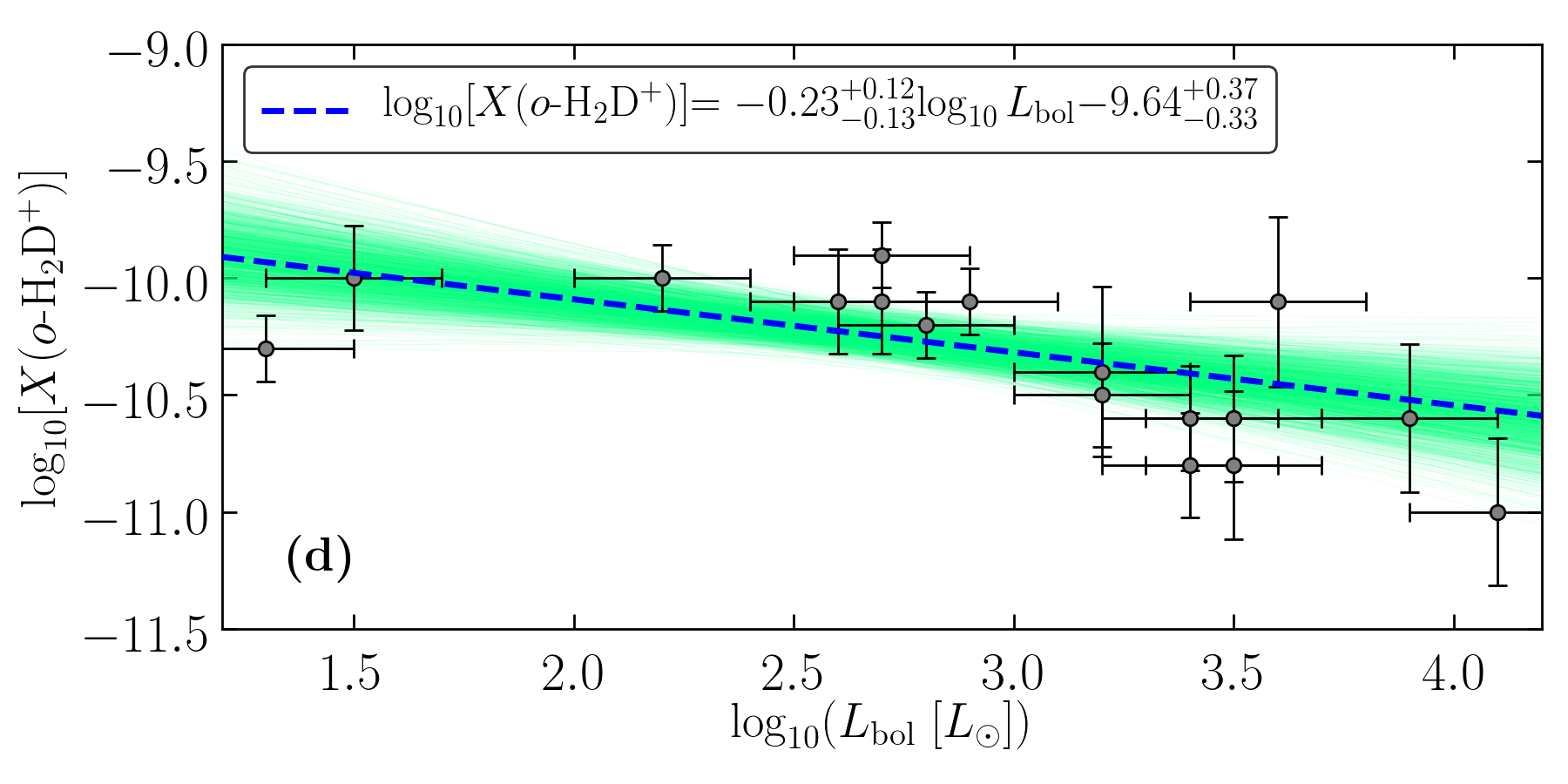}
   \includegraphics[width=0.48\hsize]{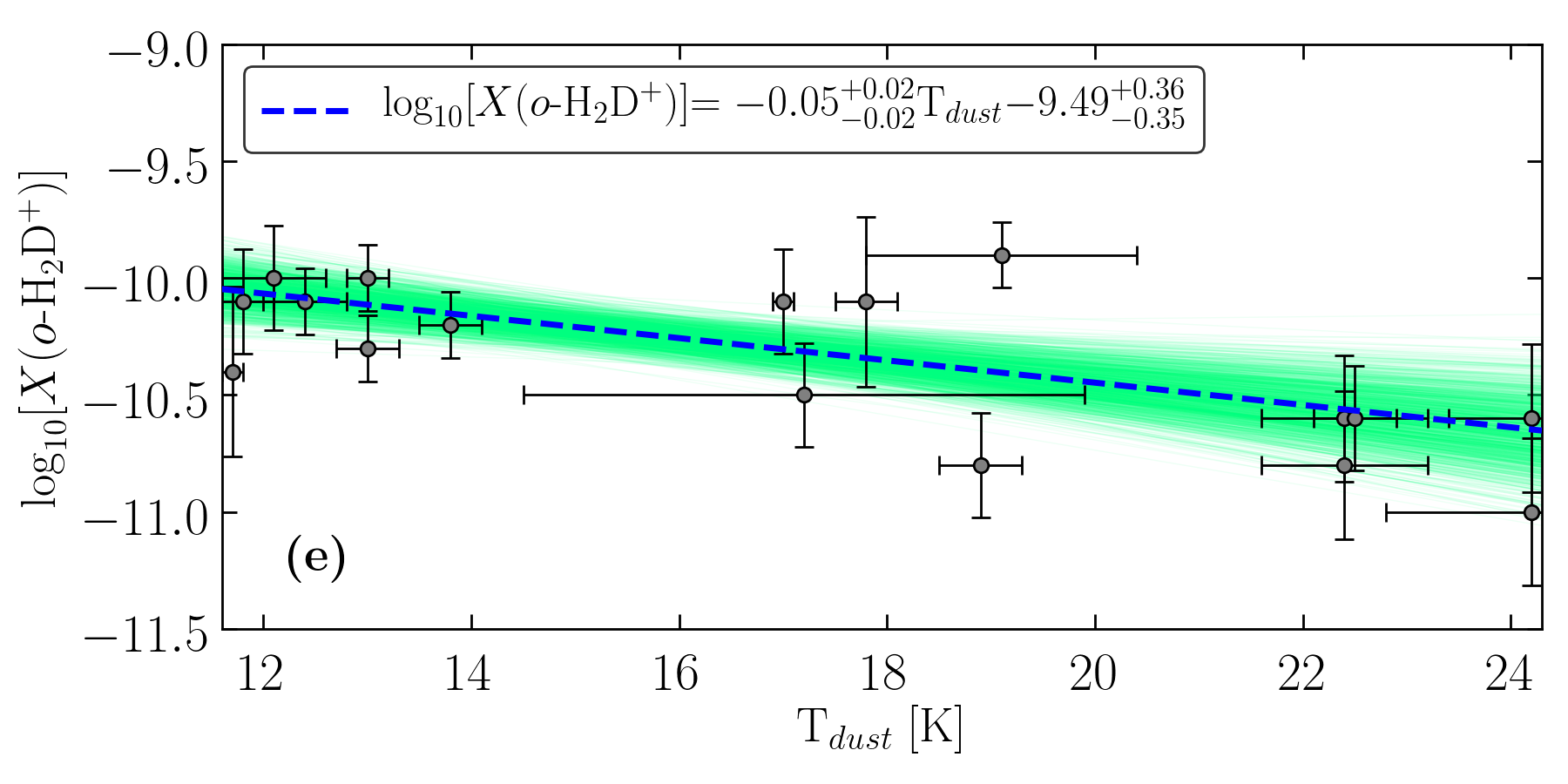}
   \includegraphics[width=0.48\hsize]{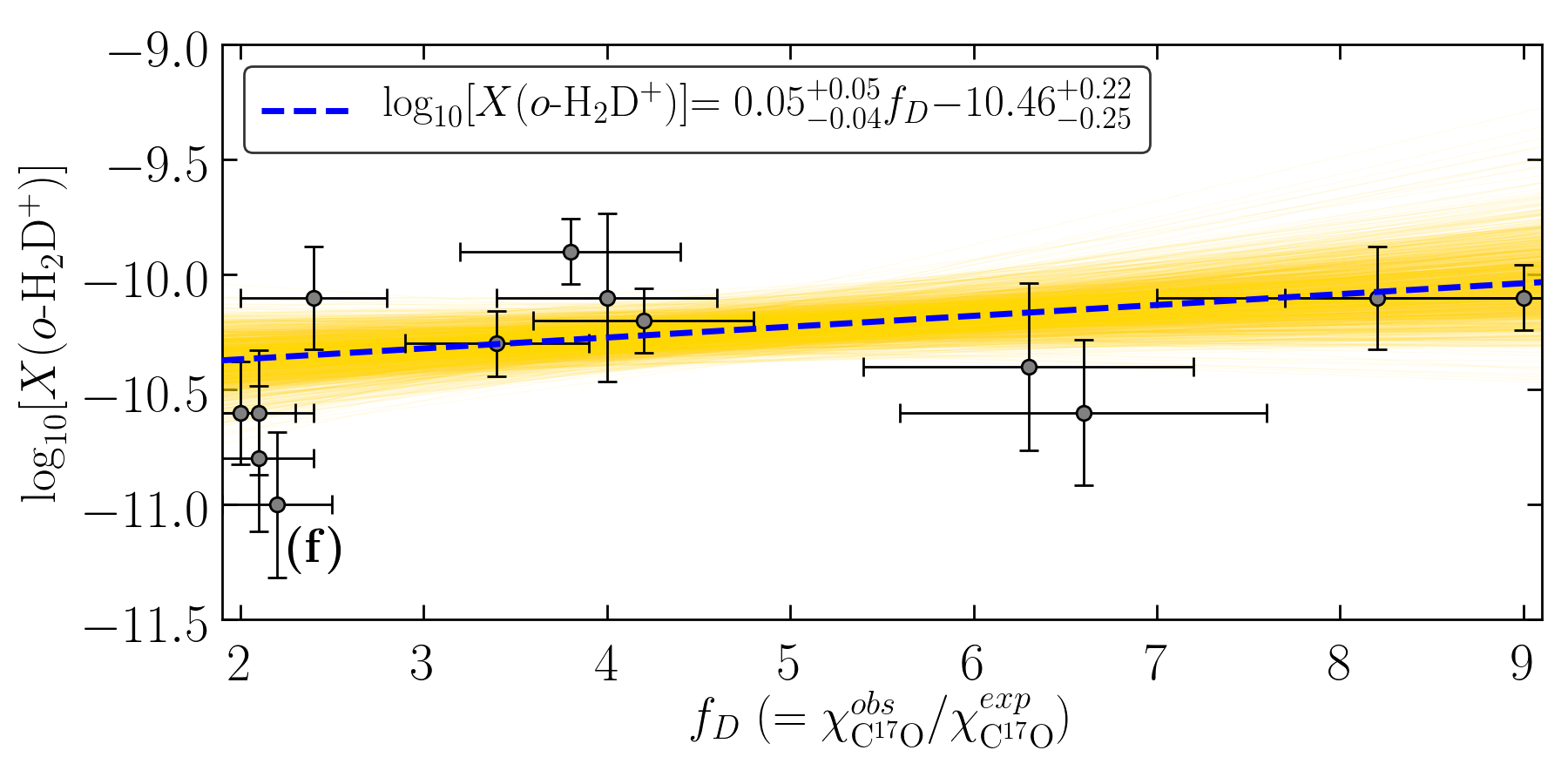}
   \includegraphics[width=0.48\hsize]{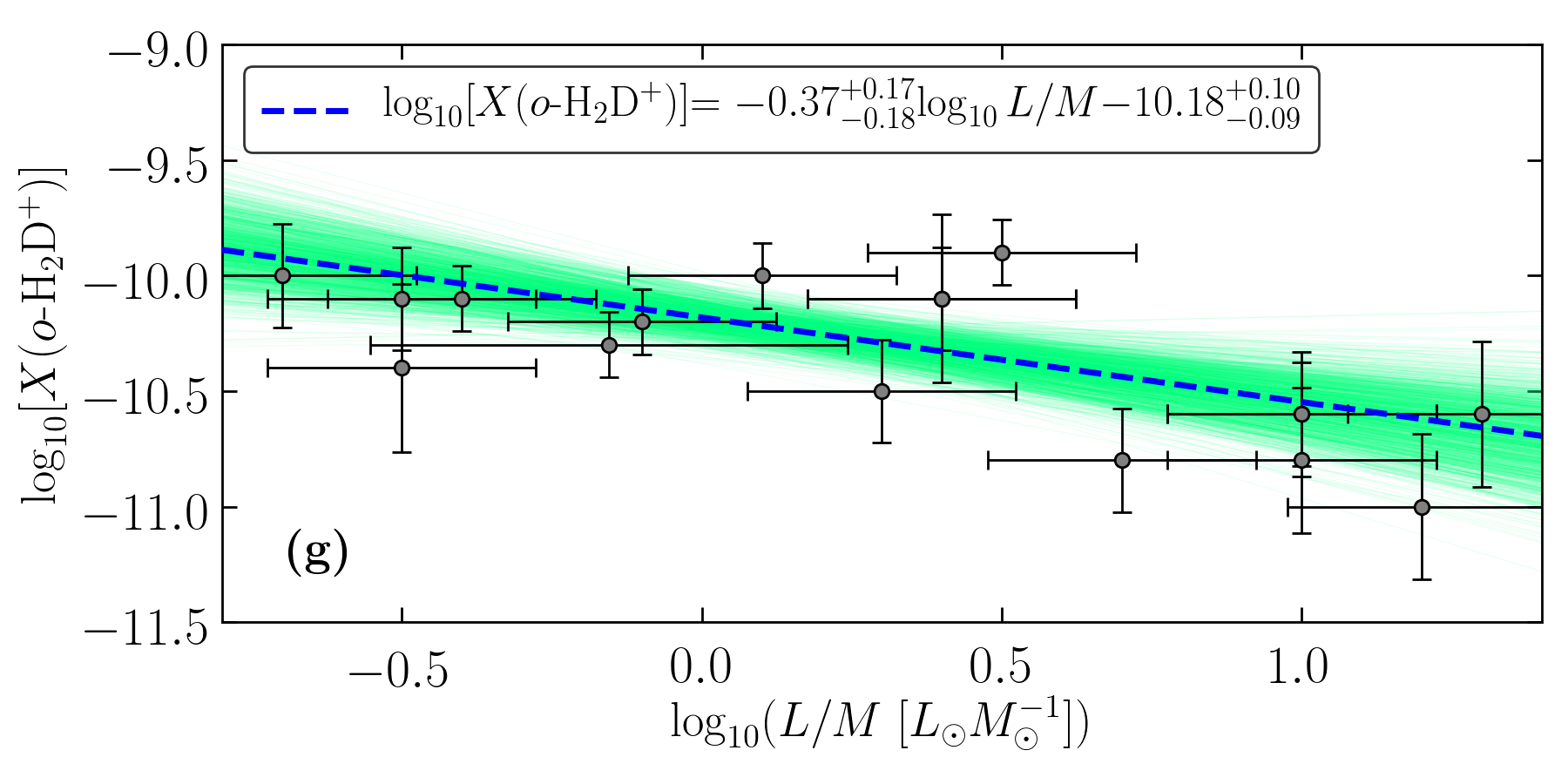}
   \includegraphics[width=0.48\hsize]{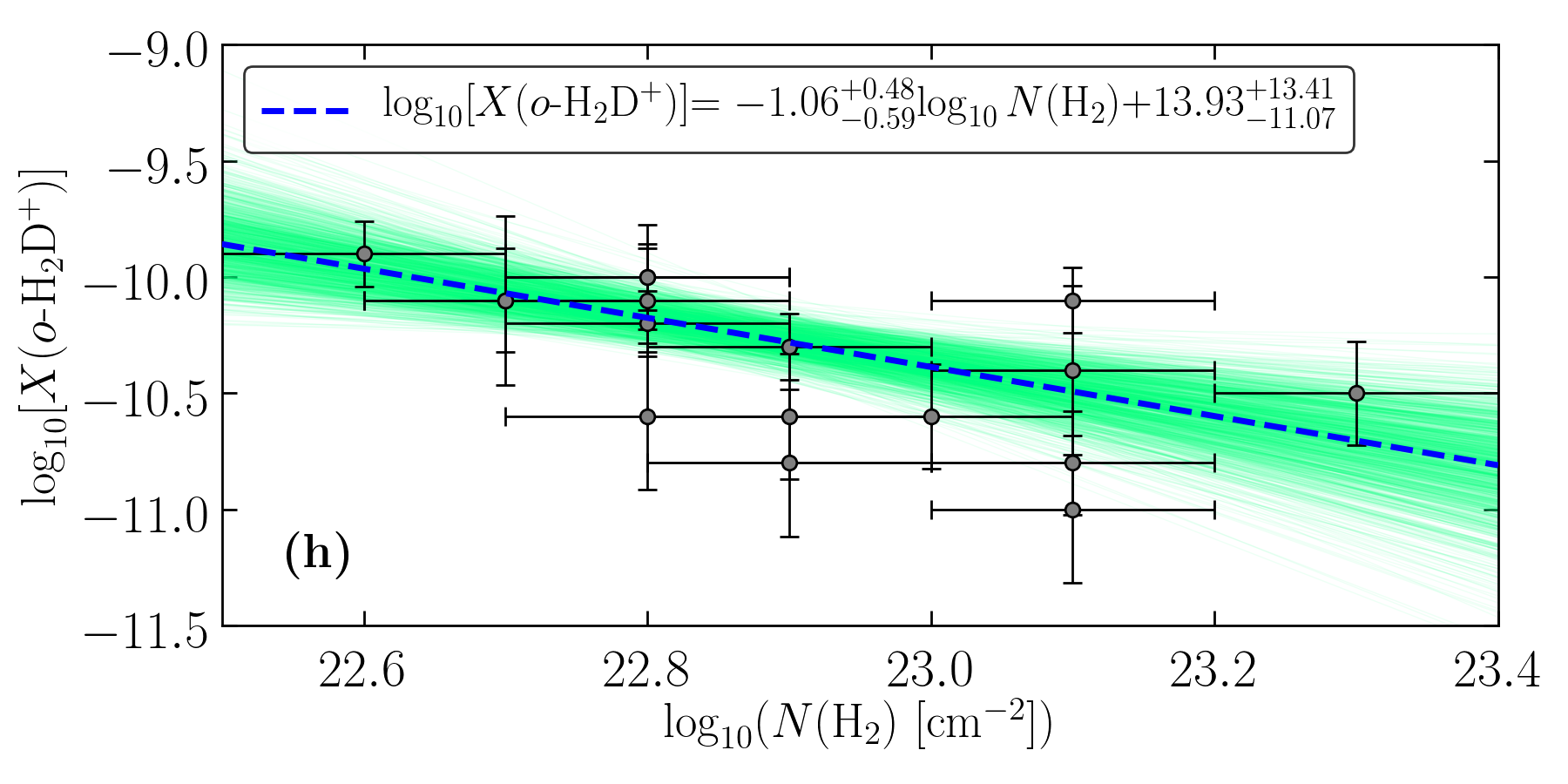}
   \includegraphics[width=0.48\hsize]{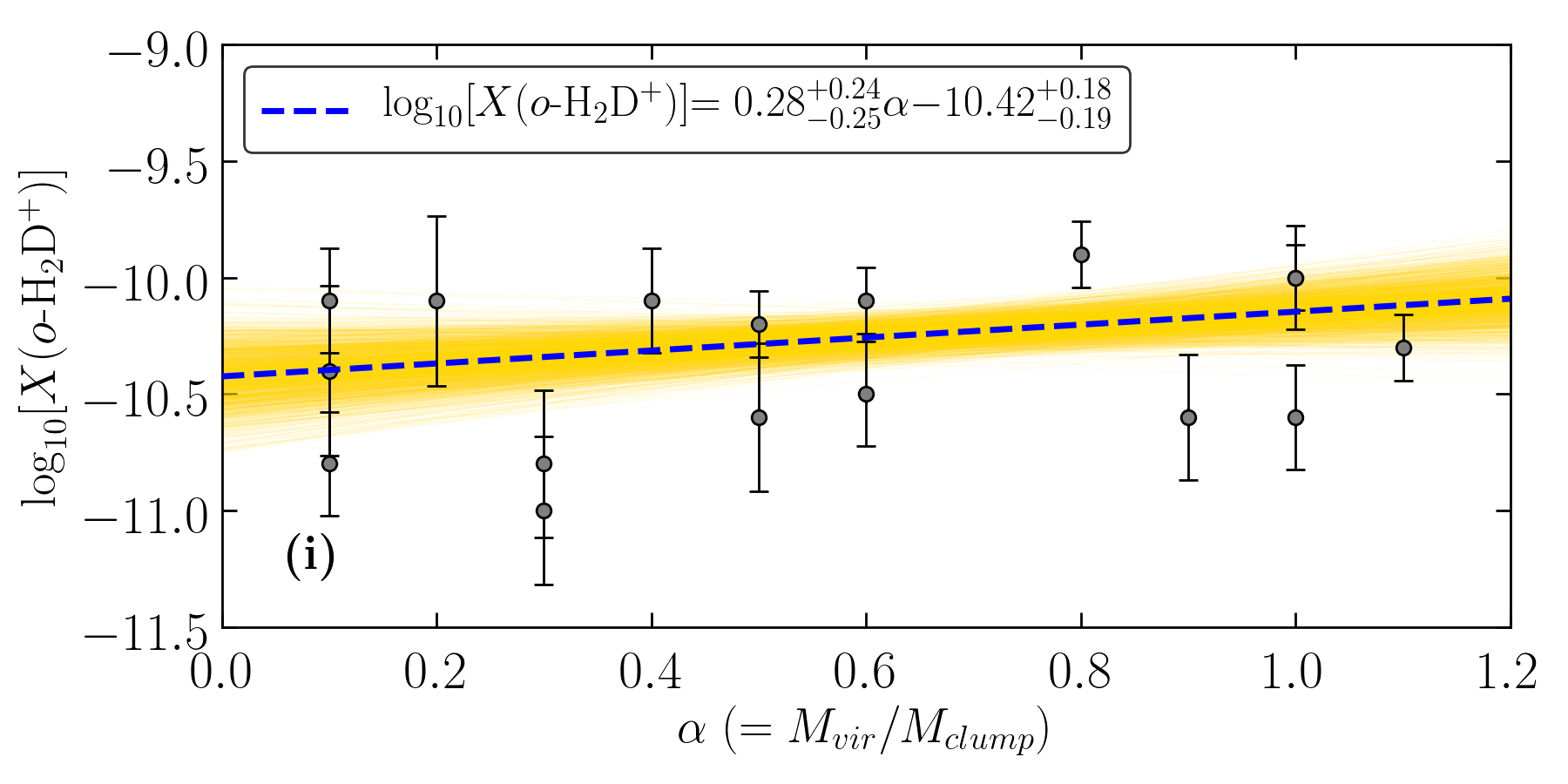}
   \includegraphics[width=0.48\hsize]{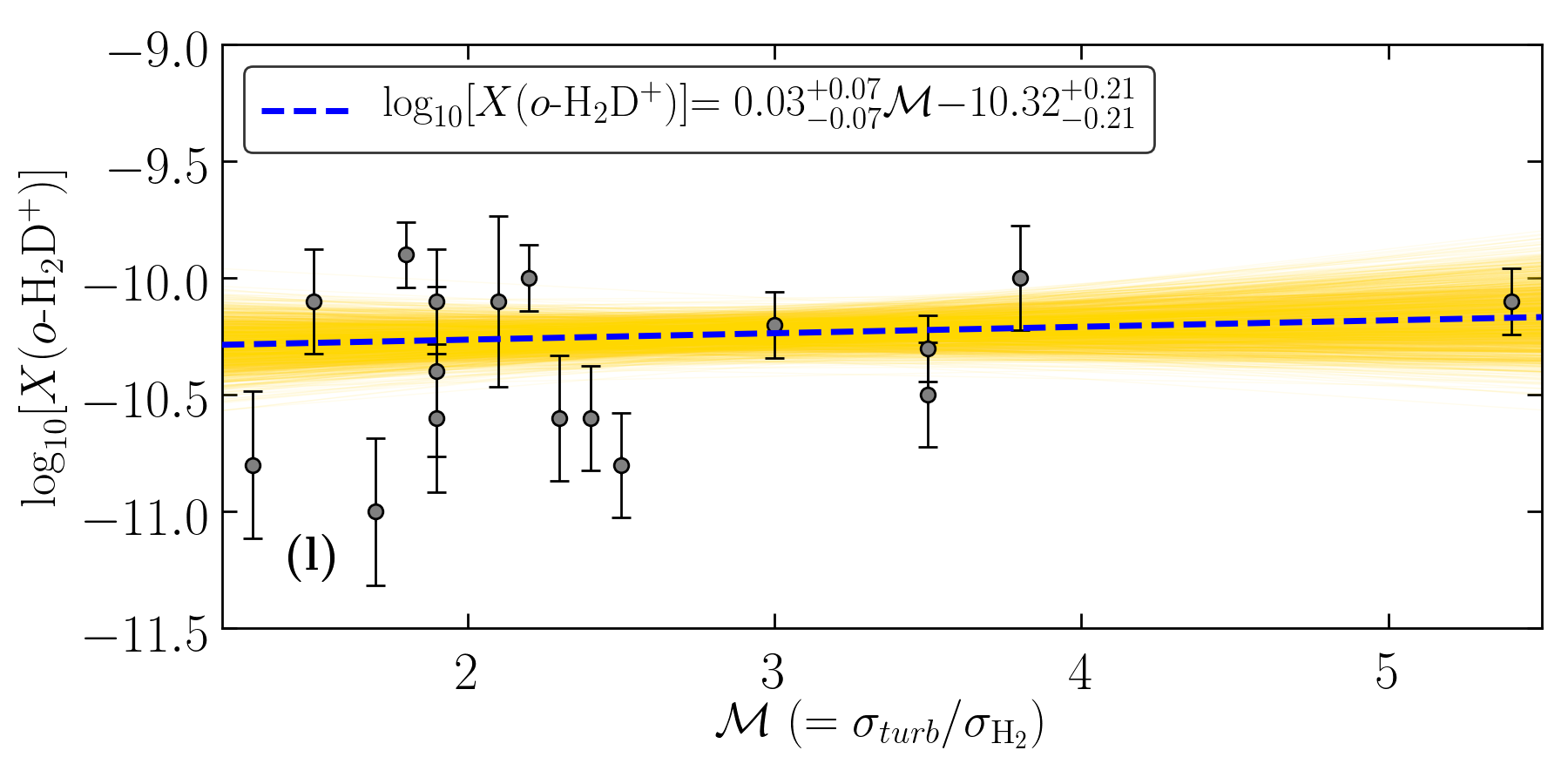}
   \caption{Collection of the different correlations between \X and the quantities summarised in Tables~\ref{tab:obsprop}~and~\ref{tab:dervprop}. Grey dots are associated with each source, while uncertainties are shown as black bars. Orange and green %\textcolor{green}{AG: (very minor) usually red means something bad} 
   shaded regions are the 3$\sigma$ results of the MCMC linear fit respectively for the unreliable ($\mathcal{B}_{2,1}<1$) and reliable ($\mathcal{B}_{2,1}>1$) correlations we found for the whole sample.
   Blue dashed lines represent the {\it fiducial model} (i.e.  with the highest likelihood among the ones explored by MCMC). The fit parameters are shown in the caption of each correlation. All plots are in log-linear scale, except for panels (c), (d), (g), and (h) where the axes are set in log-log scale.}
              \label{fig:correlations}%
    \end{figure*}
    
\section{Results}\label{sec5:results}
  We have explored possible correlations between the \mbox{\ohhdp} relative abundance obtained from our survey and the other physical quantities in Tables~\ref{tab:obsprop} and~\ref{tab:dervprop}. In the following, we divide the discussion into three main parts: correlations with ($i$) the source parameters, to examine the possible connection between the amount of \ohhdp and the quantities that characterise the clump structures and their distribution in the Galactic plane; ($ii$) the evolutionary parameters, to highlight trends during the star formation process; and ($iii$) the dynamical parameters, for possible dependencies on factors such as the concentration and the turbulence of the gas of the source.

\subsection{New correlations with $X$($o$-H$_2$D$^+$)}

\begin{table}
\caption{Bayes factors, $\mathcal{B}_{2,1}$, found for each correlation in Fig.~\ref{fig:correlations}. The correlations are computed with respect to the  abundance of $o$-H$_2$D$^+$ derived from the column densities in Tables~\ref{tab:obsprop} and \ref{tab:dervprop} (i.e. N[$o$-H$_2$D$^+$]/N[H$_2$]).}\label{tab:bayesfactors}
\setlength{\tabcolsep}{15pt}
\renewcommand{\arraystretch}{1.2}
\centering
\begin{tabular}{c|ll}
\hline\hline
Panel       & Correlation with  & Bayes factors\\
\hline                                                                                             
(a) & D$_{GC}$                     & $5.5 \times 10^{-5}$ \\
(b) & R$_{eff}$                    & $1.0 \times 10^{-3}$ \\
(c) & log$_{10}$($M_{clump}$)      & $1.0 \times 10^{-3}$ \\
(d) & log$_{10}$($L_{bol}$)        & $9.7$ \\
(e) & log$_{10}$($T_{dust}$)       & $4.5$ \\
(f) & $f_{\rm D}$                  & $4.1 \times 10^{-4}$ \\
(g) & log$_{10}$($L/M$)            & $51.1$ \\
(h) & log$_{10}$[$N$(${\rm H_2}$)] & $1.9 \times 10^{4}$ \\
(i) & $\alpha$                     & $2.4 \times 10^{-1}$ \\
(l) & $\mathcal{M}$                & $4.5 \times 10^{-5}$ \\
\hline
\end{tabular}
\end{table}

Figure~\ref{fig:correlations} shows the linear correlations we find between the $\logten$\X and the set of physical quantities of each clump listed in Tables~\ref{tab:obsprop} and \ref{tab:dervprop}. Each linear correlation was obtained using a Bayesian approach relying on Markov chain Monte Carlo (MCMC) algorithms. While the full description of the method is left to Appendix~\ref{sec:appA}, where we also show an example of the  posterior distributions of the model parameters, in each panel of Fig.~\ref{fig:correlations} we report the best fit and the models within $3\sigma$. In addition, to give statistical significance to the correlations, we  performed a Bayesian analysis by fixing the slope of the fitting function to zero and calculating the Bayes factor $\mathcal{B}_{2,1}$ (see Table~\ref{tab:bayesfactors}). According to the Jeffreys scale (\citealt{Jeffreys61}) if $\mathcal{B}_{2,1}<1$, there is no statistical evidence to justify the use of a model with two free parameters (yellow shaded regions in Fig.~\ref{fig:correlations}). Alternatively, if $\mathcal{B}_{2,1}>1$, a model with two free parameters is justified and we can consider the correlation  reliable (green shaded regions in Fig.~\ref{fig:correlations}).
We tested the same fit approach on the TOP100 sub-sample, always finding an agreement with the correlations shown in Fig.~\ref{fig:correlations}. This result excludes possible bias depending on the sample selection or on the procedures followed to derive the other values reported in Tables~\ref{tab:obsprop} and \ref{tab:dervprop}.

In the following paragraphs we  discuss each group of correlations, paying particular attention to whether the dependence  of each pair can be physically explained or if it is simply fortuitous.

\subsubsection{Source parameters}
The dependence of \X on the physical properties of clumps does not seem particularly pronounced. % and we do not expect to be \sbov{(Stefano)}. 
In the first three plots  in Fig.~\ref{fig:correlations} (panels a, b, and c), \X appears roughly uniform as the other quantities vary. In these cases the correlations we found are in agreement with a purely flat regime, with small fluctuations of the fiducial models slopes, probably caused by regions with sparse density of observed points.

Variations in the deuterium abundance with respect to hydrogen, [D/H], through the Galactic disk are expected considering that all deuterium atoms were formed at the birth of the Universe and then progressively consumed by thermonuclear reactions within the stellar cores (e.g. \citealt{Lubowich00} and \citealt{Ceccarelli14}; see also \citealt{Galli13} for a review). Following this scenario, [D/H] should vary with location and the same would be expected for deuterated species.
On the other hand, it is well known that the major boost in deuterium fractionation is driven by the star formation process itself:  the temperatures, the global chemistry, and energetics of the stars can play a central role, (dis-)favouring the deuteration process (e.g. \citealt{Caselli02, Bacmann03, Lis06, Pillai12, Ceccarelli14} and \citealt{Giannetti19}). The last dependence seems particularly evident in panel (d) of Fig.~\ref{fig:correlations}, which shows that \X clearly decreases as $L_{bol}$ rises (i.e. $\mathcal{B}_{2,1}\gg1$). The \ohhdp abundance is found to change by $\sim 0.5$ orders of magnitude for a factor of $\sim4$ change in log$_{10}$($L_{bol}$), with a fiducial model's slope of $\sim$ $-$0.23.

\subsubsection{Evolutionary parameters}
Among the quantities we defined as evolutionary tracers we have included the following: the CO-depletion factor, $f_D$, as the indicator of the depletion degree of each source, and defined as the ratio of the expected CO abundance with respect to H$_2$ to the observed value (see \citealt{Caselli99, Fontani12}); the dust temperature, which is determined by the activity of the protostars that are forming in each source; and  $L/M$ as the main indicator for the evolutionary stage of star formation processes in the high-mass regime (e.g. \citealt{Saraceno96, Molinari08, Koertgen17, Urquhart18, Giannetti19}). The values for each clump are summarised in Tables~\ref{tab:obsprop} and~\ref{tab:dervprop}. Depletion factors are derived from \cite{Giannetti14} and \cite{Sabatini19} using the C$^{17}$O column density distributions generated by taking into account optical depth effects, while $T_{dust}$ and  $L/M$ values  are taken from \cite{Giannetti17_june}, \cite{Konig17}, \cite{Urquhart18}, and \cite{Giannetti19} to which we refer for a more detailed discussion of how these quantities were estimated. Depletion factors are also the only incomplete values      due to the lack of observational data for the sources not in the TOP100. Therefore, we note that panel (f) of Fig.~\ref{fig:correlations} is the only one obtained with a reduced number of data points (i.e. we could not include G12.50-0.22, G14.23--0.51, G15.72--0.59, G316.76-0.01).

In panel (e) of Fig.~\ref{fig:correlations}, \X correlates with $T_{dust}$ with small slopes, changing by $\sim 0.5$ orders of magnitude in the range of $T_{dust}$ between 12 and 24 K. The fit with the model with two free parameters is statistically supported by a $\mathcal{B}_{2,1}>1$.
However, we note that the scale chosen for the plot and the different ranges covered by the quantities involved may confuse the interpretation of these correlations, leading one to conclude that small slopes could mean soft correlation. 

The correlation between \X and $f_D$ in Fig.~\ref{fig:correlations} (f) appears similar to that found for $T_{dust}$ (i.e. a clear correlation with a small slope). However, we find that $\mathcal{B}_{2,1}<1$, which excludes any statistical evidence to justify the use of a model with two free parameters to establish a correlation between \X and $f_D$. We note  in this case that  the reduced number of sources may complicate the fit with two free parameters, while a correlation between \X and $f_D$ is expected since the highly CO-depleted environments are those in which deuteration is favoured (e.g. \citealt{Dalgarno84} and \citealt{Caselli12b}).
The log-lin scale in Fig.~\ref{fig:correlations} (e) and (f) was chosen to allow direct comparison with \cite{Caselli08}, where a survey of the \ohhdp  towards a sample of 16 low-mass sources - between starless and protostellar cores - was presented. This study gives us the opportunity to test the connection between low- and high-mass star-forming regions, with two samples statistically comparable. \cite{Caselli08} found a progressively decreasing trend of \X at increasing temperatures. Here, we find the same behaviour even if less pronounced, but supported by an inverse correlation with $f_D$ (even if not statistically confirmed), also in agreement with \cite{Vastel06}.

We also find a strong  anti-correlation (i.e. $\mathcal{B}_{2,1}\gg1$) of \X with $L/M$; \X is found to decrease by one order of magnitude for one order of magnitude change in $L/M$, with a slope of $-0.37$. This confirms the hypothesis outlined by \cite{Giannetti19} that \textit{o}-H$_2$D$^+$ can be considered a good chemical clock during the star formation process (see Sect.~\ref{sec6.1:evolution} for a detailed discussion).

\begin{table*}[b!]
\caption{Summary of the quantities to calculate the CRIR.}\label{tab:CRIR}
\setlength{\tabcolsep}{2pt}
\renewcommand{\arraystretch}{1.2}
\centering
\begin{tabular}{l|cccccccc}
\hline\hline
ATLASGAL-ID$^a$ & N(C$^{17}$O)\tablefootmark{a}    &$\chi$(CO)\tablefootmark{a}&  N(H$^{13}$CO$^+$)  &    N(DCO$^+$)       &R$_{D}$& $\zeta_2^A$ & $\zeta_2^{\rm ref}$& $\zeta_2^B$ \\
             &             &          &               &              &                   & $\ell = 0.5{\rm R}_{eff}$ & $\ell = {\rm R}_{eff}$ & $\ell = 2{\rm R}_{eff}$\\

           &log$_{10}$(cm$^{-2}$)& ($10^{-5}$)  &log$_{10}$(cm$^{-2}$)&log$_{10}$(cm$^{-2}$)&          &   ($10^{-17}$ [s$^{-1}$]) &   ($10^{-17}$ [s$^{-1}$]) &   ($10^{-17}$ [s$^{-1}$])     \\
\hline  
G13.18+0.06  & 15.5        & 8.0      & 13.6$\pm 0.1$ & 12.4$\pm 0.2$&  0.002$\pm 0.001$ &   6.92&  3.46 & 1.73 \\   
G14.11--0.57 & 15.7        & 11.5     & 13.8$\pm 0.1$ & 13.6$\pm 0.1$&  0.011$\pm 0.004$ &   3.18&  1.59 & 0.80 \\   
G14.23--0.51 & {\bf 15.7}  & {\bf 4.6}& 14.0$\pm 0.1$ & 13.8$\pm 0.1$&  0.011$\pm 0.004$ &   3.67&  1.84 & 0.92 \\
G14.49--0.14 & 15.6        & 4.7      & 13.3$\pm 0.1$ & 13.0$\pm 0.1$&  0.011$\pm 0.003$ &   6.68&  3.34 & 1.67 \\
G14.63--0.58 & 15.5        & 5.8      & 13.6$\pm 0.1$ & 12.8$\pm 0.1$&  0.003$\pm 0.001$ &   11.62& 5.81 & 2.91 \\
G15.72--0.59 & {\bf 15.6}  & {\bf 11.1}& 13.2$\pm 0.1$ & 13.2$\pm 0.1$&  0.018$\pm 0.006$ &  3.06&  1.53 & 0.77 \\
G18.61--0.07 & 15.3        & 4.6      & 13.5$\pm 0.1$ & 12.9$\pm 0.1$&  0.005$\pm 0.002$ &   1.33&  0.67 & 0.34 \\
G19.88--0.54 & 15.7        & 5.8      & 13.8$\pm 0.1$ & 13.2$\pm 0.1$&  0.005$\pm 0.002$ &   3.00&  1.50 & 0.75 \\
%Csegeri' L: 0.4, 0.4,0.3, 0.7, 0.3, 0.3, 0.6, 0.5 pc -> leading to <zeta_2> lower by a factor ~2.3
\hline
\end{tabular}

\tablefoot{$\zeta_2$ values are calculated following the recent results of \citet{Bovino20} and assuming three values of $\ell$ in eq.~\ref{eq:zeta2};
\tablefoottext{a}{data and calculations from \citet{Giannetti14}. We assumed a constant [$^{18}$O]/[$^{17}$O] $=4.16$, as in \citet{Wouterloot08}. Values in boldface were taken from \citet{Csengeri16}}.

%\tablefoottext{c}{}
}
\end{table*}

\subsubsection{Dynamical parameters}
We analyse here the possible correlations between \X and the total H$_2$ column density, the virial parameter, and the Mach number, which we consider quantities related to the dynamical stage of the clumps. The correlation we find between \X and N(H$_2$), shown in Fig.~\ref{fig:correlations} (h), suggests that an increased density could also play a relevant role in the deuteration process. The \ohhdp abundance is found to change by one order of magnitude for less than one order of magnitude change in log$_{10}$[$N({\rm H_2}$)], with a fiducial model slope of $\sim$ $-$1.06.
Frequent collisions are expected in denser environments (i.e. fast chemical kinetics) that would accelerate the formation of \textit{o}-H$_2$D$^+$. On the other hand, when moving to a denser environment,  the conversion of H$_2$D$^+$ in heavier isotopologues (D$_2$H$^+$ and D$_3^+$) or the deuterium transfer from H$_2$D$^+$ to other chemical species (e.g. N$_2$ and CO) can also be boosted, as suggested by \cite{Giannetti19} .

The last two panels of Fig.~\ref{fig:correlations}, (i) and (l), present the correlations with $\alpha$ and $\mathcal{M}$ calculated from the \ohhdp detected lines, as described in Sect.~\ref{sect:dinamical_q}. Although the virial parameter and the Mach number depend on quantities for which a correlation with \X~was established, such as the line FWHM, on which both $M_{vir}$ and $\sigma_{\nu}$ depend, or the dust temperature, we do not find any particularly relevant trend with \X. We find that \X does not show a significant variation with $\alpha$ and $\mathcal{M}$, which means that  the slopes are in agreement (within the uncertainties) with zero in these log-linear plots.\\

To summarise, we find strong correlations between \X and different physical quantities of the clumps, some expected and  others less trivial. In Fig.~\ref{fig:correlations} the correlations with $\mathcal{B}_{2,1}>1$ are highlighted in green. To avoid any possible bias we repeated the MCMC fit approach looking for correlations with the \ohhdp column densities, and we found that only the correlation between N($o$-H$_2$D$^+$) and N(H$_2$) is not confirmed.  All the other parameters in which a correlation is confirmed for both \X and N($o$-H$_2$D$^+$) are more or less directly connected to the evolution of the clumps. We conclude that the deuterium fractionation seems to be driven more by a temporal evolution  than by a dynamical evolution of the star-forming regions, once the minimum physical conditions required to boost deuteration are fulfilled (i.e. cold, dense, and highly CO-depleted regions).

\subsection{Estimates of the CRIR}\label{sec:CRIR}
Recently, a new way to estimate the cosmic-ray ionisation rate (CRIR) of hydrogen molecules, $\zeta_2$, was presented by \cite{Bovino20}, based on H$_2$D$^+$ and other H$^+_3$ isotopologues. Our survey provides the opportunity to test, for the first time, this new method on a large sample of massive star-forming regions with simultaneous detections of \ohhdp, DCO$^+$, H$^{13}$CO$^+$, and C$^{17}$O. 
According to \cite{Bovino20} the CRIR can be obtained from

\begin{equation}\label{eq:zeta2}
    \zeta_2 = {\bar \alpha}\:k^{{\rm H^+_3}}_{{\rm CO}}\:\frac{\chi{\rm (CO)}\:{\rm N(H^+_3)}}{\ell}
,\end{equation} 

\noindent where ${\bar \alpha} = 0.77$ is the correction factor that takes into account the errors due to approximations in the method \citep[see][for details]{Bovino20}; $k^{{\rm H^+_3}}_{{\rm CO}}$ is the rate at which CO destroys ${\rm H^+_3}$; $\ell$ is the path length over which the column densities are estimated; and $\chi$(CO) is the relative abundance of CO with respect to H$_2$, which we derive by the simple formula: 

\begin{equation}
   \chi({\rm CO}) =\phi_{18/17} \frac{{\rm N(C^{17}O)}}{{\rm N(H_2)}}  \frac{{\rm [^{16}O]}}{{\rm [^{18}O]}}.
\end{equation}

\noindent Here $\phi_{18/17}$ is the isotopic ratio of ${\rm [^{18}O]/[^{17}O]} = 4.16$ adopted from \cite{Wouterloot08}. For the TOP100 sources, the C$^{17}$O column densities were taken from \cite{Giannetti14}, while for ATLASGAL sources not contained in the TOP100 we used the data described by \cite{Csengeri16} (bold values in Table~\ref{tab:CRIR}).

\begin{figure}
\centering
{\includegraphics[width=1\columnwidth]{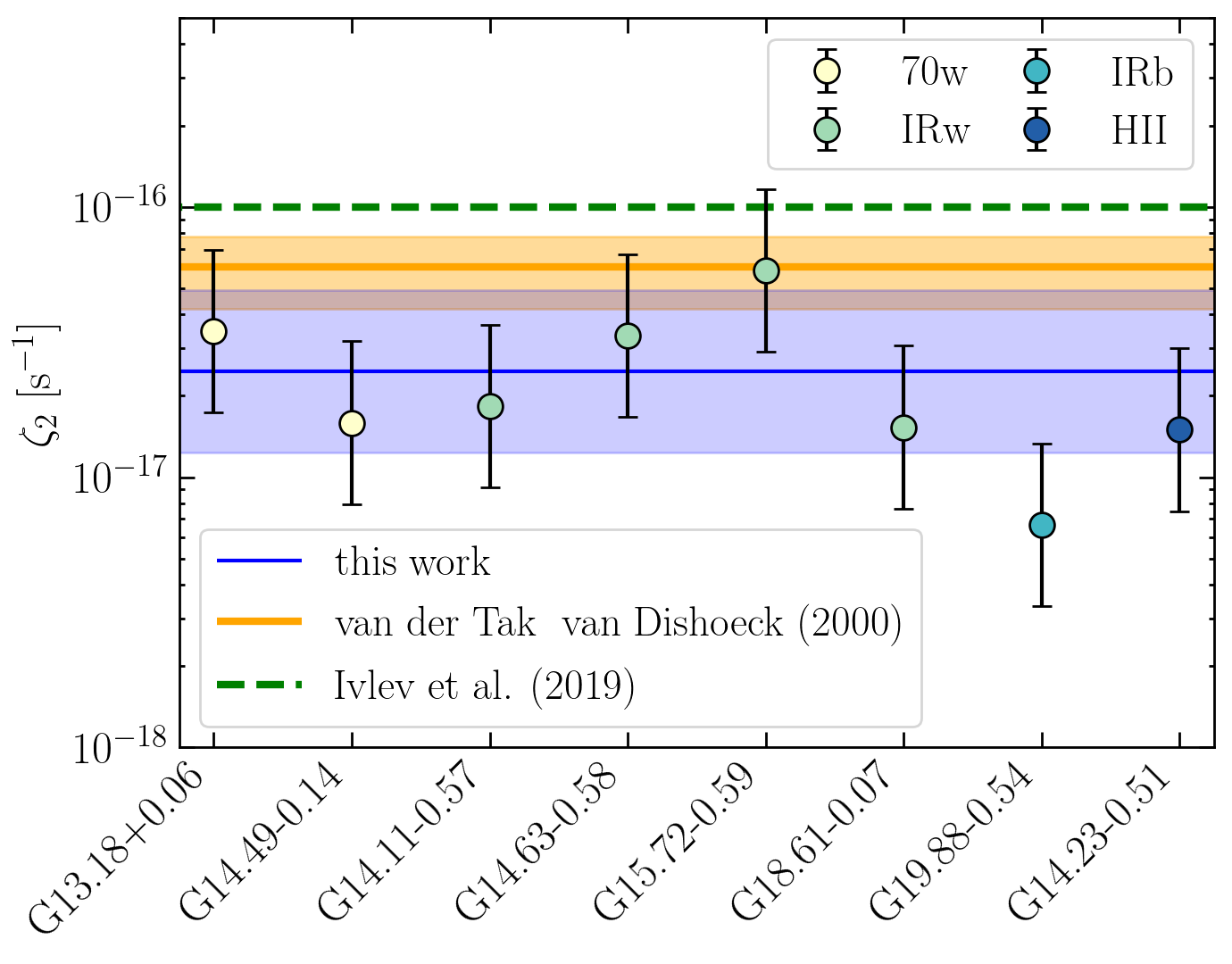}}
\caption{Estimates of the CRIR of hydrogen molecules, $\zeta_2$, derived for a sub-sample of sources for which H$_2$D$^+$, DCO$^+$, H$^{13}$CO$^+$, and C$^{17}$O observations are available, and assuming $\ell={\rm R}_{eff}$ in eq.~\ref{eq:zeta2},  representing our reference case (dots). The sources are arranged from left to right following their evolutionary phase (bottom left legend). The bars associated with each point represent the variability of our results with $\ell$ (i.e. $\ell=0.5{\rm R}_{eff}$ and $\ell=2{\rm R}_{eff}$ for the upper  and lower limit, respectively). The blue line is our mean value for $\zeta_2$ for the reference case, while the blue shaded area is its variability with $\ell$. The orange line and shaded area represents the estimates of \citet{VanDerTak00}, obtained by converting the CRIR for hydrogen atom, $\zeta_{{\rm H}}$, via $\zeta_2 = 2.3 \times \zeta_{{\rm H}}$. The green line is the recent upper limit for L1544 reported by \citet{Ivlev19}.}\label{fig:CRIR}
\end{figure}

\begin{figure}
\centering
{\includegraphics[width=1\columnwidth]{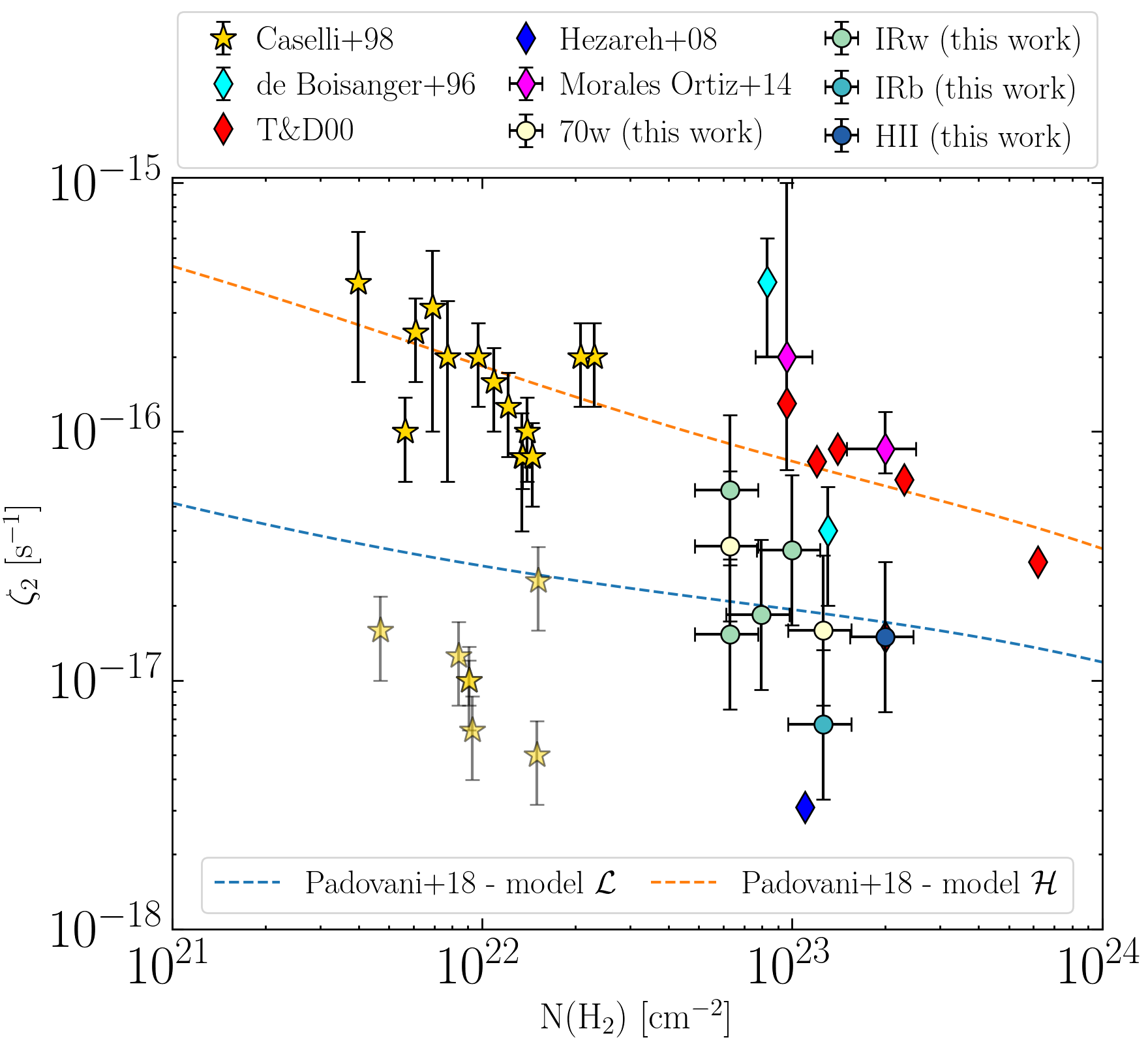}}
\caption{$\zeta_2$ variation as a function of N(H$_2$). Circles refer to our estimates of $\zeta_2$ in high-mass star-forming regions in different evolutionary stages (same colour-coding as in Fig.~\ref{fig:CRIR}). $Yellow~stars$ show the estimates of \cite{Caselli98} for a large sample of low-mass cores, while $diamonds$ represent high-mass cores ($cyan$ from \citealt{deBoisanger96}; $red$ from \citealt{VanDerTak00} (T\&D00) for  $\zeta_2$ estimates  and from \citealt{vanderTak00b} and \citealt{Doty02} for  N(H$_2$) values; $blue$ from \citealt{Hezareh08}; $magenta$ from \citealt{MoralesOrtiz14}). Dashed lines show the models discussed in \cite{Padovani18} assuming different slopes for the CR proton spectrum. Different levels of transparency have been used to separate what we considered outliers (pale color) from the other estimates (full colour). This plot was readapted from \cite{Padovani09}.}
\label{fig:NH2-CRIR}
\end{figure}

 In eq. 5 N(${\rm H^+_3}$) is the column density of ${\rm H^+_3}$ derived from the \mbox{\ohhdp} and the HCO$^+$ deuterium fractionation, $R_\mathrm{D}$, that we {calculate} from the DCO$^+$ and H$^{13}$CO$^+$ column densities reported in Table~\ref{tab:CRIR}. Column densities were derived consistently with the procedure employed for \ohhdp\ (i.e. using MCWeeds and by considering how the $^{12}$C/$^{13}$C ratio varies as a function of D$_{GC}$; see discussion in \citealt{Giannetti14}): $[^{12}$C$]/[^{13}$C$] = 6.1^{+1.1}_{-1.8}\:$D$_{GC}+14.3^{+7.7}_{-7.2}$. Since the real size of the region associated with the \ohhdp emission is unknown, we assume three multiples of R$_{eff}$ as representative cases to calculate $\zeta_2$: (a) $\ell = 0.5 {\rm R}_{eff}$, which approximately covers a range of $\ell$ comparable with the mean core size associated with the \ohhdp emission reported by \cite{Pillai12} (i.e. $\sim$0.2 pc); (b) $\ell = {\rm R}_{eff}$  and (c) $\ell = 2 {\rm R}_{eff}$, which implicitly means that the \ohhdp is emitting on the full clump size defined at 870 $\mu$m. 

Our estimates of $\zeta_2$ are summarised in Table~\ref{tab:CRIR}. 
In Fig.~\ref{fig:CRIR} we present these results arranged, from left to right, according to the evolutionary stage of each source. The blue line represents the mean value of $\zeta_2 = 2.5 \times 10^{-17} s^{-1}$ for $\ell = {\rm R}_{eff}$, while the  blue shaded area is its variation with $\ell$. It is worth noting that the range of values spans more or less an order of magnitude, which is in line with typical errors reported from models (see e.g.  \citealt{Padovani18} and reference therein). We compare this result with recent estimates from \cite{Ivlev19}, who derive an upper limit of $\zeta_2\sim 10^{-16} s^{-1}$ from a self-consistent model for the equilibrium gas temperature and size-dependent dust temperature in the prestellar core L1544. 
A second comparison is made with the estimate obtained by \cite{VanDerTak00} in a sample of seven young massive stars using models based on H$^{13}$CO$^+$ observations. They found an average value of $\zeta_2\sim(6.0 \pm 1.8) \times 10^{-17} s^{-1}$. Looking at our reference case, most of the sources show values lower than the average obtained by \cite{VanDerTak00}. 

The analysis we report has the advantage of being model-independent, but it is purely qualitative, first because the method proposed in \citet{Bovino20} includes strong approximations, and second because its validity has been shown for very small regions (i.e. $R =\ell/2 \leq 0.05$ pc, where $R$ is the distance from the centre of the clump). 
However, as the error here is driven by the uncertainties in the column density calculations, we can consider the CRIR estimates  valid and robust within the variability shown by the different $\ell$. In addition, this represents the only viable and model-independent way to estimate the CRIR in dense regions.

Figure~\ref{fig:NH2-CRIR} shows how the cosmic-ray ionisation rate varies as a function of N(H$_2$) for a sample of low- and high-mass star-forming regions. Considering the associated uncertainties, the data cover a range of $\zeta_2$ between $\sim 3 \times 10^{-18}$ and $\sim 10^{-15}$ s$^{-1}$, while the molecular hydrogen column densities are in the range $\sim (0.4-80) \times 10^{22}$ cm$^{-2}$. The yellow stars are the $\zeta_2$ values computed in low-mass cores from \cite{Caselli98}, while the values for N(H$_2$) are taken from \cite{Butner95}. Diamonds indicate massive protostellar envelopes (see the caption). The circles represent the estimates obtained in this work for the reference $\ell$ (same colour-coding as in Fig.~\ref{fig:CRIR}).
The dashed lines are the models discussed in \cite{Padovani18}, considering a single cosmic-ray (CR) electron spectrum and two different CR proton spectra, namely model $\mathcal{L}$ (in blue) and model $\mathcal{H}$ (in orange), and depending on the slope of the CR proton spectrum (see also \citealt{Ivlev15} for more details about the CR spectrum model).

The distribution of $\zeta_2$ in Fig.~\ref{fig:NH2-CRIR}, and in particular that of the full-coloured data points, seems to suggest the general trend where the CRIR decreases while the H$_2$ column density increases. We note that the estimates of $\zeta_2$ in high-mass star-forming regions are smaller by a factor of $\sim 3-4$ with respect to their low-mass counterparts, which is in agreement with the fact that IRDCs are usually much denser compared to the low-mass regime. Our estimates fall  between the two  models of \cite{Padovani18}, while a small sub-sample of low-mass cores ($\sim$ 1/3 of the total sample) and a few high-mass star-forming regions find agreement with the model with lower CR proton energy. The apparent bimodality (yellow and pale yellow stars in Fig. 5) in the low-mass regime can be explained by the different $f_D$ values assumed by \cite{Caselli98} to estimate $\zeta_2$. 
Here, by following \cite{Padovani09}, we report the best $\zeta_2$ estimates from \cite{Caselli98}, obtained from HC$_3$N/CO data assuming $f_D=3-5$. In particular, values of $\zeta_2$ around $\sim 10^{-16}$ s$^{-1}$ are produced by both of  the $f_D$ values, while those around $\sim 10^{-17}$ s$^{-1}$ correspond to $f_D=3$. On the other hand,  the morphology of the magnetic field lines can {also} play a major role in determining the CRIR. \cite{Padovani11} and \cite{Padovani13} have shown that even in high-density regions, if the magnetic field lines are very concentrated around the accreting protostar, $\zeta_2$ can be reduced by up to a factor of 10 or more.\\
\begin{figure}
\centering
{\includegraphics[width=0.99\columnwidth]{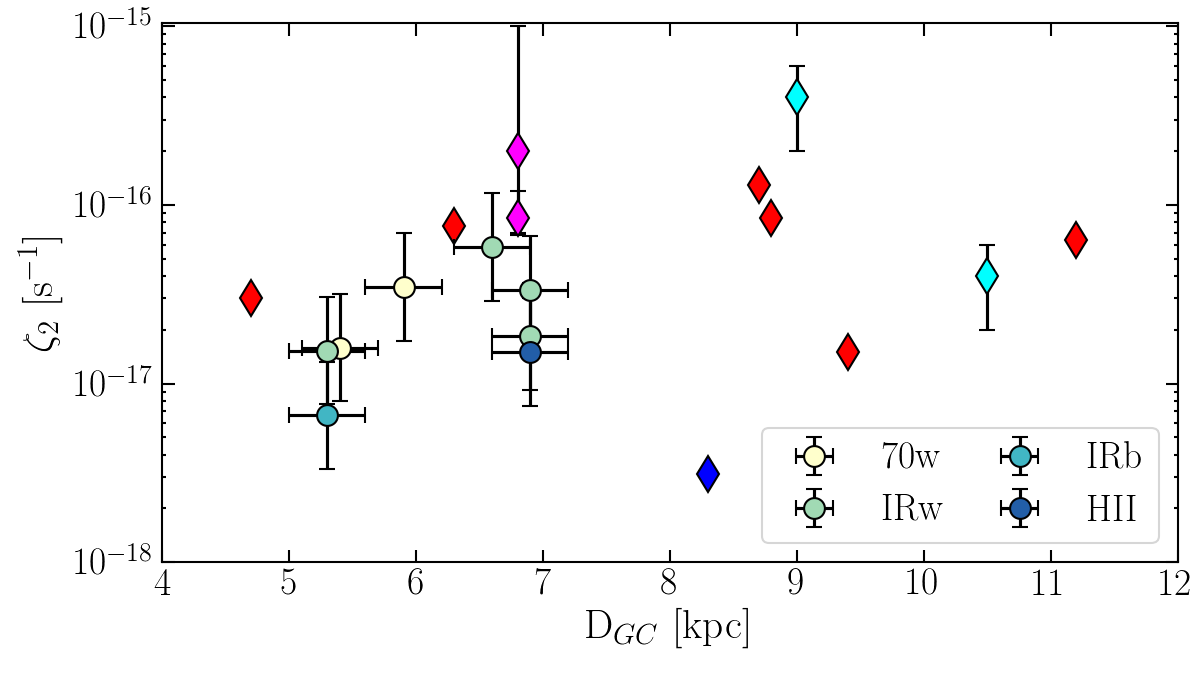}}
\caption{$\zeta_2$ variation as a function of D$_{GC}$ for the high-mass cores shown in Fig.~\ref{fig:NH2-CRIR}. Circles refer to our estimates of $\zeta_2$ in high-mass star-forming regions in different evolutionary stages (see legend). The same colour-coding as in Fig.~\ref{fig:NH2-CRIR} is assumed for diamonds. The galactocentric radii are taken from Table~\ref{tab:obsprop} (circles), \cite{VanDerTak00} ($red$), \citealt{MoralesOrtiz14} ($magenta$), \citealt{Winkel17} ($blue$), \citealt{Spina17}, and \citealt{Yan19} ($cyan$). D$_{CG}$ are scaled assuming a distance to the Galactic Center of 8.35 kpc.}
\label{fig:dcg-CRIR}
\end{figure}
\indent Finally, in Fig.~\ref{fig:dcg-CRIR} we report the cosmic ray ionisation rate as a function of the  galactocentric distances for different high-mass star-forming clumps. We found that there is no significant variation of $\zeta_2$ with D$_{CG}$ for distances  between 4.7 and 11.2 kpc.  This result is  in agreement with those of \cite{Indriolo15}, suggesting that the CRIR is  uniform for sources located at distances above 5 kpc from the Galactic centre. Therefore, the observed spread in Fig.~\ref{fig:NH2-CRIR} could be the signature of local effects,  for instance different magnetic field topology of each source.

\section{Discussion}\label{sec6:dicussion}
\subsection{\ohhdp as an evolutionary tracer for massive clumps}\label{sec6.1:evolution}
\begin{figure}
\centering
{\includegraphics[width=\columnwidth]{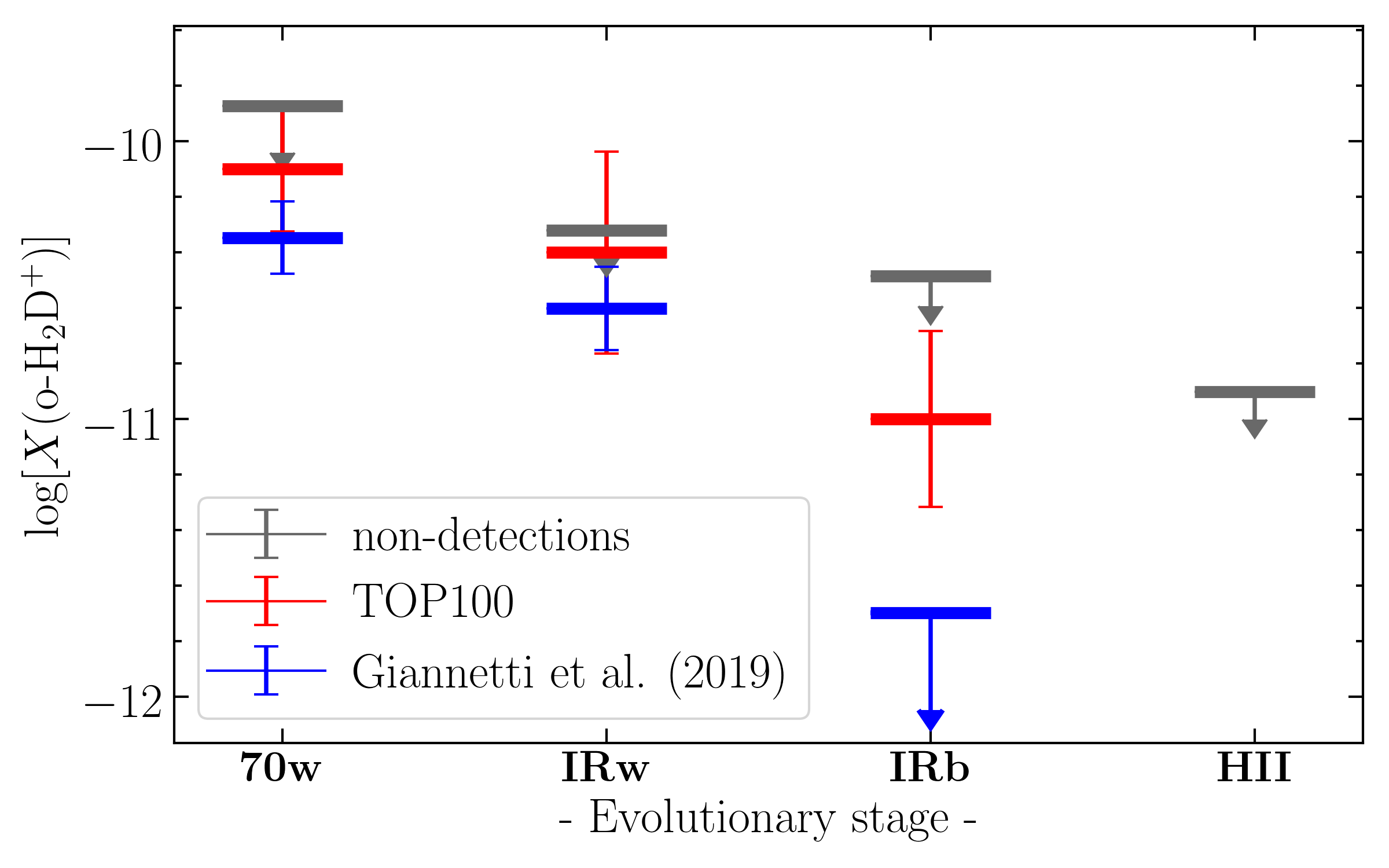}}
\caption{Average \X abundance as a function of the evolutionary classes in our sample. Blue markers indicate the recent estimates of \cite{Giannetti19}, while the red and grey ones indicate the mean of detection and the median \X derived from detection limits, respectively. Uncertainties are derived as mean and median as well, considering the errors on the individual source.}
\label{fig:trend}
\end{figure}

Figure~\ref{fig:trend} shows the abundance of \ohhdp as a function of the evolutionary class of the TOP100 clumps (red markers and error-bars). To mitigate the possible bias caused by extreme sources with peculiar initial conditions, and to make our result as general as possible, we report the median \X values for each evolutionary class. \X varies by more than one order of magnitude between the least evolved and the most advanced stage of evolution, suggesting that the deuterium fractionation of H$_3^+$ is favoured during the initial phases of star formation, while its chemical products, in this case $o$-H$_2$D$^+$, slowly disappear as massive clumps evolve. We repeated this procedure  for the detection limits discussed in Sect.~\ref{sec4:analisys} to exclude possible sensitivity effects. The results are reported in Fig.~\ref{fig:trend} as grey markers. It is clear that the trend mentioned above holds: \ohhdp decreases as the evolution progresses.

Between the two solutions, we found a discrepancy (always in agreement within the error bars) of at most a factor of 6, which is expected if we consider the different noise levels of the spectra. The trend found in Fig.~\ref{fig:trend} is a different manifestation of the result already shown in panel (g) of Fig.~\ref{fig:correlations}, where the evolution of the clumps is indicated by a progressive increase of the luminosity-to-mass ratio (e.g. \citealt{Saraceno96, Molinari08}) and \X shows the same decreasing behaviour passing from the lowest to the highest $L/M$ values (i.e. from less to more evolved sources). The same result also emerges from Fig.~\ref{fig:correlations}~(d) since the bolometric luminosity traces the evolution of clumps if we assume that during the star formation process the clump mass stays constant. This latter assumption seems reliable as \X correlates with all the other evolutionary indicators, while no correlation was found between \X and $M_{clump}$ (Fig.~\ref{fig:correlations}c), in agreement with the results of \cite{Konig17} and \cite{Urquhart18}, who find no trend in the clump mass between the evolutionary classes.

We compared the results found in Fig.~\ref{fig:trend} with those of \cite{Giannetti19}, where the same clear downward trend was observed in the \X~of three massive clumps associated with different evolutionary stages and harboured in the IRDC G351.77–0.51. Although less pronounced, similar results were recently reported by \cite{Miettinen20} for three prestellar and three protostellar low-mass cores in Orion-B9. 
In the less evolved sources with detection (i.e. 70w and IRw), the agreement with \cite{Giannetti19} is within the associated error limits, while comparing our IRb detection with the upper limit provided by \cite{Giannetti19}, we found \X higher by a factor of $\sim$5, which might depend on different initial conditions (both physical and chemical). These results confirm that through the evolutionary sequence of massive star-forming regions a general downward trend {for $o$-H$_2$D$^+$ is clearly observable with} upper limits associated with the HII class. The same trend is also visible if the average abundances reported by \cite{Miettinen20} are considered.

Figure~\ref{fig:trend}   suggests that the  observed \X trend is not influenced by the nature of the samples as a similar behaviour has been reported both within the same star formation complex \citep{Giannetti19} and within a heterogeneous sample (this work). Our new confirmation of the same trend is completely independent from  the initial chemical conditions of the ISM in which the star-forming complexes were formed and from any particular or stochastic episodes during the star formation process (i.e. possible outflows or different mass accretion rates).

\subsection{\ohhdp in a broader scenario of star formation}\label{sec6.2:SF}
The fact that \ohhdp can be considered  a valid chemical clock for the star formation process is a consequence of how strongly its chemistry is affected by changes in density and temperature during cloud contraction. The star formation process is conventionally   assumed to start in starless or  prestellar cores, that are dense, $n$(H$_2$) $\sim$ {$10^4$-$10^5$} cm$^{-3}$, and cold, $T <20$ K, clouds of gas and dust in which gravitational collapse has not yet necessarily started (e.g. \citealt{BerginTafalla07}). In these dense regions dust extinction becomes higher than $\sim$10 mag, preventing molecules from  being photodissociated or photoionised  by UV photons, and the only way to form positive ions is through cosmic ray-induced chemical processes. An example of this process is the formation of H$_3^+$, the molecular precursor of H$_2$D$^+$, produced by the interactions of a CR particle ($c.r.$) that ionises an H$_2$ molecule, and followed by a fast charge exchange reaction with H$_2$:
\begin{equation}\label{reaction1}
 {\rm H_2} + c.r. \rightarrow {\rm H^+_2 + e^-}, 
\end{equation}
\begin{equation}\label{reaction2}
 {\rm H_2^+} + {\rm H_2} \rightarrow {\rm H^+_3 + H}. 
\end{equation}

\noindent
The efficiency of H$^+_3$ production in reaction~(\ref{reaction2}) only depends on the CRIR. The deuterium fractionation can then continue with the formation of H$_2$D$^+$ via the proton-transfer reaction:
\begin{equation}\label{reaction3}
  {\rm H_3^+ + HD \rightleftharpoons H_2D^+ + H_2 + \Delta E}. 
\end{equation}

\noindent This reaction is exothermic in the forward direction unless there is a substantial fraction of o-H$_2$ (e.g. \citealt{Gerlich02}), with a released energy ${\rm  \Delta E}$ that depends on the H$^+_3$ and H$_2$D$^+$ isomer states involved (e.g. \citealt{Hugo07}). In addition, for temperatures below 20 K, the reaction tends to favour the formation of the $ortho$ state of H$_2$D$^+$ compared to its $para$ form, as demonstrated by \cite{Pagani92_model}. The thermal state of the clump represents a crucial {parameter to understand} the evolution of \ohhdp along the different evolutionary classes.

According to \cite{Urquhart18}, each evolutionary class of the ATLASGAL survey  shows a clear separation in temperature that increases from $\sim$10 to $\sim$40 K, in line with the expected evolutionary sequence. The same behaviour was also noted in the TOP100, a further confirmation of the statistical relevance of this flux-limited sub-sample as representative of the whole ATLASGAL (e.g. \citealt{Konig17}). \citet{Urquhart18} also pointed to positive strong correlations that link the gas temperature of the clumps to their embedded massive protostar luminosity or to the luminosity-to-mass ratio of each clump. 
%The latter is also an empirical correlation with a distance-independent parameter, which eliminates one of the most tricky astronomical factors of uncertainty. 
Our results resemble a similar evolutionary behaviour in the \X-$T_{dust}$ correlation reported in panel (e) of Fig.~\ref{fig:correlations}, where it is visible that higher values of \X are reached in colder environments, while as $T_{dust}$ increases, \X become progressively lower.
The temperature is also directly responsible for the evolution of CO in these regions. Low concentrations of CO molecules in the gas phase have been observed in cold environments (e.g. \citealt{Roberts00, Bacmann03, Ceccarelli14}), and correlations of the CO abundance with the evolutionary stages have been reported in the low-mass regime (e.g. \citealt{Caselli98,Bacmann02}). The same behaviour was also confirmed  in high-mass star-forming regions, where $f_D$ was found to decrease with the increase of $L/M$, as the indirect confirmation that the clumps become warmer with time (see \citealt{Giannetti14, Giannetti17_june}).\\
In highly CO-depleted environments the formation of H$_2$D$^+$ is made more efficient  by two events. On the one hand, the H$_3^+$ reservoir is kept available for reaction (\ref{reaction3}) as the following proton-transfer reaction is slowed down:
    \begin{equation}\label{reaction4}
    {\rm H^+_3 + CO \rightarrow HCO^+ + H_2}.
    \end{equation}
  %  \begin{equation}\label{reaction5}
 %   {\rm H^+_3 + N_2 \rightarrow N_2H^+ + H_2}, 
 %   \end{equation}

\noindent On the other hand, due to the same effect, H$_2$D$^+$ is prevented from reacting with CO and, to a lesser extent, with N$_2$ (which normally depletes slowly) to form DCO$^+$ and N$_2$D$^+$, remaining abundant and observable in gas phase. The $f_D$-\X correlation shown in Fig.~\ref{fig:correlations} (f), qualitatively supports this scenario. We note that in this case the {highest} \X correspond to the {highest} $f_D$ values, with an opposite trend respect to what is found for $T_{dust}$. This anti-correlation is expected as cold (and dense) environments are those in which CO-depletion is boosted (e.g. \citealt{Kramer99, Caselli99, Crapsi05, Fontani12, Wiles16, Sabatini19}).

It is also clear, from the obtained results, that the answer to how (and if) the observed \X are linked to the clumps' dynamical picture across their evolution is far from being exhaustive. In agreement with the results of \cite{Konig17}, we do not find any evident trend of $\alpha$ and $\mathcal{M}$ with the variation of the evolutionary stage in the TOP100 sources. We only note that for the entire sample $\alpha$ is found to be $\lesssim 1$, which indicates dynamically unstable sources and very fast collapses (see \citealt{Kauffmann13} for a more detailed description). 
Interestingly, similar results have been obtained by numerical simulations \citep{Koertgen17,Koertgen18} in which the influence of dynamical parameters like the  Mach number ($\mathcal{M}$), the magnetic field magnitude and distribution, and the cloud gas surface density have been shown to affect the results much less than the time evolution of the clumps. 
This prevents us from determining whether the correlations found in panels (i) and (l) of Fig.~\ref{fig:correlations} support the scenario in which \X and the clumps' dynamical quantities are not empirically correlated, and the possibility that our sample may be affected by some type of selection bias. However, the latter is not supported by the other correlations we report, while it is more likely that the link between the emitting region  associated with \ohhdp and the clump dynamics is not  so trivial. In addition, we should also consider   that the H$_2$D$^+$ velocity dispersion cannot give us full information on the clump dynamics because close to the centre of the clumps it is quickly replaced by N$_2$D$^+$ or D$_2$H$^+$ or destroyed by the presence of freshly desorbed CO. High spectral resolution observations are necessary in order to estimate with sufficient accuracy the line FWHMs, and consequently $\alpha$ and $\mathcal{M}$, together with high spatial resolution ($< 0.2$ pc; e.g. \citealt{Pillai12}), to be able to distinguish the dynamics of the sub-clumps associated with the \ohhdp emission. 

Overall, our results suggest that the \ohhdp evolution follows the physical conditions associated with the different evolutionary stages:  high abundances in the cold starless or prestellar stages (70w and IRw), and a clear decrease while the evolution of the protostellar object proceeds from its young phases to more evolved situations, including HII regions. This is in particular confirmed by the correlations reported in Fig.~\ref{fig:correlations}, which support the scenario in which deuteration proceeds faster in the  first stages favoured by a high degree of CO freeze-out and drops at later stages mainly due to the increase in temperature induced by the  presence of a luminous YSO and the subsequent release of CO back into gas-phase.

\section{Summary and conclusions}\label{sec7:conclusion}
With the aim of confirming that $o$-H$_2$D$^+$ can be used as a chemical clock to follow the evolution of massive clumps, in this work we presented the first survey of \ohhdp in high-mass star-forming regions. We collected more than $\sim 10^4$ spectra in 106 sources of the ATLASGAL-survey and almost entirely belonging to the TOP100 sample. We found 16 sources with reliable detections of \ohhdp (with {iS/N} $\gg 3$), from which we retrieved column densities and relative abundances of $o$-H$_2$D$^+$ with respect to H$_2$ by fitting the line profiles under the assumption of LTE. From the line fit outputs, we also calculated the dynamical parameters of each clump (see Tables~\ref{tab:obsprop} and \ref{tab:dervprop}). The results of this work can be summarised as follows:

\begin{enumerate}
    \item We confirm, also in the high-mass regime of star formation, the empirical correlation between \X and $T_{dust}$, and we find new correlations with $L_{bol}$ and $L/M$. \X has been found to correlate also with N(H$_2$), but we interpret this result as less relevant for the H$_3^+$ deuterium fractionation process since it is not supported by an evident correlation with N($o$-H$_2$D$^+$), which instead has been found with $f_D$ and with all the other quantities connected to the  evolutionary stage of the clump(s). It is  likely that a denser clump might have already enabled the conversion of \ohhdp into other deuterated  species (e.g. D$_2$H$^+$ and N$_2$D$^+$) due to fast kinetics, with the effect of reducing the abundance of $o$-H$_2$D$^+$. 
    \item From the additional detections of DCO$^+$, H$^{13}$CO$^+$, and C$^{17}$O available in our sample, we have added six new estimates of $\zeta_2$ in massive star-forming regions by following the analytical formulae recently reported by \citet{Bovino20}. We find a variation in the estimated $\zeta_2$ with the H$_2$ column density. Nevertheless, while being connected to the morphology of the magnetic field lines of each source, as discussed by \cite{Padovani18}, we did not find any signature that the $\zeta_2$ can be interpreted as an evolutionary indicator of the star formation activity. We estimate a mean $\zeta_2 = 2.5 \times 10^{-17}$ s$^{-1}$, assuming that the region associated with the \ohhdp emission is comparable to the clump's effective radius in Table~\ref{tab:obsprop}.
    \item We confirm that \X shows a general downward trend with massive clumps evolution, as pointed out by \cite{Giannetti19}. We extend this trend to the most advanced phases of HII regions through upper limit estimations. This new result, together with that of \cite{Giannetti19},  establish the role of this tracer as a chemical clock, and provides a useful reference for future observations of \ohhdp in massive star-forming regions.
    \item We note that the connection between \X and the dynamical parameters of clumps is not as evident as in the case of other quantities linked to the evolution of the clumps (e.g. $L/M$). This result is in agreement with the findings reported by  \cite{Koertgen17} and \cite{Bovino19} suggesting that the deuterium fractionation of H$_3^+$ seems to be driven more by the thermal than by the  dynamical evolution of the high-mass star-forming regions. It is also true that the whole process depends on the balance between the density distribution and the thermal evolution as  very dense fragments will provide a larger degree of CO freeze-out with consequent larger deuterium fractionation (see \citealt{Bovino19}).
\end{enumerate}

In this context, it could be interesting to verify the anti-correlation between \ohhdp and N$_2$D$^+$ pointed out by \cite{Giannetti19} in the IRDC G351.77–0.51, which has not yet been confirmed for the low-mass regime (\citealt{Miettinen20}). This challenging goal, if confirmed and supported by a detailed description of a chemical model in a sample of independent sources, would provide an additional tool to follow the star formation process, especially in the case of the high-mass regime.

\begin{acknowledgements}
     The authors wish to thank an anonymous Referee for his/her
     comments and suggestions that have helped clarify many aspects of this work, and are grateful to M. Padovani, J. Brand and E. Redaelli for fruitful scientific discussions and feedbacks.\\
     This work was partly funded by the Marco Polo program (Universit\`a di Bologna), making it possible for GS to spend three months at the Departamento de Astronom\'ia (Universidad de Concepci\'on) in Concepci\'on; and also partly carried out within the Collaborative Research Council 956, sub-project A6, funded by the Deutsche Forschungsgemeinschaft (DFG). This work was also based on data acquired with the Atacama Pathfinder EXperiment (APEX). APEX is a collaboration between the Max Planck Institute for Radioastronomy, the European Southern Observatory, and the Onsala Space Observatory.\\
     GS wish to thank the GILDAS-team for their support.
     SB is financially supported by CONICYT Fondecyt Iniciaci\'on (project 11170268), CONICYT programa de Astronomia Fondo Quimal 2017 QUIMAL170001, and BASAL Centro de Astrofisica y Tecnologias Afines (CATA) AFB-17002. TC has received financial support from the French State in the framework of the IdEx Université de Bordeaux Investments for the future Program.\\
     This research has made use of the IRAM GILDAS software (\url{http://www.iram.fr/IRAMFR/GILDAS}), the Cologne Database for Molecular Spectroscopy (CDMS), the NASA’s Astrophysics Data System Bibliographic Services (ADS), the Astropy (\citealt{Astropy13, Astropy18}; see also \url{http://www.astropy.org}) and Matplotlib (\citealt{Matplotlib07}).
\end{acknowledgements}

\bibliographystyle{aa} % style aa.bst
\bibliography{mybib_GAL}

\begin{appendix}

\section{Details on the \ohhdp detection limits}
\label{sec:appB}

In this section we report the details of the sources used to calculate upper limits for the \ohhdp detection limits, following the procedure described in Sect.~\ref{sec4.1:column_densities}. All data related to those sources are summarised in Table~\ref{tab:limits}. Columns 2, 3, and 7 respectively give  $T_{dust}$, N(H$_2$), and the evolutionary class; the data are from \cite{Giannetti17_june} and \cite{Urquhart18} (see Tables~\ref{tab:obsprop} and \ref{tab:dervprop}). In column 4 we report the average rms noise of each spectrum provided by GILDAS-class, while columns 5 and 6 {respectively give N(\ohhdp) and \X upper limits (derived as in Sect.~\ref{sec4.1:column_densities})}. The resulting median \X upper limits for each evolutionary class are shown as grey markers in Fig.~\ref{fig:trend}.

    \begin{table*}
        \caption{\label{tab:limits}Summary of the ATLASGAL sources used to compute the \ohhdp detection limits.}
        \setlength{\tabcolsep}{7pt}
    \renewcommand{\arraystretch}{1.1}
    \centering
        \begin{tabular}{l|cccccccccc}
        \hline\hline
    ATLASGAL-ID&$T_{dust}$&N(H$_2$)&rms noise\tablefootmark{c}&N(\ohhdp)\tablefootmark{d}&\X\tablefootmark{d}&Class\\
          &(K)&log$_{10}$(cm$^{-2}$)&(K)&log$_{10}$(cm$^{-2}$)&log$_{10}$(N[$o$-H$_2$D$^+$]/N[H$_2$])&\\
    \hline            
G30.85-0.08\tablefootmark{a}  & 16.7  & 22.72 & 0.07 & $<12.77$ & $<-9.95\pm0.14\:\:$ & 70w\\ 
G320.88-0.40\tablefootmark{a} & 16.8  & 22.69 & 0.09 & $<12.86$ & $<-9.83\pm0.13\:\:$ & 70w\\ 
G337.28+0.01\tablefootmark{a} & 10.7  & 22.92 & 0.09 & $<13.07$ & $<-9.86\pm0.12\:\:$ & 70w\\ 
G338.07+0.01\tablefootmark{a} & 18.5  & 22.51 & 0.07 & $<12.75$ & $<-9.76\pm0.15\:\:$ & 70w\\ 
G338.78+0.48\tablefootmark{a} & 12.2  & 22.85 & 0.07 & $<12.90$ & $<-9.95\pm0.13\:\:$ & 70w\\ 
G351.13+0.77\tablefootmark{a} & 18.6  & 22.49 & 0.06 & $<12.69$ & $<-9.80\pm0.14\:\:$ & 70w\\ 
G351.57+0.76\tablefootmark{a} & 17.0  & 22.67 & 0.03 & $<12.40$ & $<-10.27\pm0.14$ & 70w\\ 
G353.42-0.08\tablefootmark{a} & 17.1  & 22.38 & 0.04 & $<12.49$ & $<-9.89\pm0.14\:\:$ & 70w\\ 
\hline
G08.68-0.37\tablefootmark{a}  & 24.2  & 22.94 & 0.07 & $<12.67$ & $<-10.27\pm0.16$ & IRw\\ 
G10.45-0.02\tablefootmark{a}  & 20.7  & 22.70 & 0.05 & $<12.58$ & $<-10.12\pm0.15$ & IRw\\ 
G18.73-0.23\tablefootmark{a}  & 21.9  & 22.90 & 0.06 & $<12.66$ & $<-10.24\pm0.15$ & IRw\\ 
G18.89-0.47\tablefootmark{a}  & 14.4  & 23.17 & 0.02 & $<12.25$ & $<-10.92\pm0.13$ & IRw\\ 
G23.21-0.38\tablefootmark{a}  & 22.1  & 23.20 & 0.03 & $<12.36$ & $<-10.84\pm0.15$ & IRw\\ 
G24.63+0.17\tablefootmark{a}  & 18.1  & 22.57 & 0.06 & $<12.71$ & $<-9.86\pm0.14\:\:$ & IRw\\ 
G317.87-0.15\tablefootmark{a} & 19.3  & 23.03 & 0.06 & $<12.68$ & $<-10.35\pm0.14$ & IRw\\ 
G318.78-0.14\tablefootmark{a} & 24.9  & 22.60 & 0.07 & $<12.68$ & $<-9.92\pm0.15\:\:$ & IRw\\ 
G331.71+0.60\tablefootmark{a} & 21.0  & 22.88 & 0.05 & $<12.56$ & $<-10.32\pm0.15$ & IRw\\ 
G335.79+0.17\tablefootmark{a} & 24.7  & 23.18 & 0.06 & $<12.61$ & $<-10.57\pm0.16$ & IRw\\ 
G337.26-0.10\tablefootmark{a} & 21.7  & 22.72 & 0.06 & $<12.62$ & $<-10.10\pm0.16$ & IRw\\ 
G338.92+0.56\tablefootmark{a} & 24.2  & 23.48 & 0.07 & $<12.68$ & $<-10.80\pm0.16$ & IRw\\ 
G340.78-0.10\tablefootmark{a} & 26.2  & 22.82 & 0.08 & $<12.70$ & $<-10.12\pm0.16$ & IRw\\ 
G342.48+0.18\tablefootmark{a} & 23.6  & 22.82 & 0.05 & $<12.50$ & $<-10.32\pm0.16$ & IRw\\ 
G343.75-0.16\tablefootmark{a} & 24.3  & 23.30 & 0.06 & $<12.57$ & $<-10.73\pm0.16$ & IRw\\ 
G351.45+0.66\tablefootmark{a} & 21.4  & 23.88 & 0.03 & $<12.34$ & $<-11.54\pm0.16$ & IRw\\ 
\hline
G19.88-0.54\tablefootmark{a}  & 24.2  & 23.12 & 0.03 & $<12.30$ & $<-10.82\pm0.16$ & IRb\\ 
G34.41+0.23\tablefootmark{a}  & 26.1  & 23.25 & 0.05 & $<12.50$ & $<-10.75\pm0.16$ & IRb\\ 
G37.55+0.20\tablefootmark{a}  & 28.4  & 22.73 & 0.05 & $<12.50$ & $<-10.23\pm0.16$ & IRb\\
G53.14+0.07\tablefootmark{a}  & 25.4  & 22.90 & 0.04 & $<12.42$ & $<-10.48\pm0.16$ & IRb\\ 
G332.09-0.42\tablefootmark{a} & 30.8  & 22.94 & 0.05 & $<12.46$ & $<-10.48\pm0.16$ & IRb\\ 
G333.31+0.11\tablefootmark{a} & 25.9  & 22.76 & 0.04 & $<12.44$ & $<-10.32\pm0.16$ & IRb\\ 
G341.22-0.21\tablefootmark{a} & 27.0  & 22.86 & 0.06 & $<12.55$ & $<-10.31\pm0.16$ & IRb\\ 
G351.16+0.70\tablefootmark{a} & 21.9  & 23.67 & 0.06 & $<12.62$ & $<-11.05\pm0.15$ & IRb\\ 
G351.25+0.67\tablefootmark{a} & 32.5  & 23.34 & 0.06 & $<12.56$ & $<-10.78\pm0.16$ & IRb\\ 

\hline
G10.62-0.38\tablefootmark{a}  & 34.5  & 23.57 & 0.08 & $<12.67$ & $<-10.90\pm0.17$ & HII\\ 
G12.81-0.20\tablefootmark{a}  & 35.1  & 23.56 & 0.07 & $<12.60$ & $<-10.96\pm0.17$ & HII\\ 
G14.33-0.64\tablefootmark{b}  & 21.6  & 23.51 & 0.06 & $<12.65$ & $<-10.86\pm0.15$ & HII\\ 
G31.41+0.31\tablefootmark{a}  & 26.3  & 23.57 & 0.06 & $<12.62$ & $<-10.95\pm0.16$ & HII\\ 
G34.26+0.15\tablefootmark{a}  & 31.0  & 23.85 & 0.07 & $<12.66$ & $<-11.19\pm0.16$ & HII\\ 
G34.40+0.23\tablefootmark{a}  & 22.8  & 23.11 & 0.06 & $<12.63$ & $<-10.48\pm0.15$ & HII\\ 
G330.95-0.18\tablefootmark{a} & 33.0  & 23.76 & 0.04 & $<12.37$ & $<-11.39\pm0.16$ & HII\\ 
G333.13-0.43\tablefootmark{a} & 35.2  & 23.40 & 0.05 & $<12.50$ & $<-10.90\pm0.16$ & HII\\ 
G333.28-0.39\tablefootmark{a} & 30.4  & 23.24 & 0.05 & $<12.51$ & $<-10.73\pm0.16$ & HII\\ 
G333.60-0.21\tablefootmark{a} & 41.1  & 23.49 & 0.07 & $<12.60$ & $<-10.89\pm0.17$ & HII\\ 
G337.40-0.40\tablefootmark{a} & 31.8  & 23.36 & 0.07 & $<12.70$ & $<-10.65\pm0.16$ & HII\\ 
G337.70-0.02\tablefootmark{a} & 25.6  & 23.35 & 0.08 & $<12.66$ & $<-10.70\pm0.16$ & HII\\ 
G343.13-0.06\tablefootmark{a} & 30.9  & 23.39 & 0.05 & $<12.48$ & $<-10.91\pm0.16$ & HII\\
    \hline
        \end{tabular}
        \tablefoot{Derived physical properties of the sample of massive clumps used to derive the \ohhdp detection limits described in Sect.~\ref{sec4.1:column_densities}. The sources are separated, from top to bottom, into their evolutionary classes. The error associated with N(H$_2$) is  20\% for each source; \\
        \tablefoottext{a}{$T_{dust}$, N(H$_2$), and Class from  \citet{Giannetti17_june};}
        \tablefoottext{b}{$T_{dust}$, N(H$_2$), and Class from  \citet{Urquhart18};}
        \tablefoottext{c}{The temperatures are reported on the main-beam temperature scale;}
        \tablefoottext{d}{derived in this work.}}
        \end{table*}

\section{Details on linear fit}
\label{sec:appA}

Each data set comprises $N$ measurements $\DD\equiv\{\xk,\dxk,\yk,\dyk\}$, with $k=1,...,N$.  Each point $k$ of a data set corresponds to a different massive star-forming region of the ATLASGAL survey; $\xk$ and $\yk$ are the quantities over which we look for a linear correlation, while $\dxk$ and $\dyk$ are the corresponding errors. We test our data sets against linear correlations of the form $y=mx+q$, where $m$ and $q$ are, respectively, the slope and the normalisation. In our specific case we always consider the $\yk$ as measures of $\logten$\X, while the $\xk$ can be  the target's galactocentric radius, $\dGC$; effective radius, $\Reff$; clump total mass, $\logten \Mclump$; bolometric luminosity, $\logten \Lbol$; dust temperature, $\Tdust$; CO-depletion factor, $f_{\rm D}$; H$_2$ column density, $\logten{\rm H}_2$; luminosity-to-mass ratio, $\logten L/M$; virial factor, $\alpha$; {or} the Mach number, $\mathcal{M}$ (see Tables~\ref{tab:obsprop} and \ref{tab:dervprop} and Sect.~\ref{sec2:sample}).

We assume that the errors $\dxk$ and $\dyk$ follow a Gaussian distribution and we define the model's likelihood $\ln\LL(\btheta|\DD)$, defined by the parameters $\btheta\equiv\{m,q\}$, given the data $\DD$
\begin{equation}\label{for:logl}
    \ln\LL \equiv \frac{1}{2}\sum_{k=1}^{N}\biggl[-\biggl(\frac{\yk-m\xk-q}{\sigmak}\biggr)^2 - \ln(2\pi\sigmak^2)\biggr]
,\end{equation}
where $\sigma = \sqrt{m^2\dxk^2 + \dyk^2}$.

We rely on a Bayesian approach and we run a parameter space search using a Markov {Chain} Monte Carlo (MCMC) algorithm, using uninformative flat priors over the models' free parameters. For each correlation, we run ten chains, each evolved for 10000 steps, using a classical Metropolis-Hasting sampler to sample from the posterior (\citealt{Metropolis53} and \citealt{Hastings70}). All the chains converge quickly towards the highest probability region of the parameter space and we eliminate, as a conservative choice, the first 3000 steps from each chain, considered to be  a reliable burn-in. The remaining steps are used to compute the posterior distribution over the model's free parameters. We test the correlations we obtain from the MCMC also considering  smaller burn-in and we note that our results do not depend significantly on this choice as long as we eliminate at least the first 1000 steps.

To test the performances of our models in fitting data, for each correlation we evaluate the Bayesian evidence $\mathcal{E}$, defined as the average of the likelihood (eq. \ref{for:logl}) under the considered priors. Specifically,
\begin{equation}
    \mathcal{E} = \int \LL(\btheta|\DD)P(\btheta)\:\dd\btheta,
\end{equation}
where  $P$ denotes the priors that we  used. Afterwards, we compute the Bayesian evidence after fitting each data set with a model where the slope has been fixed to zero. In $\mathcal{M}_1$ and $\mathcal{M}_2$, which are respectively the models with $m=0$ and the general model with two free parameters, the Bayes factor,
\begin{equation}
 \mathcal{B}_{2,1} = \frac{\mathcal{E}_2}{\mathcal{E}_1},
\end{equation}
describes whether     $\mathcal{M}_1$ provides a better description of the data than $\mathcal{M}_2$ ($\mathcal{B}_{2,1}<1$), or the other way around ($\mathcal{B}_{2,1}>1$). The Bayesian evidence is computed using the software package PyMultiNest \citep{Buchner14}, as implemented in its publicly available version.

\begin{figure}
\centering
{\includegraphics[width=0.49\textwidth]{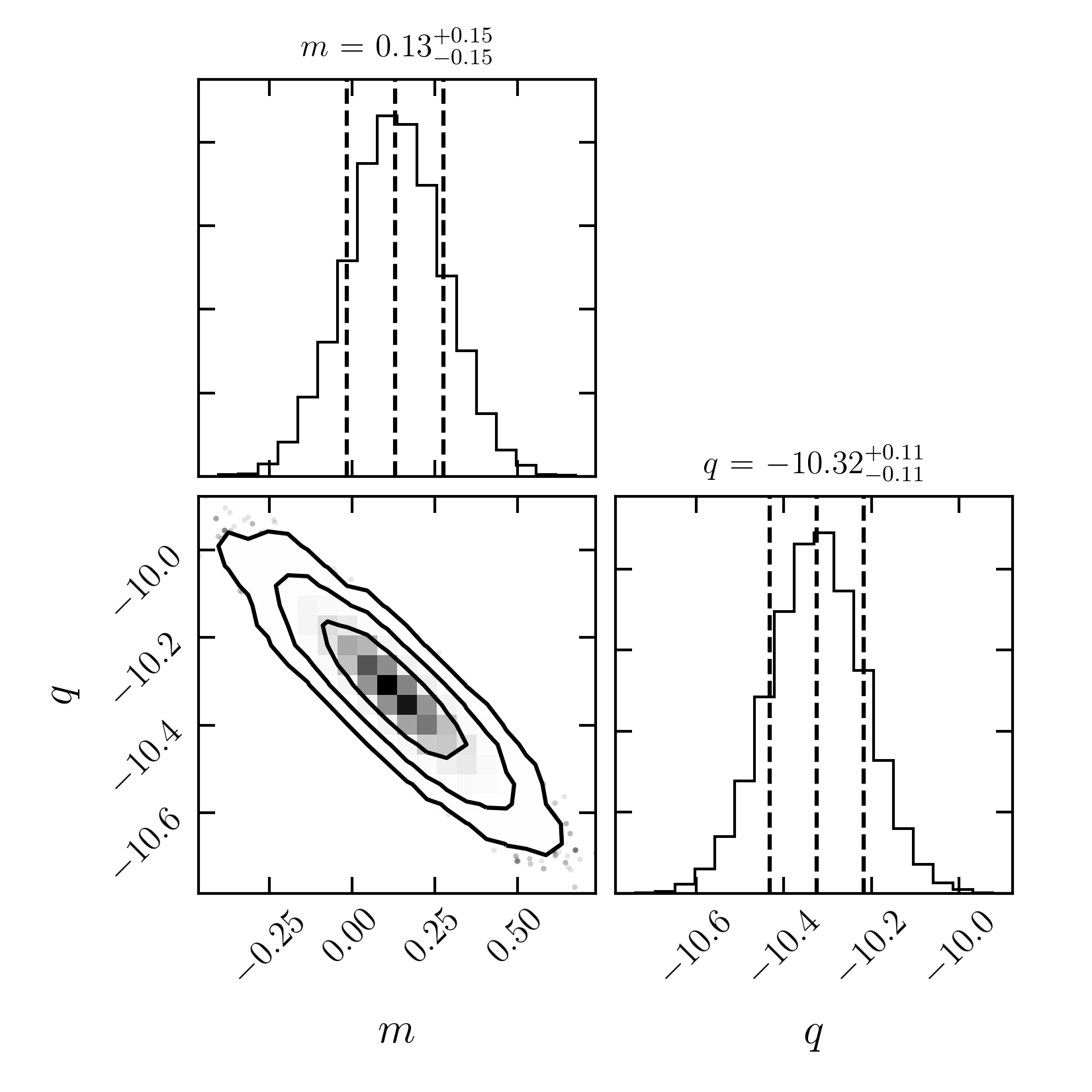}}
\caption{Example of {one- and two-dimensional marginalized posterior distributions of the free parameters in our model.} The black curves in the two-dimensional distribution correspond to regions enclosing  68\%, 95\%, and 99.7\% of the total probability, while the black vertical lines in the one-dimensional marginalised distributions correspond to the $16{\rm th}$, $50{\rm th}$, and $84{\rm th}$ percentiles, used to estimate the uncertainties on the models' parameters.}
\label{fig:posteriors}
\end{figure}

Figure~\ref{fig:posteriors} shows an example of the posterior distribution obtained for each correlation. The black curves in the two-dimensional distributions indicate regions enclosing   68\%, 95\%, and 99.7\% of the total probability (i.e. 1$\sigma$, 2$\sigma$, and 3$\sigma$, respectively). The black vertical lines in the one-dimensional marginalised distributions  instead indicate the 16${\rm th}$, $50{\rm th}$, and $84{\rm th}$ percentiles, used to estimate the 1$\sigma$ error bars. To derive the 2$\sigma$ error bars we use instead the interval between the 5${\rm th}$ and 95${\rm th}$ percentiles of the corresponding one-dimensional marginalised posterior distributions.

\end{appendix}

\end{document}